\documentclass[a4paper,12pt]{article}
\usepackage[top=2cm,bottom=2cm,left=2.5cm,right=2.5cm]{geometry}
\usepackage{amsmath,amssymb}
\usepackage[colorlinks=true]{hyperref}
\usepackage{graphicx}
\usepackage[dvipsnames]{xcolor}
\usepackage{float}
\usepackage{sidecap}
\usepackage{setspace}

\newcommand{\la}{\lambda}

\newcommand{\bs}{\boldsymbol}
\newcommand{\rmd}{\mathrm{d}}
\newcommand{\aver}[1]{\left\langle #1 \right\rangle}

\DeclareMathOperator{\re}{Re}
\DeclareMathOperator{\im}{Im}

\newtheorem{remark}{Remark}[section]

\title{Riemann problem for polychromatic soliton gases:\\ a testbed for the spectral kinetic theory}
\author{Thibault Congy, Henry T. Carr,  Giacomo Roberti \& Gennady A. El}
\date{}

\begin{document}

\maketitle

\begin{abstract}
We use Riemann problem for soliton gas as a benchmark  for a detailed numerical validation of the spectral kinetic theory for the Korteweg-de Vries (KdV) and the focusing nonlinear Schr\"odinger (fNLS) equations.  We construct weak solutions to the kinetic equation for soliton gas describing  collision of two dense ``polychromatic'' soliton gases composed of a finite number of   ``monochromatic''  components, each consisting of solitons with nearly identical spectral parameters of the scattering operator in the  Lax pair.  The interaction between the gas components plays the key role in the emergent, large-scale hydrodynamic evolution. We then use the solutions of the spectral kinetic equation to evaluate macroscopic physical observables in KdV and fNLS soliton gases and compare them with the respective ensemble averages  extracted from the ``exact'' soliton gas numerical solutions of the KdV and fNLS equations. To numerically synthesise dense  polychromatic soliton gases we develop a new method which combines recent advances in the spectral theory of the so-called soliton condensates and the effective algorithms for the numerical realisation of $n$-soliton solutions with large $n$.
	\end{abstract}

\section{Introduction and problem description}
 Soliton gases (SGs) in integrable systems  have recently attracted significant attention both as the fundamental  structures underlying  important physical phenomena such as spontaneous modulational instability and the rogue wave formation and as a  promising mathematical framework for some novel problems at the interface of integrability and randomness.  
 
 The  concept of SG  was originally introduced  by V.E.  Zakharov in  1971   \cite{zakharov1971kinetic} by considering  large ensembles of well-separated  solitons of the Korteweg-de Vries (KdV) equation randomly distributed on $\mathbb{R}$ with certain density and also having some given distribution over amplitudes. Given the integrable nature of the KdV equation, the most natural description of solitons and their ensembles is achieved within the inverse scattering transform (IST) formalism where each soliton in an $n$-soliton solution corresponds to a point $\zeta_i <0$, $i=1, \dots, n$, of  discrete spectrum  of the Lax (linear Schr\"odinger) operator  associated with the KdV equation \cite{ablowitz_solitons_1981, novikov_theory_1984}.  In a SG, the discrete spectrum points are assumed to be distributed with some density on a given compact interval $\Gamma$ so one introduces a continuous spectral variable, $\zeta_i \to \zeta=-\eta^2$, where $\eta>0$.  The key quantity in the SG theory is the density of states (DOS) $f(\eta;x,t)$ defined as the number of soliton states  per unit interval of the spectral parameter $\eta $ and per unit interval of space $x$. For a spatially uniform (homogeneous) SG $f \equiv f(\eta)$.  The  characteristic scales of $x,t$-variations of the DOS in a weakly non-homogeneous SG are much larger than the characteristic spatio-temporal scales associated with individual solitons.
 
The motion of a chosen ``tracer'' soliton in a SG  is affected by its collisions with other solitons, each collision being accompanied  the phase/position shift, resulting in the effective (average) soliton velocity $s(\eta;x,t)$ over sufficiently large distances of propagation being different from the  ``free'' soliton velocity in a ``vacuum'' $s_0(\eta)$.  For a rarefied, or diluted, SG the effective  velocity adjustment represents a small correction which is readily evaluated in terms of the DOS  $f(\eta;x,t)$, the ``soliton counting function'' determining the collision rate in the gas. The expression for the effective velocity $s(\eta;x,t)$ of a tracer soliton,  combined with the continuity equation for the DOS (a consequence of the isospectrality of integrable evolution) form an approximate kinetic equation describing macroscopic evolution of a weakly non-homogeneous  rarefied SG  \cite{zakharov1971kinetic}:
\begin{equation}\label{kin_zakh}
\begin{split}
&\partial_tf(\eta;x,t) + \partial_x[f(\eta;x,t)s(\eta;x,t)]=0, \\
&s(\eta;x,t) \approx s_0(\eta) +  \int_{\Gamma} G(\eta, \mu) f(\mu; x,t) [s_0(\eta) -  s_0(\mu)]\rmd \mu \, .
\end{split}
\end{equation}
Here $s_0(\eta) = 4 \eta^2$ is the velocity of a free KdV soliton with the spectral parameter $\eta$ and $\Gamma$ is the spectral support of the DOS (in the original Zakharov paper it was assumed that $\Gamma = [0, \infty)$ although subsequent developments of the theory use a compact support $\Gamma \subset \mathbb{R}^+$ consistent with the boundedness of the requisite KdV solutions, see \cite{el_soliton_2021}). The integral kernel $G(\eta, \mu)=\frac{1}{\eta}\ln \left|
\frac{\eta+\mu}{\eta-\mu} \right| $ is the well-known expression for the phase/position shift in the KdV two-soliton collision \cite{novikov_theory_1984}. The condition of  the applicability of the rarefied SG approximation  \eqref{kin_zakh} is  $\kappa \equiv \int_{\Gamma} f(\eta) \rmd \eta \ll \eta_0$, where $\kappa$ is the gas' spatial density and $\eta_0^{-1}$ is the width of a typical soliton within the gas with the  spectral parameter $\eta_0$. 

The above construction of the kinetic equation for rarefied SG is applicable to other integrable dispersive equations---one just needs to insert appropriate expressions for the free soliton velocity $s_0(\eta)$ and the phase shift kernel $G(\eta, \mu)$. We also note that in the KdV equation there is no distinction between the  soliton phase and its position  while, e.g.,  in the focusing nonlinear Schr\"odinger (fNLS) equation they are independent quantities with different shifts in two-soliton collisions \cite{novikov_theory_1984}.  Nevertheless, with a slight abuse of terminology we shall be using the conventionally accepted term ``phase shift'' for the position shift in both KdV and fNLS cases. 

In the  statistical-mechanics construction of a rarefied SG solitons are viewed as point-like quasi-particles exhibiting short-range   interactions accompanied by  well-defined scattering shifts.   In a dense gas, however, the individual solitons exhibit significant overlap and are continuously involved in a strong nonlinear interaction with each other. Thus, in a dense gas the ``particle'' interpretation of individual solitons  becomes less  relevant and the wave aspect of the collective soliton dynamics comes to the fore rendering the phase-shift arguments behind equation \eqref{kin_zakh} inapplicable. Indeed, a consistent generalisation of Zakharov's kinetic equation for  KdV solitons  to the case of  a dense SG  has been achieved in \cite{el_thermodynamic_2003} in the framework of the nonlinear wave modulation (Whitham) theory \cite{whitham_linear_1999}.  It was proposed in \cite{el_thermodynamic_2003}  that the KdV SG can be modelled by the  thermodynamic type solitonic limit of the  multiphase, finite-gap KdV solutions and their modulations \cite{flaschka_multiphase_1980}. The resulting full spectral kinetic equation  has the form of a nonlinear integro-differential equation 
\begin{subequations}\label{dense_kin}
\begin{align}
& \partial_t f(\eta;x,t) + \partial_x [f(\eta;x,t)s(\eta;x,t)] = 0, \\
& s(\eta;x,t) = s_0(\eta) +  \int_{\Gamma} G(\eta, \mu) f(\mu;x,t) [s(\eta;x,t) -  s(\mu;x,t)]\rmd \mu \, .
\end{align}
\end{subequations}

The continuity equation (\ref{dense_kin}a) for the DOS   is same as in the Zakharov model \eqref{kin_zakh},  but the approximate expression  for the effective velocity $s(\eta; x,t)$ in \eqref{kin_zakh} is now replaced by a linear integral equation -- the equation of state (\ref{dense_kin}b). Quite remarkably, this equation of state  implies that the net effect of soliton interactions  on the average soliton velocity in a dense gas is  {\it as if} the solitons were  localised quasi-particles engaged in the short-range interactions characterised by the  two-particle factorisable scattering (no multi-particle effects), despite the fact that the very concept of the two-soliton scattering/phase shift is based on the asymptotic analysis of $n$-soliton solutions that are well separated before and after the interaction  \cite{novikov_theory_1984}. This observation enabled the general phenomenological construction of the kinetic equation for dense SGs  in \cite{el_kinetic_2005} with application to the fNLS equation. This construction was recently used  in \cite{congy_soliton_2021} for the the description of bidirectional SGs of the defocusing NLS and the Kaup-Boussinesq equations.  In a different, but related, context  a similar  construction of the kinetic equation has been successfully used in generalised hydrodynamics (GHD) of integrable quantum and classical many-body systems \cite{doyon_lecture_2020, spohn_hydrodynamic_2023}.
In the present, nonlinear wave, context  the formal construction of \eqref{dense_kin} for the fNLS SG was recently justified by the spectral thermodynamic limit  derivation in  \cite{el_spectral_2020}, supporting the fundamental concept that the kinetic equation \eqref{dense_kin} provides in fact a universal description of SGs in integrable systems (see also \cite{spohn_hydrodynamic_2023} for the GHD context). We note that  the spectral parameter in the Zakharov-Shabat scattering problem associated with the fNLS equation within the IST formalism \cite{shabat_exact_1972} lives in the complex plane so the integral in the equation of state (\ref{dense_kin}b) is generally taken over some two-dimensional compact domain $\Lambda \in \mathbb{C}^{+}$. The kinetic equation \eqref{dense_kin} is subject to an important constraint 
\begin{equation} \label{dos_constr}
\int_{\Gamma} G(\eta, \mu) f(\mu) \rmd \mu  <1
\end{equation}
imposing restriction on the admissible SG density and related to the so-called {\it soliton condensate} limit \cite{el_spectral_2020, congy_dispersive_2023} (see  Appendix \ref{sec:condensate}).

The derivation of the kinetic equation \eqref{dense_kin} involves a number of  specific assumptions about the asymptotic structure of the  IST spectrum and the spatio-temporal scales separation, raising natural questions about its applicability  to the description of `real' SGs, i.e. random soliton ensembles that can be practically realised in numerical simulations and physical experiments.  For that, a comparison of some benchmark exact solutions to the kinetic equation with the appropriate parameters extracted from direct numerical simulations of SGs would be highly desirable. Such a comparison is the main goal of this paper.

\medskip

 An important class of exact solutions to the kinetic equation \eqref{dense_kin} is obtained by introducing multicomponent delta-function ansatz for the DOS, 
 \begin{equation} \label{delta-ansatz0}
 f(\eta; x,t)= \sum _{j=1}^M w_j(x,t) \delta(\eta - \eta_j),
 \end{equation}
which reduces  kinetic equation \eqref{dense_kin} to a system of $M$ hyperbolic hydrodynamic conservation laws for $w_i(x,t)$ that can be solved analytically \cite{el_kinetic_2011}. Each component in such a gas, considered separately, can be viewed as a spectrally ``monochromatic'' SG (of course, the monochromatic  delta-function DOS  is a mathematical idealisation intended to describe a SG with a narrow distribution of soliton eigenvalues around a given spectral point $\zeta_j = -\eta_j^2$). The interaction of different components in a  ``polychromatic''  gas results in non-trivial hydrodynamics which was originally introduced for a two-component fNLS SG in \cite{el_kinetic_2005} and then studied  in detail for multi-component case for a generic kinetic equation  \eqref{dense_kin} under some conditions for the structure of the phase shift scattering matrix $G(\eta_i, \eta_j)$, which are typically satisfied for soliton equations. It was shown in \cite{el_kinetic_2011} that the resulting system of $M$  hydrodynamic conservation laws for polychromatic SGs is a linearly degenerate integrable system for any $M$ (the integrability is understood in the sense of the generalised hodograph transform \cite{tsarev_poisson_1985}). A further detailed study of integrability properties of multi-component hydrodynamic reductions of the kinetic equation in the GHD context  was undertaken in \cite{bulchandani_classical_2017}.
 Very recently  the integrability theory of polychromatic hydrodynamic reductions has been extended to a more general class of ``non-isospectral''  reductions introduced in \cite{pavlov_generalized_2012} and exhibiting the Jordan block structure, see \cite{ferapontov_kinetic_2022, vergallo_hamiltonian_2023, vergallo_hamiltonian_2024}. 
  
  \medskip
Comparisons of  analytical solutions to the kinetic equation with direct  numerical simulations of SGs for several dispersive hydrodynamic models (KdV, defocusing NLS and Kaup-Bussinesq equations)  have been previously performed in \cite{carbone_macroscopic_2016, congy_soliton_2021}  for the Riemann problems describing the collision of two relatively rarefied monochromatic  SGs,  with a very good agreement observed between the analytical and numerical results.   We also mention two recent experimental works \cite{suret_soliton_2023} and \cite{fache_interaction_2023}  where some aspects of the kinetic SG theory for the focusing NLS equation have been successfully verified in the fibre optics and deep-water tank experiments respectively.   

While the above comparison results induce some confidence in the applicability of the SG kinetic theory, a more systematic study comparing analytical solutions of the ``macroscopic'' kinetic equation with direct numerical simulations of the ``microscopic'' models such the KdV or NLS equations for a broad range of parameters (particularly for higher SG densities, where the rarefied gas theory is not applicable)  is necessary. In this paper, we undertake a detailed comparison of the exact solutions to the Riemann problem for polychromatic reductions of the kinetic equation with the ensemble averages of numerically synthesised dense SGs for the KdV and fNLS equations using the recently developed highly efficient algorithms for the numerical construction of  $n$-soliton solutions with large $n$  based on the Darboux transformation and the implementation of  high precision arithmetic routine to overcome the numerical accuracy problems \cite{Gelash:18}. 

One of the major challenges encountered in the realisation of the above programme is the numerical implementation of dense, spatially uniform monochromatic and polychromatic gases. It turns out that in the numerical implementation of the delta-function DOS \eqref{delta-ansatz0}  the `natural' distributions of the spectral eigenvalues like, e.g.,  uniform or Gaussian distributions centred at some given distinct spectral points do not allow one to achieve the desired SG densities while preserving spatial uniformity of the SG. To overcome this obstacle, in this paper we have taken advantage of the recently developed spectral theory of soliton condensates \cite{congy_dispersive_2023} suggesting special DOS configurations that enabled us to achieve  higher  polychromatic SG densities  than those achievable by applying ad hoc narrow  DOS distributions. We stress that our numerical results are based on the  implementations of exact $n$-soliton solutions and do not involve demanding direct KdV or NLS simulations, via e.g. finite difference or spectral methods, that incur inevitable propagation errors.

\medskip
The paper is organised as follows. In Sec. \ref{sec:key_ingred} we present the main ingredients of the SG kinetic theory  for the KdV and fNLS equations. In Sec. \ref{sec:polychromatic} we introduce polychromatic hydrodynamic reductions of the SG kinetic equation and present  analytical expressions for some important ensemble averages of the nonlinear wave fields in the KdV and fNLS polychromatic gases. In Sec. \ref{sec:riemann} we derive weak solutions to the Riemann problem  for polychromatic hydrodynamic reductions from Sec. \ref{sec:polychromatic}. Sec. \ref{sec:synthesis} develops  an IST based algorithm for the synthesis of dense polychromatic SGs for the KdV and fNLS equations. In Sec. \ref{sec:num2}, which contains the main results of the paper, we present numerical implementations of the Riemann problem for the KdV and fNLS polychromatic SGs and perform detailed comparisons of the analytical predictions of the kinetic theory with the corresponding parameters of the numerical solutions. Sec. \ref{sec:concl} contains conclusions and outlook of future directions. Finally, in the Appendix  we present the supplemental material concerning technical details of the numerical implementation of SGs.

\section{KdV and fNLS SGs: key ingredients from the kinetic theory}
\label{sec:key_ingred}
The general form of the kinetic equation for a dense SG is given by  the integro-differential equation  \eqref{dense_kin} which consists of the universal continuity equation (\ref{dense_kin}a) for the DOS $f(\eta; x,t)$  and the equation of state (\ref{dense_kin}b) for the effective transport velocity $s(\eta;x,t)$. The kinetic equation contains  two  system-specific ingredients: the free soliton velocity $s_0(\eta)$ and the two-soliton interaction kernel $G(\eta, \mu)$. In this section we present the specific form of  the equations of state  for the KdV and fNLS SGs and also  present explicit expressions for some moments (ensemble averages) of the wave field for the KdV and fNLS SGs in terms of the respective DOS' which will  be used later for the comparison with numerical simulations.

We consider the KdV equation in the form
\begin{equation}
	\label{eq:kdv}
	u_t+6uu_x+u_{xxx}=0.
\end{equation}
The soliton solution of eq.~\eqref{eq:kdv} corresponding to the discrete spectrum  eigenvalue $\zeta=-\eta^2$, $\eta>0$, has the form
\begin{equation}
u_{\rm s}(x,t; \eta) = 2  \eta^2 \hbox{sech}^2 [ \eta(x-4 \eta^2 t - x^0)],
\end{equation}
where $x^0$ is the initial soliton position.

The equation of state (\ref{dense_kin}b)  for the KdV SG has the form:
\begin{equation} \label{eq_state_kdv}
s(\eta;x,t) = 4\eta^2 + \frac{1}{\eta}  \int_{\Gamma}  \ln \left|
\frac{\eta+\mu}{\eta-\mu} \right| f(\mu;x,t) [s(\eta;x,t) -  s(\mu;x,t)]\rmd \mu ,
\end{equation}
where  $\Gamma \subset \mathbb{R}^{+}$ is the spectral support of DOS. 
Equation 
\eqref{eq_state_kdv} is subject to the general constraint \eqref{dos_constr} which assumes the form
\begin{equation} \label{constr_kdv}
\frac{1}{\eta}\int_{\Gamma}  \ln \left|
\frac{\eta+\mu}{\eta-\mu} \right| f(\mu;x,t) \rmd \mu <  1 \, .
\end{equation}
The two first moments (ensemble averages) of the random wave field in the KdV SG are evaluated in terms of the DOS as \cite{el_thermodynamic_2003, el_soliton_2021}
\begin{equation}\label{mom_kdv}
	\aver{u} = 4 \int_{\Gamma} \eta f(\eta) \rmd\eta,\quad
	\aver{u^2} = \frac{16}{3} \int _{\Gamma} \eta^3 f(\eta) \rmd\eta \, .
\end{equation}

For the fNLS equation
\begin{align}
\label{eq:nls}
i\psi_t + \psi_{xx} + 2 |\psi|^2 \psi =0, \quad \psi \in \mathbb{C},
 \end{align}
 the discrete spectrum  in the associated linear Zakharov-Shabat scattering problem \cite{shabat_exact_1972} is complex valued, so  $\eta \in \mathbb{R}^+$ in the kinetic equation \eqref{dense_kin} should be replaced with $z  \in \mathbb{C}^{+}\setminus i\mathbb{R}^{+}$.
 The single-soliton solution of the fNLS equation \eqref{eq:nls} corresponding to the discrete spectral eigenvalue $z = \xi+ i \eta$ and its c.c. has the form
  \begin{equation}\label{fnls_soliton1}
  \psi_{\rm s} (x,t)= 2 \eta \, \hbox{sech}[2\eta (x+4\xi  t-x^0)]   e^{-2i[\xi x + 2(\xi^2-\eta^2)t]+i\phi^0} ,
   \end{equation}
where  $x^0$ is the initial soliton's position and and  $\phi^0$  the initial phase. Importantly, unlike in the KdV solitons, the amplitude $\im z$ and the velocity $-4 \re z$ are two independent parameters in the fNLS soliton solution.

 The equation of state  for the fNLS SG is given by
 \begin{equation} \label{eq_st_nls}
 s(z; x, t) = -4 \re z  + 
\frac{1}{\im z}   \iint \limits_{\Lambda}  \ln \left|\frac{y- z^*}{y-z}\right| [s(y; x, t) - s(y; x, t)] f(y; x, t) \rmd  (\re y) \rmd (\im y), 
 \end{equation}
where  $y,z  \in \Lambda \subset \mathbb{C}^{+} \setminus i\mathbb{R}^{+}$ and $z^*$ is the complex conjugate of $z$. The DOS spectral support $\Lambda$ is generally a 2D compact  set in the upper half of the complex plane. The special case $z \in i \mathbb{R}^{+}$ corresponds to the so-called bound state SGs in which all solitons are at rest \cite{el_spectral_2020}. 

The constraint \eqref{dos_constr} applied to the fNLS SG assumes the form:
\begin{equation} \label{constr_nls}
\frac{1}{ \im z}   \iint \limits_{\Lambda}  \ln \left|\frac{y- z^*}{y-z}\right|  f(y; x, t)  \rmd  (\re y) \rmd  (\im y) <1,
\end{equation}
which reduces to the KdV inequality \eqref{constr_kdv} for the bound state SG with $ \re z =0$ and $\im z = \eta$.

The expressions for the ensemble averages of the conserved densities  and the corresponding average flux densities  for the fNLS SG  were derived in \cite{tovbis_periodic_2022}. In particular, one has
\begin{align}\label{mom_nls}
\begin{split}
	&\aver{|\psi|^2} = \iint \limits_{\Lambda} 4 \im z f(z) \rmd  (\re z) \rmd  (\im z)  ,\\
   &\aver{|\psi|^4}
   = \iint \limits_{\Lambda} \im \left(-\frac{32}{3} z^3-4 z^2 s(z) \right) f(z) \rmd  (\re z) \rmd  (\im z).
   \end{split}
 \end{align}
Note that the expression for $\aver{|\psi|^4}$ in \eqref{mom_nls} is obtained by manipulating the averages of the conserved density $p_3 = |\psi|^4 -
  |\psi_x|^2$ and the flux
  $q_2 = |\psi|^4 -
  2|\psi_x|^2$, see  \cite{congy_statistics_2023}.

\section{Polychromatic SGs: hydrodynamic reductions of the kinetic equation}
\label{sec:polychromatic}

Although the existence of solutions to the  integral equations of state \eqref{eq_state_kdv}, \eqref{eq_st_nls}  has been established \cite{kuijlaars_minimal_2021}, such solutions at present are only available in several special cases. We introduce the notation:
\begin{align}
\label{eq:Gkdv}
& s_{0{\rm KdV}} (\eta) = 4\eta^2,\quad G_{\rm KdV}(\eta, \mu)\equiv \frac{1}{\eta}\ln \left|
\frac{\eta+\mu}{\eta-\mu} \right|,\quad \eta,\mu \in \Gamma \subset \mathbb{R}^{+},\\
\label{eq:Gnls}
& s_{0{\rm fNLS}}(z) = -4\re z,\quad G_{\rm fNLS}(z, y)= \frac{1}{\im z} \ln \left|\frac{y-z^*}{y-z}\right|,\quad  y,z \in \Lambda \subset \mathbb{C}^{+}.
\end{align}
It was shown in \cite{el_kinetic_2011} that  the  kinetic equation \eqref{dense_kin} greatly simplifies if discretisation of the DOS with respect to the soliton spectral parameter is admissible. We adopt this simplification in the following, and consider the SGs that are composed of a finite number of spectrally distinct  components, termed monochromatic, or cold, components. In the following the kinetic description is written with the notations for the KdV SG; the simple substitution $\eta \in \Gamma \to z \in \Lambda$ and $G_{\rm KdV} \to G_{\rm fNLS}$ yields the corresponding description for the fNLS SG.
We consider the DOS in the form of a linear combination of the Dirac delta-functions centered at distinct spectral points $\eta_j \in \Gamma$, 
\begin{equation}\label{u_delta1}
f(\eta; x, t) = \sum \limits_{j=1}^{M} w_j (x,t)\delta(\eta - \eta_j),
\end{equation}
where $w_j(x,t)>0$ are the components' spatial densities, and  
$\{\eta_{j}\}_{j=1}^M \subset \Gamma$, $(\eta_j \ne \eta_k \iff  j \ne k)$.  We shall call  SG with the DOS \eqref{u_delta1} a {\it polychromatic} SG. As pointed out in \cite{el_kinetic_2011, el_spectral_2020}, the multicomponent ansatz \eqref{u_delta1} is a mathematical idealisation; physically one would replace the $\delta$ functions by narrow distributions around the spectral points $\eta_j$. 
Substitution of  ansatz \eqref{u_delta1} into the kinetic equation  \eqref{dense_kin}  reduces it to a system of hydrodynamic conservation laws
\begin{equation}\label{cont_1}
	(w_j)_{t}+(w_{j}s_{j})_{x} =0,\qquad j=1,\dots,M,  
\end{equation}
where the component densities $w_j(x,t)$ and the transport velocities $s_{j}(x,t)\equiv s(\eta_j; x, t)$ are related 
algebraically:
\begin{equation}\label{s_alg}
	\bigg(1-\sum \limits_{\substack{k=1\\ k \ne j}}^M G_{jk} w_k\bigg)s_{j}+ \sum \limits_{\substack{k=1\\ k \ne j}}^M G_{jk} w_k s_{k} = s_{0j}, \quad j=1, \dots, M.
\end{equation}
Here we used the notation
\begin{equation}
s_{0j} \equiv s_{0{\rm KdV}}(\eta_{j}) , \quad G_{jk} \equiv G_{\rm KdV}(\eta_{j}, \eta_{k}), \quad j \ne k.  
\end{equation}
Note that the condition \eqref{constr_kdv} implies that the densities $w_j$ are constrained by
\begin{equation}
	\label{eq:constr}
	\sum \limits_{\substack{k \ne j}} G_{jk} w_k <1,\quad j=1, \dots, M.\end{equation}
It was shown in \cite{el_kinetic_2011} that, under some natural constraints for the interaction kernel $G(\eta, \mu)$ (satisfied for both KdV and fNLS) the conservation law system \eqref{cont_1}, \eqref{s_alg} can be reduced to the Riemann invariant form for any $M \in \mathbb{N}$ and represents a linearly degenerate integrable system \cite{ferapontov_integration_1991} so that its general solutions can be obtained via the generalised hodograph method \cite{tsarev_poisson_1985}. 
Linear degeneracy implies that the characteristic velocities of  the system \eqref{cont_1}, \eqref{s_alg} coincide with the component transport velocities $s_i$ \cite{pavlov_hamiltonian_1987}. We emphasise that the  system \eqref{cont_1}, \eqref{s_alg}  is a universal hydrodynamic reduction of the kinetic equation \eqref{dense_kin} applicable to both KdV and fNLS SGs.

\medskip
For $M=2$ system \eqref{s_alg} can be solved  to give explicit expressions for $s_{1,2}(w_1, w_2)$:
\begin{equation} \label{s_12}
s_1= s_{01} - \frac{ G_{12} w_2 (s_{02}-s_{01})}{1-(G_{12} w_2+ G_{21} w_1)},
\quad s_{2}= s_{02}+\frac{ G_{21} w_1 (s_{02} - s_{01})}{1-(G_{12} w_2+ G_{21} w_1)}. 
\end{equation}
It is important to stress that the polychromatic reductions are  subject to the constraint \eqref{eq:constr}  which implies  that the denominators in \eqref{s_12} must be positive.  
Note that it follows from \eqref{s_12} that
\begin{equation}
	\label{eq:order}
	s_2-s_1 = \frac{s_{02}-s_{01}}{1-(G_{12}w_2+G_{21}w_1)}, 
\end{equation}
so 
 $s_{01}<s_{02}$ implies $s_1<s_2$.

Substituting the polychromatic ansatz \eqref{u_delta1} in the expressions \eqref{mom_kdv}, \eqref{mom_nls} for the KdV and fNLS SG first moments we obtain
\begin{equation}
	\label{eq:averu}
	\aver{u} = \sum_{j=1}^M 4 \eta_j w_j,\quad
	\aver{u^2} = \sum_{j=1}^M \frac{16}{3} \eta_j^3 w_j,
\end{equation}
for the KdV SG and
\begin{align} \label{mom_poly_NLS}
	\aver{|\psi|^2} = \sum_{j=1}^M 4 \im z_j \, w_j,\quad
   \aver{|\psi|^4}
   = -\sum_{j=1}^M \left(\frac{32}{3} \im(z_j^3)+4\im(z_j^2) s_j\right) w_j.
 \end{align}
 for the fNLS SG. Using $z_j = \xi_j + i \eta_j$  expressions \eqref{mom_poly_NLS} reduce to 
  \begin{align}\label{mom_poly_NLS_bound}
	\aver{|\psi|^2} = \sum_{j=1}^M 4 \eta_j w_j,\quad
   \aver{|\psi|^4} 
   =\sum_{j=1}^M  \left(\frac{32}{3} \eta_j^3-32\xi_j^2 \eta_j - 8\xi_j \eta_j s_j \right)w_j.
 \end{align}

\section{Riemann problem for polychromatic gases}
\label{sec:riemann}

Having obtained the hydrodynamic description of polychromatic SG  we  consider the Riemann problem that plays the fundamental role in classical gas  and fluid dynamics \cite{landau_fluid_1987, smoller_shock_1994}. 
The classical Riemann problem consists in finding solution to a system of hyperbolic conservation laws subject to piecewise constant initial conditions exhibiting discontinuity at $x=0$.   The distribution solution of the Riemann problem generally consists of a combination of constant states, simple (rarefaction) waves and strong discontinuities (shocks or contact discontinuities) \cite{lax_hyperbolic_1973}. The discontinuous solutions satisfy the Rankine-Hugoniot jump conditions.
If a hyperbolic system is linearly degenerate it does not support nonlinear simple waves and classical shocks, and the solution of Riemann problem contains only constant states and contact discontinuities \cite{rozhdestvenskii_systems_1983}. Here, following \cite{carbone_macroscopic_2016}, \cite{congy_soliton_2021}, we present such solutions for different types of SGs.

The Riemann problem for polychromatic gas is defined by 
\begin{align}
	\label{eq:F}
	&(w_j)_t + (w_j s_j(\bs w) )_x = 0,\quad j=1, \dots, M,\\
	\label{eq:init}
	&w_j(x,0) =
	\begin{cases}
		w_j^{(-)}, &\text{if } x<0,\\[2mm]
		w_j^{(+)}, &\text{if } x> 0,
	\end{cases}
\end{align}
where the dependence of $s_j$ on the densities $\bs w =(w_1,\dots,w_M)$ is given by the resolution of the linear system \eqref{s_alg}.
Due to the scale invariance of the problem (the kinetic
equation \eqref{eq:F} and the initial condition \eqref{eq:init} are
both invariant with respect to the scaling transformation $x \to \alpha x$,
$t \to \alpha t$), the solution is a self-similar distribution ${\bs
	w}(x/t)$. Because of  linear degeneracy of the hydrodynamic 
system \eqref{eq:F} the only admissible similarity solutions are constant states
separated by propagating contact discontinuities, cf. for
instance \cite{rozhdestvenskii_systems_1983}. Discontinuous weak
solutions are physically admissible here since the kinetic equation
describes the conservation of the number of solitons within any given
spectral interval, and the Rankine-Hugoniot type conditions can be imposed
to ensure the conservation of the number of solitons across the
discontinuities. As a result, the solution of the Riemann problem \eqref{eq:F}, \eqref{eq:init} for each component $w_j(x,t)$ is composed of
$M+1$ constant states, or plateaus, separated by $M$ discontinuities
(see e.g. \cite{lax_hyperbolic_1973}):
\begin{equation}
	\label{eq:sol}
	w_j(x,t) =
	\begin{cases}
		w_j^{(1)}  = w_j^{(-)}, &x/t<c_1,\\
		\dots\\
		w_j^{(k)}, & c_{k-1} < x/t < c_k,\\
		\dots\\
		w_j^{(M+1)} = w_j^{(+)}, &c_M < x/t,
	\end{cases}\quad
\end{equation}
where the lower index $j$ indicates the $j$-th component of the vector
$\bs{w}$, and the upper index $(k)$ is the
index of the plateau. The contact discontinuities propagate at the characteristic
velocities so that \cite{lax_hyperbolic_1973}
\begin{equation}
	\label{eq:z}
	c_j =s_{j} \left(\bs w^{(j)}\right) = s_{j} \left(\bs w^{(j+1)} \right),
\end{equation}
where the $M(M-1)$ unknown plateaus' values $w_j^{(k)}$ are obtained
from the Rankine-Hugoniot jump conditions \cite{whitham_linear_1999, smoller_shock_1994}:
\begin{align}
	\label{eq:RH}
	-c_k \Big[ w_j^{(k+1)} -w_j^{(k)}\Big] + \Big[
	s_{j}\left(\bs w^{(k+1)} \right)
	w_j^{(k+1)} -  s_{j}\left(\bs w^{(k)} \right)
	w_j^{(k)} \Big] = 0,
\end{align}
where $j,k=1, \dots ,M$ and $j\neq k$. We assume here that the labelling of the components is such that we have the ordering
\begin{equation}\label{eq:ordering}
s_{j}(\bs w^{(j)})<s_{j+1}(\bs w^{(j+1)}).
\end{equation}
For realistic densities $w_j$, this condition is equivalent to the condition $s_{0j}<s_{0,j+1}$ (see for instance the formula \eqref{eq:order} for the case $M=2$).
The Rankine-Hugoniot conditions \eqref{eq:RH} with $j=k$ are trivially satisfied by the
definition of the contact discontinuity \eqref{eq:z}. According to the formulae \eqref{eq:averu} (KdV) and \eqref{mom_poly_NLS_bound} (fNLS), the moments are also composed of $M+1$ constant states separated by $M$ discontinuities, e.g. for KdV we have
\begin{equation}
	\label{eq:sol_moments}
	\aver{u(x,t)} =
	\begin{cases}
		\aver{u}^{(1)}, &x/t<c_1,\\
		\dots\\
		\aver{u}^{(k)}, & c_{k-1} < x/t < c_k,\\
		\dots\\
		\aver{u}^{(M+1)}, &c_M < x/t,
	\end{cases}\quad \aver{u}^{(k)} = \sum_{j=1}^M 4\eta_j w_j^{(k)}.
\end{equation}

If the solitons were not interacting, the initial step
distribution $w_j(x,0)$ for the component $\eta = \eta_j$ would have
propagated at the free soliton velocity $s_{0j}$  as a single discontinuity:
\begin{equation}
	\label{eq:sol_free}
	w_{{\rm free},j}(x,t) = w_j(x-s_{0j} t,0)=
	\begin{cases}
		w_j^{(-)}, &x/t< s_{0j},\\[2mm]
		w_j^{(+)}, &x/t > s_{0j},
	\end{cases}
	\quad j=1, \dots, M,
\end{equation}
which is very different  compared to the solution \eqref{eq:sol} which exhibits $M$ contact discontinuities. However, since the gas is composed of $M$ distinct components, the moments will still exhibit $M$ discontinuities propagating at the free soliton speeds $x= s_{0j} t$. For instance, the substitution of \eqref{eq:sol_free} in \eqref{eq:averu} yields:
\begin{equation}
	\label{eq:sol_moments_free}
	\begin{split}
	&\aver{u(x,t)}_{\rm free} =
	\begin{cases}
		\aver{u}_{\rm free}^{(1)}, &x/t<s_{01},\\
		\dots\\
		\aver{u}_{\rm free}^{(k)}, & s_{0,k-1} < x/t < s_{0k},\\
		\dots\\
		\aver{u}_{\rm free}^{(M+1)}, &s_{0M} < x/t,
	\end{cases}\\[2mm]
	&\aver{u}_{\rm free}^{(k)} = \sum_{j=k}^{M} 4\eta_j w_j^{(-)}+\sum_{j=1}^{k-1} 4\eta_j w_j^{(+)}.
	\end{split}
\end{equation}
The variation of the moments in the non-interacting case \eqref{eq:sol_moments_free} will be used as a reference in the numerical simulations presented in Sec. \ref{sec:num2}.

\

In practice we consider the specific initial condition:
\begin{equation}
	\label{eq:init2}
	\bs w^{(-)} = \left(0,\dots,0,w_{M_++1}^{(-)},\dots,w_{M_++M_-}^{(-)} \right),\quad
	\bs w^{(+)} = \left(w_1^{(+)},\dots,w_{M_+}^{(+)},0,\dots,0 \right),
\end{equation}
representing a polychromatic SG with $M_-$ ``fast'' components initially at $x<0$ colliding with a polychromatic SG with $M_+$ ``slow'' components initially at $x>0$; the total number of components at $t=0$ is $M=M_-+M_+$. 
By definition the slow components' free velocities are smaller than the fast components' free velocities.
Since $c_j<c_{j+1}$ we have: $w_j(x,t)=0$ for $x/t<c_j$ if $j \leq M_+$ and for $x/t>c_j$ if $j \geq M_++1$, i.e.
\begin{equation}
\label{eq:cancel}
	w_j^{(k)} = 0 \quad \text{for} \quad 1 \leq k \leq j \leq M_+ \quad \text{or} \quad  M_++1 \leq j \leq k+1 \leq M+1,
\end{equation}
which simplifies the resolution of the Rankine-Hugoniot condition \eqref{eq:RH}.

\

We first consider the case $M_-=M_+=1$. The comparison with  numerics is presented in Sec. \ref{sec:monokdv} (and \ref{sec:mononls} for the fNLS SG). The solution \eqref{eq:sol} simplifies:
\begin{equation}
	w_1(x,t) =
	\begin{cases}
		w_1^{(1)} = 0, &x/t<c_1,\\[2mm]
		w_1^{(2)}, & c_1 < x/t < c_2,\\[2mm]
		w_1^{(3)} = w_1^{(+)}, &c_2 < x/t, 
	\end{cases}\quad
	w_2(x,t) =
	\begin{cases}
		w_2^{(1)} = w_2^{(-)}, &x/t<c_1,\\[2mm] 
		w_2^{(2)}, & c_1 < x/t < c_2,\\[2mm]
		w_2^{(3)} = 0, &c_2 < x/t.
	\end{cases}
\end{equation}
The  speeds of contact discontinuities are given by \eqref{s_12}, \eqref{eq:z}
\begin{equation}
\label{eq:c1c2}
	\begin{split}
		&c_1 = s_1 \left(\bs w^{(1)} \right) = s_1 \left(0,w_2^{(-)} \right) =
	s_{01} - \frac{ G_{12} w_2^{(-)} (s_{02}-s_{01})}{1-G_{12} w_2^{(-)}}= \frac{ s_{01}-G_{12} w_2^{(-)} s_{02}}{1-G_{12} w_2^{(-)}},\\
	&c_2 = s_2\left(\bs w^{(3)} \right) = s_2 \left(w_1^{(+)},0 \right) = s_{02}+\frac{ G_{21} w_1^{(+)} (s_{02} - s_{01})}{1- G_{21} w_1^{(+)}} = \frac{s_{02} -G_{21} w_1^{(+)} s_{01}}{1- G_{21} w_1^{(+)}}. 
	\end{split}
\end{equation}
We also have from \eqref{eq:z} that $c_1 = s_1(\bs w^{(2)})$ and $c_2 = s_2(\bs w^{(2)})$, i.e. $c_1$ is the velocity of the slow solitons, $\eta=\eta_1$, and $c_2$ the velocity of the fast solitons, $\eta=\eta_2$, in the ``interactions region'' $c_1 t < x < c_2 t$. Since $1-G_{12} w_2^{(-)}>0$ and $1- G_{21} w_1^{(+)}>0$ (see constrain \eqref{eq:constr}), the interaction results in  slowing down the slow solitons, $c_1<s_{01}$, whereas the fast solitons propagate faster, $c_2>s_{02}$.

The values of the plateau densities $w_1^{(2)}$ and $w_2^{(2)}$ are given by the system of the Rankine-Hugoniot conditions \eqref{eq:RH} with $(j,k)=(1,2)$ and $(j,k)=(2,1)$:
\begin{equation}
	\label{eq:w12}
	\begin{cases}
		-c_2 \Big[ w_1^{(+)} -w_1^{(2)} \Big] + \Big[
			s_{1}\left(w_1^{(+)},0 \right)w_1^{(+)}-  s_{1}\left(w_1^{(2)},w_2^{(2)} \right)
			w_1^{(2)} \Big] = 0,\\[3mm]
		-c_1 \Big[ w_2^{(2)} -w_2^{(-)}\Big] + \Big[
			s_{2}\left(w_1^{(2)},w_2^{(2)} \right)
			w_2^{2} -  s_{2}\left(0,w_2^{(-)} \right)
			w_2^{(-)} \Big] = 0.
	\end{cases}
\end{equation}
The resolution of  system \eqref{eq:w12} yields:
\begin{equation}
	\label{eq:w12b}
	w_1^{(2)} = \frac{\left(1-G_{12} w_2^{(-)} \right)w_1^{(+)}}{1-G_{12} w_2^{(-)} G_{21} w_1^{(+)}},\quad w_2^{(2)} = \frac{\left(1-G_{21} w_1^{(+)} \right)w_2^{(-)}}{1-G_{12} w_2^{(-)} G_{21} w_1^{(+)}}.
\end{equation}
It is insightful to compute the mean $\aver{u}=\aver{u}^{(2)}$ in the interaction region $c_1t<x<c_2 t$ combining \eqref{eq:averu} and \eqref{eq:w12b}:
\begin{equation}
\label{eq:int2}
	\aver{u}^{(2)} = 4\eta_1 w_1^{(2)}+ 4\eta_2 w_2^{(2)} = \frac{4\eta_1 w_1^{(+)}+4\eta_2 w_2^{(-)} - 4 (\eta_1 G_{12}+\eta_2 G_{21})w_1^{(+)} w_2^{(-)}}{1-G_{12} w_2^{(-)} G_{21} w_1^{(+)}}\,.
\end{equation}
We observe that this mean is smaller than the mean obtained in the non-interacting reference case (cf. \eqref{eq:sol_moments_free})
\begin{equation}
\aver{u}_{\rm free}^{(2)} = 4\eta_1 w_1^{(+)} + 4\eta_2 w_2^{(-)}.
\end{equation}
Similarly, one can show that the total density of solitons in the interaction region $ w_1^{(2)}+w_2^{(2)}$ is smaller than the sum of the initial densities $w_1^{(+)}+w_2^{(-)}$, implying that the interaction between the two components $\eta_1$ and $\eta_2$ ``rarefies'' the SG.

\

The second problem investigated numerically in Sec. \ref{sec:bikdv} (and Sec. \ref{sec:binls} for the fNLS SG) is $(M_-,M_+)=(2,1)$. In that case, the solution has 4 plateaus separated by 3 contact discontinuities:
\begin{equation}
\label{eq:sol3comp}
	\boldsymbol{w}(x,t) =
	\begin{cases}
		\boldsymbol{w}^{(-)}, &x/t<c_1,\\
		\boldsymbol{w}^{(1)}, & c_1 < x/t < c_2,\\
		\boldsymbol{w}^{(2)}, & c_2 < x/t < c_3,\\
		\boldsymbol{w}^{(+)}, &c_3 < x/t,
	\end{cases}\quad
\end{equation}
where $\boldsymbol{w}^{(1)}=(w_1^{(1)},w_2^{(1)},w_3^{(1)})$ and $\boldsymbol{w}^{(2)}=(w_1^{(2)},w_2^{(2)},w_3^{(2)})$ are obtained by solving  the nonlinear system of  $M(M-1)=6$ Rankine-Hugoniot conditions \eqref{eq:RH} numerically. For the choice of initial conditions \eqref{eq:init2}, the density of the second component on the second plateau cancels: $w_2^{(2)}=0$ (cf. the general formula \eqref{eq:cancel}). Thus, the three components of the SG interact in the region $c_1<x/t<c_2$, whereas only the components $\eta_1$ and $\eta_3$ interact in the region $c_2<x/t<c_3$. In particular, we can obtain the analytical expression for $c_3$:
\begin{equation}\label{eq:c3}
c_3= s_3(w_1^{(2)},0,w_3^{(2)}) =  s_3(w_1^{(-)},0,0) = \frac{s_{03} -G_{31} w_1^{(+)} s_{01}}{1- G_{31} w_1^{(+)}} .
\end{equation}

\section{Numerical synthesis of polychromatic SGs}
\label{sec:synthesis}

The main objective of this paper is the numerical validation of the spectral kinetic theory of SGs using analytical solutions of  the ``polychromatic'' Riemann problem for the kinetic equation as a benchmark for the comparison with direct ``microscopic'' numerical computations of the solutions to the the KdV and fNLS equations.  Such a comparison has been  partially realised in the previous works \cite{carbone_macroscopic_2016, congy_soliton_2021} by considering collision of two  monochromatic SGs,  modelled by superpositions of  single-soliton solutions randomly distributed on  $\mathbb{R}$ with relatively small density to ensure  negligible overlap of solitons in the gas almost everywhere on $\mathbb{R}$. The modelling of a monochromatic  gas was achieved in \cite{carbone_macroscopic_2016, congy_soliton_2021} by distributing the discrete spectrum parameters  of  solitons in a narrow vicinity of some central value.    While the numerical simulations in \cite{carbone_macroscopic_2016, congy_soliton_2021} displayed a  good agreement with the theoretical predictions of the kinetic theory for several integrable equations (KdV, defocusing NLS, Kaup-Boussinesq), the obtained results are not fully satisfactory due to relatively low densities of the colliding SGs  so that the quantitative aspects of the kinetic  theory for dense gases had not been verified. In this paper, we perform a detailed numerical verification of the kinetic equation  for both KdV and fNLS polychromatic SGs for a broad range of spectral and density parameters. Along with direct numerical validation of the kinetic theory, an important outcome of our work is the development of an effective method of numerical realisation of dense polychromatic SGs.

Numerical implementation of a polychromatic Riemann problem implies fulfilment of several technically challenging and to some degree conflicting requirements which include the necessity to combine high density of the SG with  its spatial and statistical homogeneity. These issues are not new and have been the subject of a number of recent numerical works for the fNLS equation, particularly  \cite{gelash2018strongly}, where an effective IST-based numerical algorithm for the numerical realisation of a dense uniform SG on some finite interval $x\in[-\ell/2,\ell/2]$ was developed.  The method of \cite{gelash2018strongly} employs  the Darboux transformation for the generation of $n$-soliton solution with the implementation of  high precision arithmetic routine to overcome the numerical accuracy problems and
generate solutions with a number of solitons $n \gtrsim 10$. Combined with some empirical observations of  the ``optimal'' sampling of the spectral parameters and the distribution of the soliton positional phases within the $n$-soliton solution the approach of \cite{gelash2018strongly} enables one to numerically build  a maximally dense SG while maintaining its spatial homogeneity on $x\in[-\ell/2,\ell/2]$.  This algorithm  has been  used in \cite{gelash2019bound, Suret2020Nonlinear, agafontsev2021rogue} to numerically study various properties of the fNLS SGs and in \cite{roberti_numerical_2021} for the synthesis of fNLS breather gases. The method of \cite{gelash2018strongly} was recently adapted to the KdV SG case in \cite{congy_dispersive_2023, bonnemain_generalized_2022} where  the phenomenological distributions of the soliton positional phases employed in \cite{gelash2018strongly} have been quantified in the context of KdV SGs by invoking   the  theoretical results of \cite{gurevich2000statistical} on semi-classical random soliton ensembles. 

We note that the previous numerical results on dense SGs were mostly concerned with the special class of fNLS SGs termed soliton condensates \cite{el_spectral_2020} and characterised by the so-called Weyl's DOS.  The numerical realisation of dense SGs  in the polychromatic Riemann problem requires additional non-trivial considerations that would enable the implementation of narrow spectral distributions to effectively model the delta-functional DOS in the polychromatic spectral ansatz \eqref{u_delta1}. We will first present the numerical implementation of polychromatic SGs for the KdV equation, and then will introduce its natural extension to the class of the so-called bound state polychromatic fNLS gases which are spectrally analogous to the KdV SGs. 

\subsection{$\bs{n}$-soliton approximation of KdV SG}
\label{sec:nsol}

The  spectral theory of SGs is based on the thermodynamic limit of finite-gap potentials \cite{el_thermodynamic_2003, el_spectral_2020}. At the same time, the practical (numerical) implementation of SGs is more conveniently realised with $n$-soliton solutions. Although the equivalence of the solitonic and finite-gap approaches to the description of SGs appears very natural, it has not been yet rigorously established. Nevertheless, there are strong indications that these two approaches are indeed equivalent at the end. In particular, the rigorous asymptotic analysis of ``deterministic SGs'' in \cite{girotti_soliton_2023} based on $n$-soliton solutions in the so-called ``primitive potential" infinite-soliton limit  \cite{dyachenko2016primitive} yields at leading order the same results as the theory of  soliton condensates  \cite{congy_dispersive_2023} derived from the thermodynamic limit of finite-gap potentials.  Additionally, we mention an early work \cite{gurevich2000statistical} where   semi-classical random-phase $n$-soliton ensembles of the KdV equation  were considered in the framework of the  small-dispersion limit of the KdV equation.  The authors of \cite{gurevich2000statistical} employed the Lax-Levermore type minimisation procedure \cite{lax_small_1983}  to derive  linear integral equations for some spectral distribution functions from which one can  infer the equation of state \eqref{eq_state_kdv} of the  KdV SG in the uniform, $x,t$-independent case. 

We consider the KdV $n$-soliton solution in the general form
\begin{equation}
  \label{eq:soliton-sol}
  u \equiv
  u_n(x,t;\tilde \eta_1,\dots,\tilde \eta_n,x^0_1,\dots,x^0_n),\quad n \gg 1,
\end{equation}
 where the $\tilde \eta_i$'s are the usual 
spectral parameters related to the discrete spectrum of the Lax (linear Schr\"odinger) operator via $\zeta_j = - \tilde \eta_j^2$ \cite{ablowitz_solitons_1981, novikov_theory_1984} and $x^0_i$'s  are  the ``spatial'' or ``positional'' soliton phases.  We use the tilded notation $\tilde \eta_j$ for the soliton spectral parameters here to distinguish them from the monochromatic spectral components $\eta_j$ in the polychromatic DOS ansatz \eqref{u_delta1}.  The phases  $x^0_j$ can be associated with the individual soliton positions within $n$-soliton solution at $t=0$. They are related to the so-called norming constants of the discrete spectrum in the IST formalism \cite{ablowitz_solitons_1981, novikov_theory_1984} (see eq. \eqref{norm_const} in Appendix \ref{sec:app_algo}; note the subtle difference in the definitions of norming constants in the standard IST and in the Darboux transformation formalism \cite{huang1992darboux}). The positional  phases  also have a physically transparent interpretation in the GHD context as the average values of the ``asymptotic coordinates'' $x_i^+$ and $x_i^-$ of the $i$th soliton in an $n$-soliton solution:  $x^0_i = \frac{1}{2}(x_i^- + x_i^+)$ \cite{bonnemain_generalized_2022, congy_dispersive_2023}.  Here $x_i^+$ and $x_i^-$ are the $x$-intercepts of the phase-shifted trajectories of the soliton at $t \to \pm \infty$ respectively, where the solitons separate and follow their free-velocity propagation. 
  
As a matter of fact, $n$ is finite in numerical applications, and the $n$-soliton solution can be used to approximate a SG realisation only within a finite spatial region $x\in[-\ell/2,\ell/2]$, with $\ell \sim \kappa^{-1} n$, where $\kappa=\int_\Gamma f(\eta) \rmd \eta$ is the SG spatial density. A pertinent  question that needs to be addressed is the following: how should one distribute the spectral parameters $\tilde \eta_i$ and phases $x^0_i$ of the $n$-soliton solution \eqref{eq:soliton-sol} to achieve a sufficiently accurate approximation on $[-\ell/2, \ell/2]$ of a dense uniform SG with a given DOS?  To answer this question we invoke the results of  \cite{gurevich2000statistical}  where $n$-soliton KdV solutions  with random spatial phases $x^0_i$ were considered in the semi-classical approximation, when $n \gg 1$ and with $x$ and $t$ in \eqref{eq:soliton-sol} being scaled  as $\varepsilon \sim n^{-1}$. 
The results of \cite{gurevich2000statistical}, translated into the language of modern spectral SG theory \cite{el_soliton_2021} provide the two key ingredients for our numerical construction: 
(i) the spatial interval $I_s \subset \mathbb{R}$ on which the phases $x^0_i$ are distributed,  and (ii) the density distribution $\phi(\eta)$ of $\tilde \eta_i \in \Gamma$ in the $n$-soliton ensemble approximating a SG on $[-\ell/2, \ell/2]$.  These relations given by the equations \eqref{eq:kappas}-\eqref{eq:rho}  below, enable quantification of the previous empirical observations  used for the effective numerical $n$-soliton modelling of SGs in the context of the fNLS equation \cite{gelash2018strongly, gelash2019bound, Suret2020Nonlinear, agafontsev2021rogue}, and  they play an important role in the numerical algorithm we use  to evaluate
\eqref{eq:soliton-sol} with large $n$ at any $x$ and $t$, see Appendix \ref{sec:app_algo}.  This algorithm  was recently developed by co-authors of the present paper  to numerically implement KdV SGs, see  \cite{congy_dispersive_2023, bonnemain_generalized_2022} and in this paper we will also extend it to the fNLS case.

 In the spatial region $[-\ell/2, \ell/2]$, one attains an (approximately) uniform SG with $x$-independent DOS $f(\eta)$ by distributing the $n$-soliton parameters $(\tilde \eta_i,x_i^0)$ in the following way. The phases $x^0_i$  are taken to be independent random variables, each  distributed uniformly on the interval (the set $\{ x_i^0 \}$ is called the ``$S$-set'' in \cite{gurevich2000statistical}):
\begin{equation}
  \label{eq:kappas}
  I_s = \left[-\frac{n}{2\kappa_s},\frac{n}{2\kappa_s} \right],\quad \kappa_s = \int_{\Gamma } \frac{\eta}{\sigma(\eta)} \rmd \eta.
\end{equation}
Here $\kappa_s$ has the meaning of the  density of the positional phases on $I_s$, and $\sigma(\eta)>0$ is the so-called spectral scaling function, which is expressed in terms of the DOS by the  nonlinear dispersion relation of SG (see \cite{el_thermodynamic_2003, el_soliton_2021, bonnemain_generalized_2022}
and Appendix \ref{sec:condensate}):
\begin{equation}
	\label{eq:sigma}
	\sigma(\eta) = \frac{\eta}{f(\eta)} \left(1-\int_{\Gamma}G_{\rm KdV}(\eta,\mu)f(\mu) \rmd\mu\right),
\end{equation}
with $G_{\rm KdV}(\eta,\mu)$ given in \eqref{eq:Gkdv}.
The requirement of positivity of $\sigma$ implies the constraint \eqref{dos_constr} which is fundamental to the spectral theory of SG. 

Next, the spectral parameters $\tilde \eta_i$'s in the $n$-soliton solution are distributed on $\Gamma$ with the density
\begin{equation}
  \label{eq:rho}
  \phi(\eta) = \frac{1}{\kappa_s} \frac{\eta}{\sigma(\eta)} = \frac{1}{\kappa_s} \frac{f(\eta)}{1-\int_{\Gamma}G_{\rm KdV}(\eta,\mu)f(\mu) \rmd\mu},
\end{equation}
where the pre-factor $1/\kappa_s$ ensures that $\phi(\eta)$ is normalised to $1$. We note that relations \eqref{eq:kappas}-\eqref{eq:rho} have a natural interpretation within the GHD approach to SGs \cite{bonnemain_generalized_2022}.

We stress that generally, the spatial density of solitons $\kappa$ on $x \in [-\ell/2, \ell/2]$ does not coincide with the  density of the positional phases $\kappa_s$ on $I_s$. 
However, in the rarefied gas limit ($\kappa \ll 1$), the phase-shift contribution (the integral term in \eqref{eq:sigma}, \eqref{eq:rho})  can be neglected yielding $\sigma(\eta)\sim \eta/f(\eta)$ so that $\kappa_s=\kappa$. In this case $\phi(\eta)$ corresponds to the normalised DOS, $\phi(\eta)=\kappa^{-1} f(\eta)$. 
Indeed, in the rarefied gas regime the solitons of the $n$-soliton solution are almost non-overlapping, and the $n$-soliton solution  \eqref{eq:soliton-sol} can be approximated by a linear superposition of $n$ single-soliton solutions:
\begin{equation}
	\label{eq:linear}
	u_n(x,t;\tilde \eta_1,\dots,\tilde \eta_n,x^0_1,\dots,x^0_n) \sim \sum_{i=1}^n 2\tilde \eta_i^2 {\rm sech}^2[\tilde \eta_i (x-x_i(t))],\quad x_i(t)=4\tilde \eta_i^2 t+x_i^0.
\end{equation}
The position $x_i(t)$ of the $i$th soliton within $n$-soliton solution \eqref{eq:linear} is well-defined for all $t$ and  the uniform distribution  of the spatial phases $x_i^0$ on $I_s$ is equivalent in this case to the uniform distribution of the physical solitons' positions on $[-\ell/2, \ell/2]$ at $t=0$.   

\begin{remark}\label{remark_5.1}
As $\kappa$ increases, the interaction between the solitons is no longer negligible, and $x_i(t)$ defined in \eqref{eq:linear} no longer corresponds to a soliton's physical position in space. In the limiting case of soliton condensate  one has $I_s \to \{0\}$ so that $\sigma \to 0$, $\kappa_s \to \infty$, $\kappa_s \sigma = \mathcal{O}(1)$. Furthermore,
\begin{equation}
 \kappa_s \left(1-\int_{\Gamma}G_{\rm KdV}(\eta,\mu)f(\mu) \rmd\mu \right)  \to \kappa =  \mathcal{O}(1)
\end{equation}
in the condensate limit. It then follows from \eqref{eq:rho} that $f(\eta) =\kappa \phi(\eta)$ for soliton condensates. This result is similar to the relation between $\phi(\eta)$ and $f(\eta)$ in a rarefied SG with the important difference that now $\kappa  \ne \kappa_s$.
\end{remark}

\begin{remark}\label{remark_5.2} In the general case $f(\eta) \neq \kappa\, \phi(\eta)$, and the approximation of a uniform SG realisation by $n$-soliton solution as described in this section is only valid for $x\in [-\bar \ell/2,\bar \ell/2]$ with 
\begin{equation}
\bar \ell \simeq \frac{\phi(\eta_{\rm min})}{f(\eta_{\rm min}) } n < \ell,
\end{equation}
 where $\eta_{\rm min}$ is the location of the minimum of $\phi(\eta)$; see Appendix \ref{sec:implementation} for the details of the computation.
\end{remark}

The linear superposition formula \eqref{eq:linear} was used in \cite{carbone_macroscopic_2016, congy_soliton_2021} to numerically initiate SGs in the ``bi-chromatic'' Riemann problems. This  setting from the very beginning  constrains the numerical simulations  to the rarefied SG regime $\kappa \ll 1$. In contrast, the direct evaluation of the $n$-soliton solution \eqref{eq:soliton-sol} in this work enables the verification of the theoretical results in the dense gas regime  with $\kappa \neq \kappa_s$. 
An important technical advantage of the method used here is that the evolution in time of SG realisations does not rely on a numerical approximation of the KdV equation (e.g. finite difference method, spectral method, etc.) which was the case in the previous works \cite{dutykh_numerical_2014, carbone_macroscopic_2016, congy_soliton_2021}. 
Indeed, $t$ plays the role of a parameter in the exact $n$-soliton solution, which can be computed, without propagation of errors, at any value of $t$. 
In that sense, the determination of SG realisation via $n$-soliton solution is exact (providing the elimination of round-off errors in the evaluation of $u_n(x,t)$, see Appendix \ref{sec:app_algo}). It is also faster than the direct numerical resolution of the KdV equation as the evaluation of~$u_n(x,t)$ does not necessitate its evaluation at a previous time $u_n(x,t-\Delta t)$ with $\Delta t$ a time-step.

Finally we note that, to achieve a statistically robust numerical implementation of a dense random $n$-soliton ensemble with large $n$ the parameters $\tilde \eta_i$ and $x_i^0$ in \eqref{eq:soliton-sol} need not to be both randomly distributed:  it is sufficient to only implement random phases (see e.g. the numerical observations in \cite{congy_dispersive_2023}) so in practice $\tilde \eta_i$'s are sampled deterministically from $\phi(\eta)$ \eqref{eq:rho} by
\begin{equation}\label{eq:phi}
	\int^{\tilde \eta_i} \phi(\eta) \rmd \eta = \frac{i}{n}, \quad i=1, \dots, n.
\end{equation}

\subsection{Polychromatic SG}
\label{sec:RP_num}

The DOS of a monochromatic SG
\begin{equation}
	\label{eq:onedelta}
	f(\eta) = w_1 \delta(\eta-\eta_1)
\end{equation}
with all solitons having the same spectral parameter $\eta=\eta_1$ cannot be directly modelled by $n$-soliton KdV solutions due non-degeneracy of the discrete spectrum of the Lax (Schr\"odinger) operator associated with the KdV equation in the IST formalism. Thus, the $\delta$-distribution should be replaced by a narrow, compactly supported distribution around $\eta_1$ as indicated before, and the numerically implemented SG of this kind can be more accurately characterised as a quasi-monochromatic SG. We shall nevertheless be using the term monochromatic for simplicity. The challenge now is to find the optimal spectral distribution approximating the delta function and enabling the realisation of a SG which is sufficiently dense and spatially uniform at the same time

As the simplest compactly supported replacement for the delta-function one can take the ``box''  distribution: $\delta(\eta-\eta_1) = 1/(2\varepsilon)$ if $|\eta-\eta_1|<\varepsilon/2\ll 1$ and $0$ otherwise, where the spectral parameters $\tilde \eta_i$ are distributed within the close neighbourhood of $\eta_1$.
We show in Appendix \ref{sec:comparison} that this distribution is in fact not optimal as it does not generate a SG which is sufficiently dense and spatially uniform at the same time and so is not suitable for the numerical  implementation of the step distribution of the Riemann problem for dense SGs. 

\

To overcome the above difficulty we invoke the concept of   soliton condensate, which represents the critically dense SG for a given compact spectral support $\Gamma $ \cite{el_spectral_2020, congy_dispersive_2023} (see also Appendix \ref{sec:condensate}). The spectral support of KdV soliton condensates consists of a finite number of disjoint bands, $\Gamma= [0, \la_1] \cup  [\la_2, \la_3] \cup \dots \cup [\la_{2M}, \la_{2M+1}] $. Within the soliton condensate framework the modelling of the polychromatic delta-function DOS is naturally achieved by making appropriate  bands $[\la_{2i}, \la_{2i+1}]$ sufficiently narrow.  At the same time, as was shown in \cite{congy_dispersive_2023}, the KdV soliton condensates are ``deterministic gases'' that coincide with probability 1 with finite-gap potentials having their band structure defined by $\Gamma$. At the level of  numerical modelling the condensate DOS $f^{(M)}(\eta)$ can only be achieved by collapsing the interval $I_s$ \eqref{eq:kappas} into a single point \cite{congy_dispersive_2023}, see Remark \ref{remark_5.1}. Therefore, to reinstate randomness of a SG one needs to slightly ``dilute'' the condensate by widening the collapsed interval $I_s$, see \cite{congy_dispersive_2023} for a discussion of the properties of diluted KdV condensates.  

The delta-distribution \eqref{eq:onedelta} is thus replaced by
 the following diluted condensate DOS
\begin{equation}
	\label{eq:genus1}
	f(\eta) = C f^{(1)}(\eta;\lambda_1=0,\lambda_2=\eta_1-\varepsilon,\lambda_3=\eta_1+\varepsilon),\quad 0<\varepsilon \ll 1,\quad 0<C<1,
\end{equation}
where $C$ is the dilution parameter and $f^{(1)}$ is the DOS satisfying 
\begin{equation}
\label{eq:cond1}
\int_{\Gamma} G_{\rm KdV}(\eta,\mu) f^{(1)}(\mu;\lambda_1,\lambda_2,\lambda_3) \rmd\mu =1,\quad  \eta \in \Gamma = [0,\lambda_1]\cup [\lambda_2,\lambda_3].
\end{equation}
The general solution of the integral equation \eqref{eq:cond1} is given by eq.~\eqref{f1} in Appendix \ref{sec:condensate}. 
Fig. \ref{fig:examples_dos} displays a typical variation of $f^{(1)}(\eta)$ with $\lambda_1=0$ and for a narrow compact support around $\eta=\eta_1$ according to \eqref{eq:genus1}. 

Heuristically, the condition \eqref{dos_constr} for the DOS can be interpreted as a spectral constraint on the ``dense packing'' of solitons in the SG (cf. \cite{el_spectral_2020, congy_dispersive_2023}): the densest packing DOS being reached when the integral in  \eqref{dos_constr} is equal to $1$ for all $\eta \in \Gamma$.
Thus $f^{(1)}$ corresponds to the DOS of the critically dense SG with the support $\Gamma$, also called genus $1$ soliton condensate (see \cite{congy_dispersive_2023} and Appendix \ref{sec:condensate}). In that case $\sigma(\eta)=0$ i.e. 
$\kappa_s \to \infty$, so  that the phases in the corresponding $n$-soliton implementation are not random but all fixed at $x_i^0=0$, $i=1, \dots, n$, according to \eqref{eq:kappas}; note that the spatial density of solitons $\kappa$ remains finite even for soliton condensate, see Remark \ref{remark_5.1} in the previous section.

As we already mentioned, one of the results of \cite{congy_dispersive_2023} was a  conjecture, supported by careful numerical simulations, that all realisations of a genus $n$ soliton condensate  with probability 1 are given by $n$-gap KdV solutions spectrally defined by $\Gamma$ (e.g. realisations of \eqref{eq:genus1} with $C=1$ correspond to a cnoidal wave).
To reestablish the stochastic nature of the SG in the numerical implementation presented here, we ``dilute'' the soliton condensate by choosing $C<1$ in \eqref{eq:genus1}, thus slightly decreasing the  densities of both spatial phases and ``physical'' solitons.
This constitutes one of the first advantages of choosing the DOS \eqref{eq:genus1} over the box-distribution: 
we  ensure that the SG is in a dense regime by picking the value $C$ sufficiently close to $1$. 
\begin{figure}
\centering
\includegraphics[width=7.5cm]{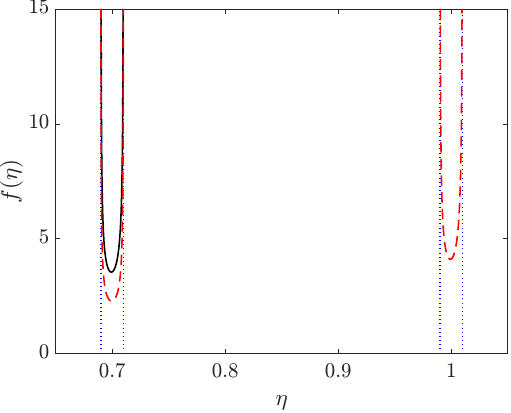}
  \caption{Variation of the diluted condensate genus 1 DOS \eqref{eq:genus1} (black solid line) and genus 2 DOS \eqref{eq:genusM} (dashed red line) used to model the polychromatic delta-function ansatz \eqref{u_delta1} for $M=1$ and $M=2$ respectively. For both distributions the parameters are: $C=0.9$, $\eta_1=0.7$, $\eta_2=1$  and $\varepsilon=\varepsilon_1=\varepsilon_2=0.01$. The endpoints of the bands $\lambda_i$ are located by dotted vertical blue lines.}
  \label{fig:examples_dos}
\end{figure}

\

The more general diluted condensate DOS \eqref{u_delta1} modelling a polychromatic SG with $w_j(x,t)={\rm cst}$, $j=1, \dots, M$, is implemented via
\begin{align}
	\label{eq:genusM}
	\begin{split}
		&f(\eta) = C f^{(M)}(\eta;\lambda_1,\lambda_2,\dots,\lambda_{2M+1}),\quad \Gamma = [0,\lambda_1] \cup [\lambda_2,\lambda_3] \cup \dots \cup [\lambda_{2M},\lambda_{2M+1}],\\
	&0<C<1,\quad \lambda_1 = 0,\quad \lambda_{2j} = \eta_j-\varepsilon_j,\quad  \lambda_{2j+1} = \eta_j+\varepsilon_j,\quad 0<\varepsilon_j\ll 1,
	\end{split}
\end{align} 
with $f^{(M)}$ is the DOS of the genus $M$ soliton condensate satisfying $\int_{\Gamma} G_{\rm KdV}(\eta,\mu) f^{(M)}(\eta) \rmd\mu =1$, see Appendix \ref{sec:condensate}. A typical variation of $f^{(2)}(\eta)$ for a narrow compact support is  shown  in Fig. \ref{fig:examples_dos} (dashed line). One could also implement the different ``monochromatic'' components of \eqref{u_delta1} with a sum of genus $1$ condensate DOS \eqref{eq:genus1}, but the choice \eqref{eq:genusM} ensures that the SG is a dilution of the densest possible gas with the support $\Gamma$ given in \eqref{eq:genusM}. Fig. \ref{fig:examples_dos}(b) displays the typical variation of $f^{(2)}(\eta)$ for a narrow-banded compact support.

By definition, $w_j$ corresponds to the spatial density of solitons with a spectral parameter between $\eta_j-\varepsilon_j$ and $\eta_j+\varepsilon_j$.
This fixes here the relation between $w_j$, $C$ and~$\varepsilon_j$:
\begin{equation}
	\label{wj}
	w_j = \int_{\eta_j-\varepsilon_j}^{\eta_j+\varepsilon_j} f(\eta) \rmd\eta = C \int_{\eta_j-\varepsilon_j}^{\eta_j+\varepsilon_j} f^{(M)}(\eta;\lambda_1,\lambda_2,\dots,\lambda_{2M+1}) \rmd\eta.
\end{equation}
Thus the value of $w_j$ can be chosen by fixing conveniently the values of $C<1$ and $\varepsilon_j \ll 1$. Substituting the DOS \eqref{eq:genusM} in the expressions \eqref{eq:kappas} and \eqref{eq:rho}, the expressions of $\kappa_s$ and $\phi(\eta)$ drastically simplify:
\begin{equation}
\label{eq:ks2}
	\kappa_s = \frac{\kappa}{1-C} =\frac{C \kappa^{(M)}}{1-C},\quad \phi(\eta) = \frac{f^{(M)}(\eta)}{\kappa^{(M)}},
\end{equation}
where $\kappa^{(M)}=\int_{\Gamma} f^{(M)}(\eta) \rmd\eta $ is the spatial density of the genus $M$ soliton condensate. In that case $\phi(\eta)=f(\eta)/\kappa$, i.e. $\phi(\eta)$ is the normalised DOS, and we show in Appendix \ref{sec:comparison} that the spatial uniformity of the SG is a direct consequence of this equality.
The expression for $\kappa^{(M)}$ can be found analytically for $M=1$ (see Appendix \ref{sec:comparison}), and we have for $\varepsilon_1=\varepsilon \ll 1$:
\begin{equation}
\kappa^{(1)} = \frac{\eta_1}{\ln(4\eta_1/\varepsilon)}+ {\cal O} (\varepsilon^2/\ln(\varepsilon)). 
\end{equation}
In practice $\varepsilon_j = {\cal O}(10^{-2})$ is sufficient to guarantee  finite spatial density $w_j$.

\begin{figure}
  \unitlength=1cm
  \begin{picture}(15,5.5)   
\put(0,0){\includegraphics[width=7.5cm]{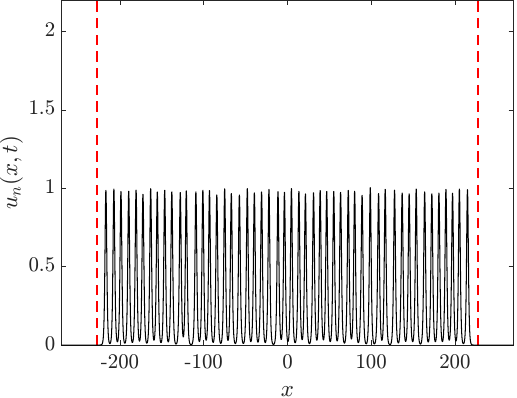}}
\put(8.5,0){\includegraphics[width=7.5cm]{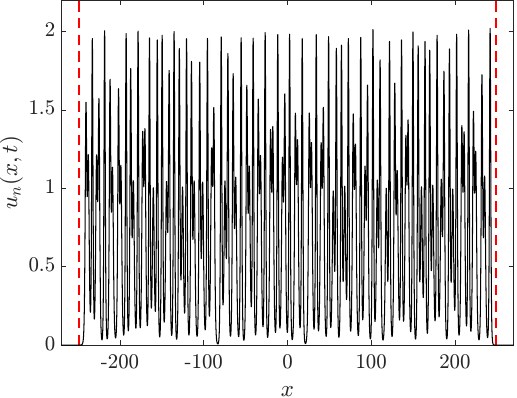}}
\put(0,0){\small (a)}
\put(8.5,0){\small (b)}
\end{picture}
  \caption{(a) Numerical realisation of the monochromatic SG \eqref{eq:onedelta} with $n=50$ solitons, $\eta_1=0.7$ and $w_1=0.11$ (implemented with the DOS shown  in Fig. \ref{fig:examples_dos}(solid line)). (b) Numerical realisation of the polychromatic SG \eqref{u_delta1} with $M=2$, $n=100$ solitons, $(\eta_1,\eta_2)=(0.7,1)$ and $(w_1,w_2)=(0.07,0.13)$ (implemented with the DOS displayed in Fig. \ref{fig:examples_dos}(dashed line)). The dashed vertical red lines indicate the positions $|x|=\ell/2=\kappa^{-1} n/2$.}
  \label{fig:examples_real}
\end{figure}

Fig. \ref{fig:examples_real} displays examples of the numerical implementation for the cases $M=1$ and $M=2$.
We observe in Appendix \ref{sec:comparison} that, using the condensate distribution \eqref{eq:genusM}, the spatial density of solitons $\kappa$ is uniform in the region $x \in (-\ell/2,\ell/2)$ with $\ell = \kappa^{-1} n=(C \kappa^{(M)})^{-1} n$. Outside of this region, the $n$-soliton solution exponentially decays to $0$. Thus, for $n$ sufficiently large, $u_n(x,t=0)$ approximates a SG realisation with a step-like spatial density distribution, which is used in this work to implement the initial condition of the Riemann problem.

\subsection{Riemann problem}
\label{sec:RP_num2}

The Riemann problem is implemented numerically by first distributing the parameters $(\tilde \eta_i, x_i^0)$ in such a way that for each set $(\tilde \eta_i, x_i^0)$ the resulting $n$-soliton solution \eqref{eq:soliton-sol} models a realisation of the SG with DOS \eqref{u_delta1}, \eqref{eq:init2} at $t=0$; the time-evolved  $n$-soliton solution is then evaluated by simply varying the ``parameter'' $t>0$.

The SG described by the step-distribution \eqref{u_delta1}, \eqref{eq:init2} is composed of $M_-$ components initially placed at $x<0$, and $M_+$ components initially placed at $x>0$, for a total of $M=M_-+M_+$ components. The initial DOS can be rewritten as the following piecewise, $M$-component distribution
\begin{equation}
\label{step2}
f(\eta;x,t=0) = \begin{cases}
\displaystyle \sum_j w^{(-)}_j \delta(\eta - \eta_j), &x<0,\\[4mm]
\displaystyle \sum_j w^{(+)}_j \delta(\eta - \eta_j), &x>0,
\end{cases}
\end{equation}
where $w_j^{(-)}=0$ for $j\leq M_+$ and $w_j^{(+)}=0$ for $j > M_+$.
This DOS can be implemented numerically with
\begin{equation}
f(\eta) = \begin{cases}
C_- f^{(M_-)}\left(\eta;\lambda_1^{(-)}=0,\lambda_2^{(-)},\dots,\lambda_{2M_-+1}^{(-)} \right), &x<0,\\[2mm]
C_+ f^{(M_+)} \left(\eta;\lambda_1^{(+)}=0,\lambda_2^{(+)},\dots,\lambda_{2M_++1}^{(+)} \right), &x>0,
\end{cases}
\end{equation}
with the spectral supports $\Gamma_{\pm}$ defined by the endpoints (see \eqref{eq:genusM})
\begin{equation}
\begin{split}
&\lambda_{2j}^{(+)}= \eta_j-\varepsilon_j^{(+)},\quad \lambda_{2j+1}^{(+)}= \eta_j+\varepsilon_j^{(+)},\quad 1 \leq j \leq M_+,\\
&\lambda_{2j}^{(-)}= \eta_{M_++j}-\varepsilon_j^{(-)},\quad \lambda_{2j+1}^{(-)}= \eta_{M_++j}+\varepsilon_j^{(-)},\quad 1 \leq j \leq M_-,
\end{split}
\end{equation}
where $C_-$, $\varepsilon_j^{(-)} $ and  $C_+$, $\varepsilon_j^{(+)}$ are conveniently chosen to approximate the ideal DOS \eqref{step2}.

\

The SG realisation is approximated by a $n$-soliton solution with $n=n_-+n_+ \gg 1$; 
where  $n_-$ solitons of the ``left fast polychromatic SG'' ($x<0$) are split from $n_+$ solitons of the ``right slow polychromatic SG'' ($x>0$). The first $n_+$ spectral parameters $\tilde \eta_i$ are given by  equation \eqref{eq:phi}, where $\phi(\eta)=f^{(M_+)}(\eta)/\kappa^{(M_+)}$ and $\kappa^{(M_+)} = \int f^{(M_+)}(\eta) \rmd\eta$. The remaining $n_-$ spectral parameters are given by \eqref{eq:phi} where $\phi(\eta)=f^{(M_-)}(\eta)/\kappa^{(M_-)}$ and $\kappa^{(M_-)} = \int f^{(M_-)}(\eta) \rmd\eta$. Notice that at this stage of the implementation, nothing differentiates the solitons that will be initially placed at $x<0$ from the ones placed at $x>0$.

The distinction between the two gases is manifested in the distribution of phases $x_i^0$: the phases of the $n_+$ right solitons are uniformly distributed in the interval
\begin{equation}
	I_+ = \left[-\frac{n_+}{2 \kappa_s}+\Delta_+,\frac{n_+}{2 \kappa_s}+ \Delta_+\right],\quad \kappa_s = \frac{C_+ \kappa^{(M_+)}}{1-C_+},
	\end{equation}
and the phases of the $n_-$ left solitons are uniformly distributed in the interval
\begin{equation}
I_- = \left[-\frac{n_-}{2 \kappa_s}+\Delta_-,\frac{n_-}{2 \kappa_s} +\Delta_-\right],\quad \kappa_s = \frac{C_- \kappa^{(M_-)}}{1-C_-}.
\end{equation}
We pick $\Delta_+$ and $\Delta_-$ such that the slow SG and the fast SG are respectively on the right and on the left of $x=0$ respectively. Because of the finite width of the solitons ($\propto \eta_j^{-1}$), the initial density step is located at $x={\cal O}(\max_j \eta_j^{-1})={\cal O}(1)$.

\subsection{Generalization to fNLS SG}
\label{sec:fNLS}

In the fNLS case the KdV spectral parameter $\eta$ is substituted by the complex number $z=\xi+i \eta$ such that $\delta(\eta-\eta_j)$ transforms into the 2D distribution $\delta(z-z_j)=\delta(\xi-\xi_j)\delta(\eta-\eta_j)$. 
Similar to the KdV SG we approximate realisations of the fNLS SG  by the $n$-soliton solution of \eqref{eq:nls}
\begin{equation}
\label{eq:psin}
  \psi \equiv
  \psi_n \left(x,t;\tilde z_1,\dots,\tilde z_n,x^0_1,\dots,x^0_n,\theta^0_1,\dots,\theta^0_n \right),\quad n \in \mathbb{N},
\end{equation}
where $\tilde z_k \in \mathbb{C}^{+}$ and $(x^0_k,\theta^0_k) \in \mathbb{R} \times [0,2\pi)$ correspond respectively to the
spectral parameter and the phase parameters of the solitons (we remind that the fNLS solitons \eqref{fnls_soliton1} are characterised by two types of phase parameters).
The Riemann problem is still implemented numerically by distributing the spectral parameters $\tilde z_k$ as well as the phase parameters $(x_k^0,\theta_k^0)$ such that the $n$-soliton solution represents a realisation of the SG with DOS \eqref{u_delta1}, \eqref{eq:init2} at $t=0$, and then evaluate the $n$-soliton solution for $t>0$, which plays the role of a parameter. Note that the systematic numerical realisation of SGs via the $n$-soliton solution has been first developed for the fNLS equation in \cite{gelash2018strongly}; in particular the latter reference  presents an efficient algorithm to evaluate numerically the function \eqref{eq:psin}. 

Similar to the KdV implementation, the position phases $x_i^0$ are uniformly distributed on the interval $I_s$ of width $n/\kappa_s$, see \eqref{eq:kappas}, where the expression for $\kappa_s$ remains to be determined. fNLS solitons have an additional ``degree of freedom'' encoded in the ``angular'' phases $\theta^0_i$, which are uniformly distributed in $[0,2\pi)$ to generate spatially uniform SGs. We describe below how to obtain $\kappa_s$ and how to distribute the spectral parameters $\tilde z_i$ to implement polychromatic fNLS SGs.

\

We consider first the case of a monochromatic SG with the DOS 
\begin{equation}
\label{eq:monoNLS}
f(z)= w_1 \delta(z-z_1) = w_1 \delta(\xi-\xi_1) \delta(\eta- \eta_1).
\end{equation}
The DOS \eqref{eq:monoNLS} can be implemented with the following ``bound state'' distribution:
\begin{equation}
\label{eq:cond12}
f(z)= C \delta(\xi-\xi_1) f^{(1)}(\eta;\lambda_1=0,\lambda_2=\eta_1-\varepsilon,\lambda_3=\eta_1+\varepsilon),\quad 0<\varepsilon \ll 1,\quad C<1,
\end{equation}
where $f^{(1)}$ is solution of the KdV-soliton condensate equation \eqref{eq:cond1} with the real spectral support $\Gamma = [\eta_1-\varepsilon,\eta_1+\varepsilon]$. Note that, since the spectral support is complex, one could take, in principle, any 2D spectral support in the neighbourhood of $z_1$ to implement \eqref{eq:monoNLS}.
For the DOS~\eqref{eq:cond12}, the spectral parameters lie on the segment $\Lambda = \xi_1 + i \Gamma$ so that the fNLS integral kernel simplifies:
\begin{equation} \label{G_kdv_nls}
G_{\rm fNLS}(z,y) = G_{\rm KdV}(\eta,\mu),\quad z=\xi_1+i\eta,\; y = \xi_1+i\mu,
\end{equation}
and the integration $\iint_{\Lambda} (\rmd \re y)(\rmd \im y) \, G_{\rm fNLS}(z,y) \dots$ reduces to $ \int_{\Gamma} \rmd \mu \, G_{\rm KdV}(\eta,\mu) \dots$
Thus \eqref{eq:cond12} with $C=1$ 
corresponds to the DOS of a fNLS genus $1$ soliton condensate with the 1D complex spectral support $\xi_1 + i \Gamma$. Similarly to the KdV implementation, this choice for the DOS (dilute soliton condensate), ensures that the SG is dense and spatially uniform. This reduction to the KdV interaction kernel and a 1D real, spectral support enables us to leverage the results derived in Secs. \ref{sec:nsol} and \ref{sec:RP_num}. The DOS is implemented by choosing $\tilde z_k = \tilde \xi_k+i \tilde \eta_k$ with $\tilde \xi_k = \xi_1$ and $\tilde \eta_k$ distributed by $\phi(\eta)$ given in \eqref{eq:ks2} with $M=1$; the corresponding spatial density of solitons~$\kappa_s$ is still given by \eqref{eq:ks2}. 

In the general case, the factorisation \eqref{eq:cond12} no longer applies for the implementation of the DOS
\begin{equation}
f(z) = \sum_{j=1}^M w_j \delta(z-z_j) = \sum_{j=1}^M w_j \delta(\xi-\xi_j) \delta(\eta- \eta_j)
\end{equation}
if $\xi_j \neq \xi_k$ for $j \neq k$. 
We will assume that initially the DOS' of the left fast SG and the right slow SG can be factorised, i.e.
\begin{equation}
f(z;x,t=0) = \begin{cases}
\displaystyle \delta(\xi-\xi_-) \sum_j w^{(-)}_j \delta(\eta - \eta_j), &x<0,\\[4mm]
\displaystyle \delta(\xi-\xi_+)\sum_j w^{(+)}_j \delta(\eta - \eta_j), &x>0,
\end{cases}
\end{equation}
such that SG can be implemented via $C_\pm \delta(\xi-\xi_\pm) f^{(M_\pm)}(\eta;\lambda_1^\pm=0,\lambda_2^\pm,\dots,\lambda_{2M_\pm+1}^\pm)$ (cf. Sec. \ref{sec:RP_num2}).

\section{Numerical solutions and comparisons}
\label{sec:num2}

In the following, we numerically implement the Riemann problem for the KdV and fNLS SGs using the method described in Sec. \ref{sec:synthesis}. We set $\varepsilon_j=10^{-2}$  in all the numerical implementations, such that each component of the SG is quasi-monochromatic and has a finite density. For $\eta_j = {\cal O}(1)$, we can thus estimate the deviation from ``monochromaticity'' of the gas implemented numerically to $\varepsilon_j/\eta_j = {\cal O}(10^{-2})$, which results in a relative error of ${\cal O}(10^{-2})$ observed in the comparison with the analytical results in this section.

 We will consider the two following cases: the interaction of two monochromatic gases, $(M_-,M_+)=(1,1)$, and the interaction of a ``bichromatic'' gas with a monochromatic gas, $(M_-,M_+)=(2,1)$.  The moments of the wave fields $\aver{u},\aver{u^2},\aver{|\psi|^2}$ are obtained with an ensemble averaging of $R\in[50,200]$ realisations, implemented via the $n$-soliton solution, as well as spatial averaging over the length $L \in [75,175]$. Details of the averaging procedure are given in Appendix \ref{sec:aver}.

\subsection{KdV}

The parameters of the initial conditions used in the numerical implementations of the KdV SG Riemann problem  are summarised in Table \ref{tab:param}. 
\begin{table}[h]
\renewcommand{\arraystretch}{1.5}
\centering
\footnotesize
\begin{tabular}{p{1.2cm}p{4.3cm}p{4.5cm}p{4cm}}
\hline \hline
&spectral parameters & left densities & right densities \\
\hline
[KdV1] & $(\eta_1,\eta_2)=(0.7,1)$ & $w_2^{(-)}=0.14$ & $w_1^{(+)}=0.1$ \\

[KdV2] & $(\eta_1,\eta_2)=(0.7,1)$ & $w_2^{(-)}=0.12$ & $w_1^{(+)}=0.08$ \\

[KdV3] & $(\eta_1,\eta_2)=(0.7,1)$ & $w_2^{(-)}\in[0.05,0.14]$ & $w_1^{(+)}=0.1$ \\

[KdV4] & $\eta_1\in[0.5,0.95],\;\eta_2=1$ & $w_2^{(-)}=0.08$ & $w_1^{(+)}=0.12$ \\
\hline
[KdV5] & $(\eta_1,\eta_2,\eta_3)=(0.7,0.9,1.1)$ & $(w_2^{(-)},w_3^{(-)})=(0.06,0.08)$ & $w_1^{(+)}=0.08$ \\

[KdV6] & $(\eta_1,\eta_2,\eta_3)=(0.7,0.9,1.1)$ & $(w_2^{(-)},w_3^{(-)})=(0.06,0.09)$ & $w_1^{(+)}=0.08$ \\

[KdV7] & $(\eta_1,\eta_2,\eta_3)=(0.7,0.9,1.1)$ & $(w_2^{(-)},w_3^{(-)})=(0.4,0.6)w^{(-)}$ \newline with $w^{(-)} \in[0.05,0.15]$ & $w_1^{(+)}=0.1$ \\

[KdV8] & $(\eta_2,\eta_3)=(0.9,1.1)$ \newline with $\eta_1\in[0.5,0.9]$ & $(w_2^{(-)},w_3^{(-)})=(0.06,0.09)$ & $w_1^{(+)}=0.08$ \\
\hline \hline
\end{tabular}
\caption{Parameters of the KdV SG Riemann problem with the step initial condition \eqref{eq:init}, \eqref{eq:init2} implemented numerically.}
\label{tab:param}
\end{table}

\subsubsection{Collision of KdV monochromatic SGs}
\label{sec:monokdv}

\begin{figure}
  \unitlength=1cm
    \begin{picture}(15,9)   
\put(-1,0){\includegraphics[width=17.5cm]{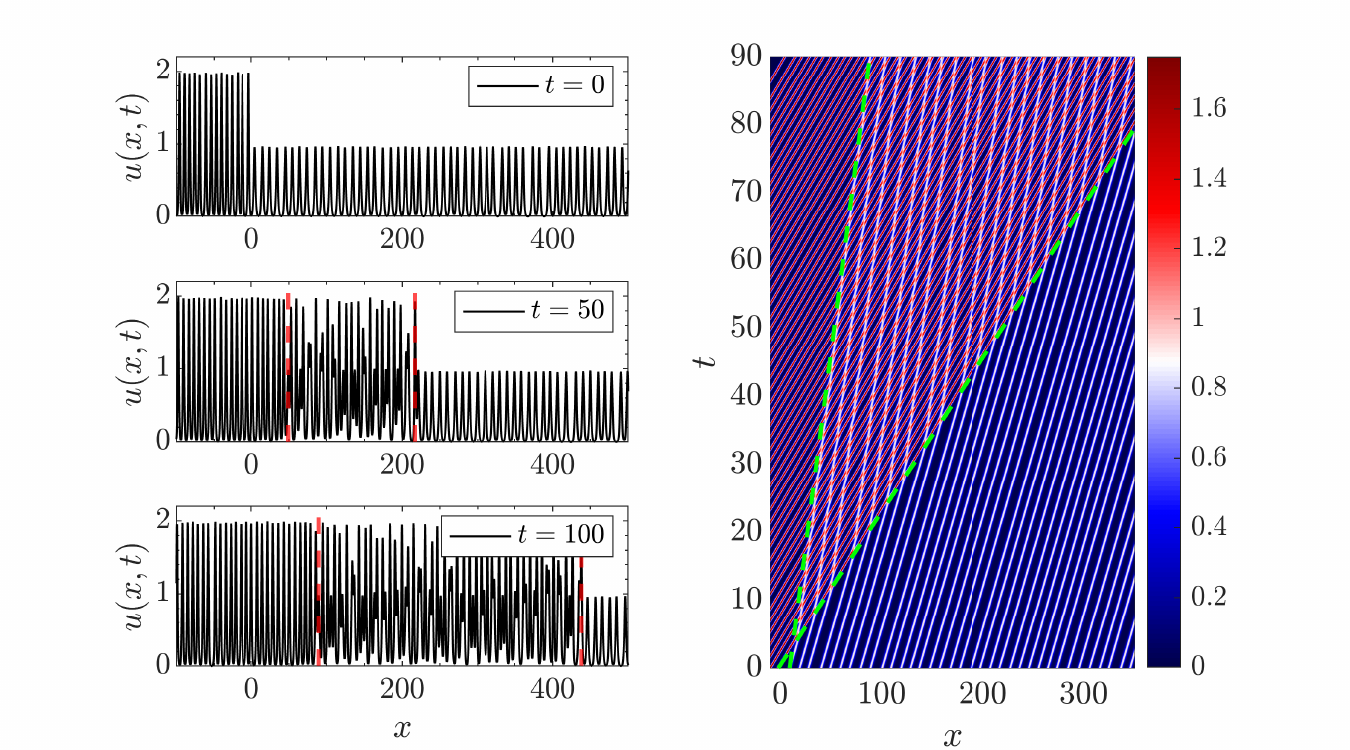}}
\put(0.5,0){\small (a)}
\put(8,0){\small (b)}
\end{picture}
  \caption{(a) Snapshots of a realisation $u(x,t)$ of the SG Riemann problem with $M_-=M_+=1$ and the set of parameters [KdV1] of Table \ref{tab:param}. The contact discontinuities separating the three distinct regions  are indicated by red dashed lines.
  (b) Corresponding spatio-temporal diagram in the $(x,t)$-plane. The colour corresponds to the intensity of the wavefield $u(x,t)$. The trajectories $x=c_j t$ of the contact discontinuities are highlighted by green dashed lines.}
  \label{fig:examples_kdv1}
\end{figure}

The typical evolution of a SG realisation of the Riemann problem with the initial condition \eqref{eq:init}, \eqref{eq:init2} and $M_-=M_+=1$, i.e.
\begin{equation}\label{eq:initmono}
\left(w_1(x,0),w_2(x,0) \right) =
	\begin{cases}
		\left(0,w_2^{(-)} \right), &\text{if } x<0,\\[2mm]
		\left(w_1^{(+)},0 \right), &\text{if } x> 0,
	\end{cases}
	\qquad s_{01}<s_{02},
\end{equation}
 is displayed in Fig. \ref{fig:examples_kdv1}. The field $u(x,t)$ represents trains of randomly distributed solitons with approximately equal amplitudes in the  regions $x<c_1 t$ and $x>c_2 t$, whereas in the  region $c_1 t< x <c_2 t$ it displays a beating pattern produced by the nonlinear interaction between the two soliton components. The individual trajectory of each soliton can be followed in the spatio-temporal plot in Fig. \ref{fig:examples_kdv1}(b): with the chosen colour scheme, 
the trajectories of slow solitons (amplitude $2\eta_1^2$) are depicted by white lines and the trajectories of fast solitons (amplitude $2\eta_2^2$) are depicted by red lines. After each soliton collision, the positions of the solitons $\eta_1$ are phase-shifted backward, and the positions of the solitons $\eta_2$ are phase-shifted forward, resulting, on average,  in a deceleration for the slow solitons displayed by the increase of the slope of their trajectories (white lines) after the interaction ($c_1 < s_{01}$) and an acceleration for the fast solitons ($c_2 > s_{02}$) displayed by the decrease of the slope of the red lines (see the analysis in Sec. \ref{sec:riemann}).

\begin{figure}
  \centering
  \includegraphics[width=0.6\textwidth]{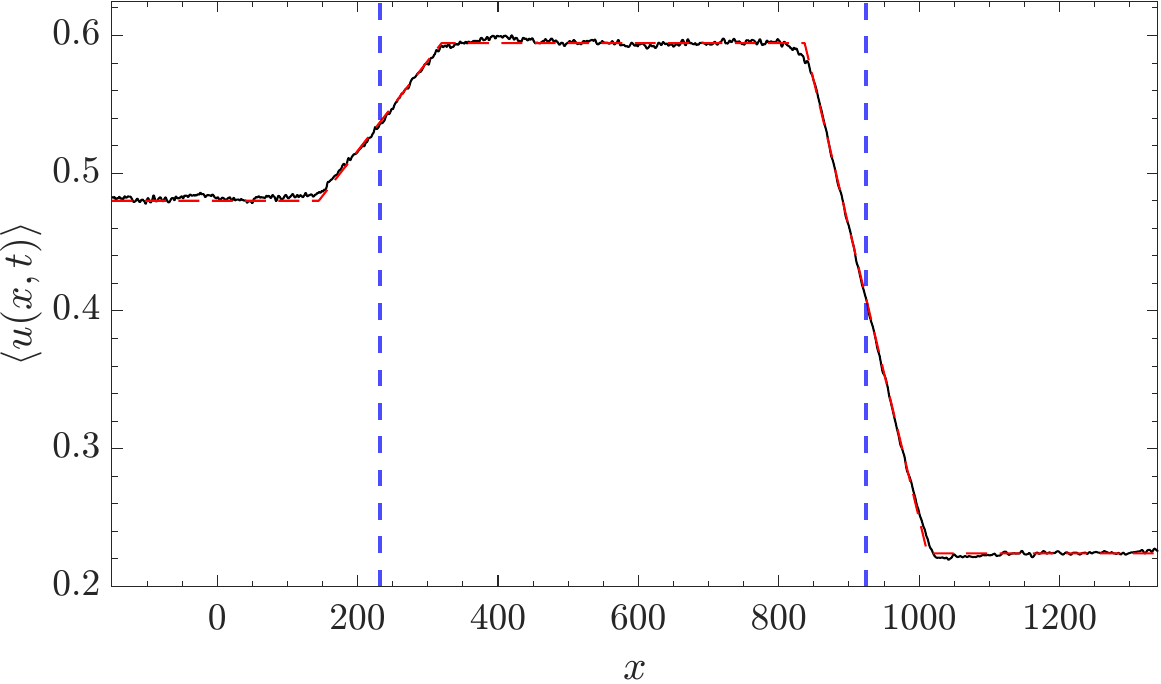}
  \caption{The  moment $\aver{u(x,t)}$ in the SG Riemann problem at $t=196$ with the set of parameters [KdV2] of Table \ref{tab:param}. 
  The black solid line corresponds the value of $\langle u \rangle$ computed numerically with $R=50$ SG realisations and spatial averaging over $L=75$. The red dashed line corresponds to the exact solution \eqref{eq:sol_moments}, \eqref{eq:w12b} augmented by a spatial average, see \eqref{eq:aver_num2}. The dashed blue lines indicate the position of the contact discontinuities $x=c_j t$.}
  \label{fig:examples_kdv2}
\end{figure}

\begin{figure}
  \unitlength=1cm
  \begin{picture}(15,6.3)   
\put(0,0){\includegraphics[width=7.5cm]{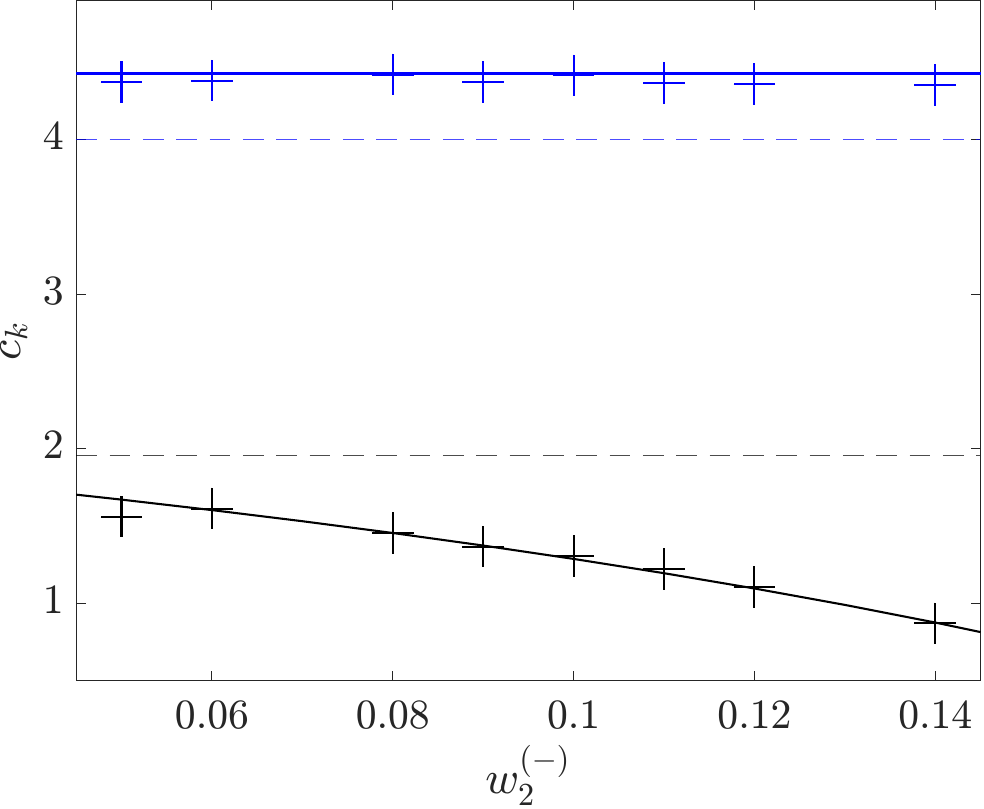}}
\put(8.,0){\includegraphics[width=7.5cm]{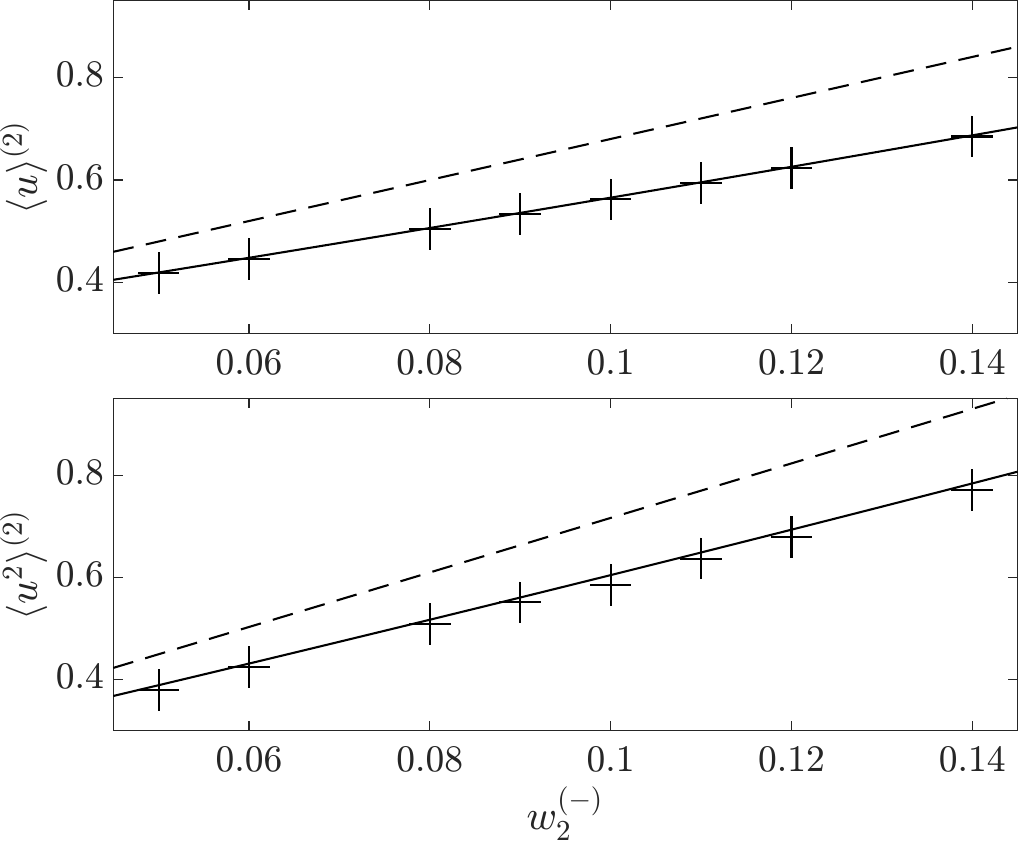}}
\put(0,0.5){\small (a)}
\put(8.3,0.3){\small (c)}
\put(8.3,3.4){\small (b)}
\end{picture}
  \caption{Variation of (a) the  speeds $c_k$ of contact discontinuities, (b) the statistical moment $\aver{u}^{(2)}$, and (c) the moment $\aver{u^2}^{(2)}$ with respect to the left initial density $w_2^{(-)}$ for the set of parameters [KdV3] of Table \ref{tab:param}. 
    The solid lines correspond to the analytical expressions \eqref{eq:c1c2} and \eqref{eq:int2}. The dashed lines correspond to the same parameters obtained with the reference (non-interacting SGs) solution \eqref{eq:sol_free}, cf. Sec. \ref{sec:riemann}. The  markers are obtained from the numerical simulations.}
  \label{fig:kdv1}
\end{figure}
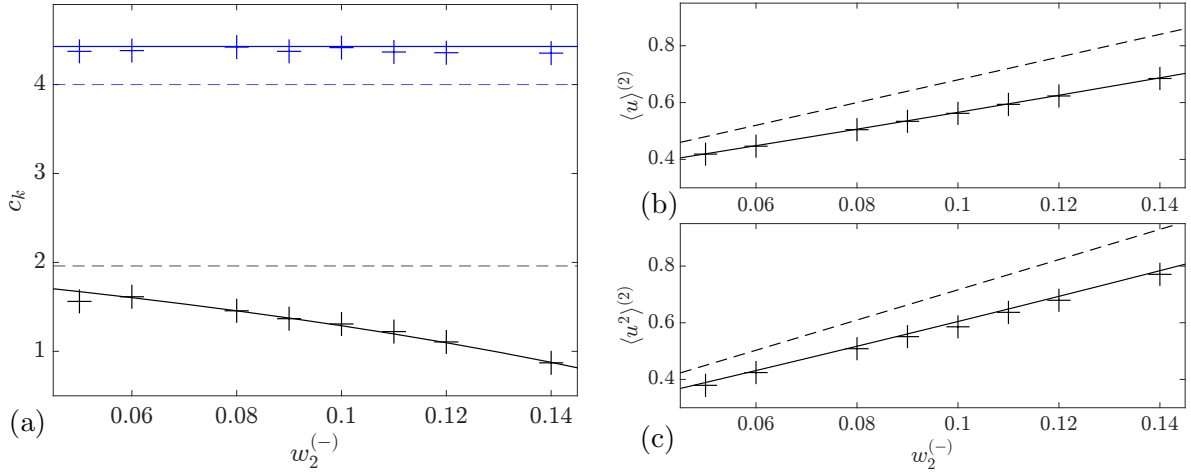

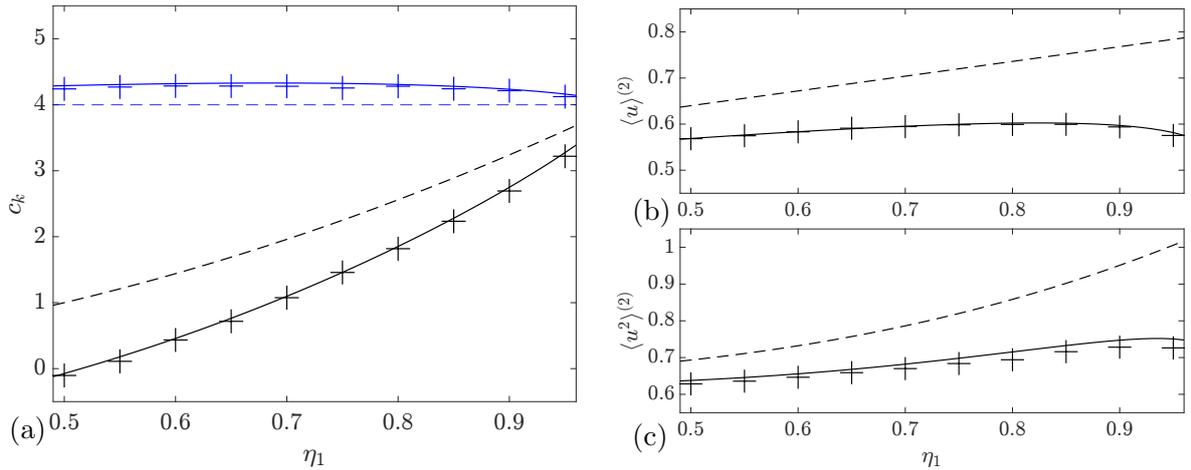
\begin{figure}
   \unitlength=1cm
  \begin{picture}(15,6.3)   
\put(0,0){\includegraphics[width=7.5cm]{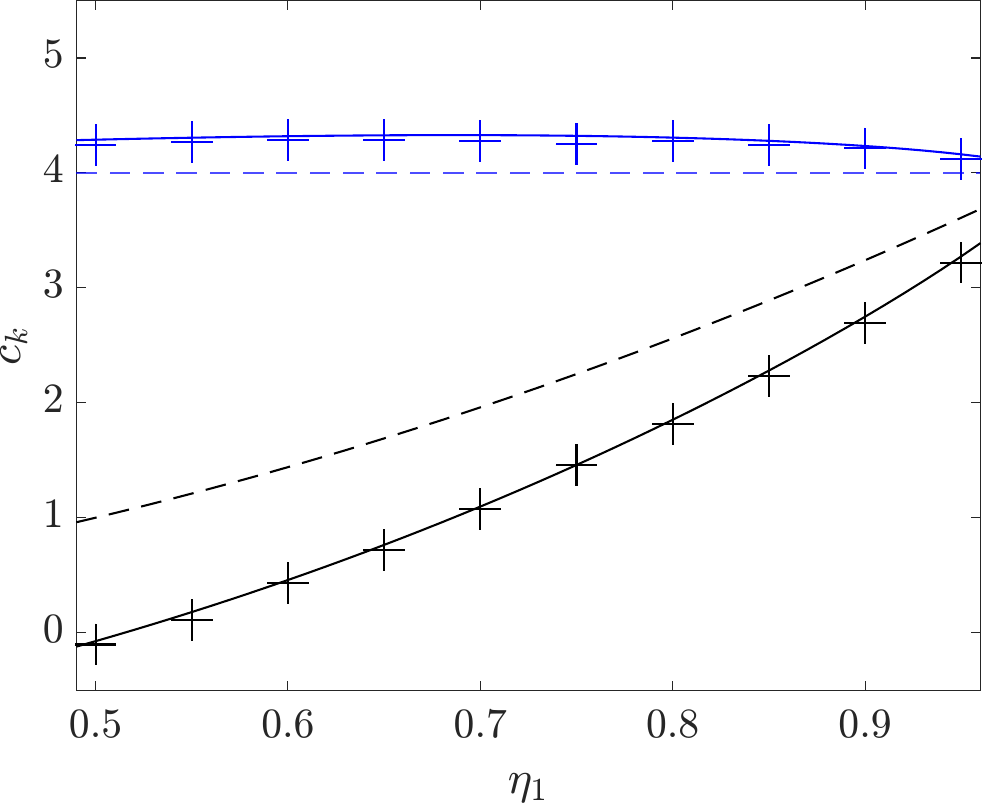}}
\put(8.,0){\includegraphics[width=7.5cm]{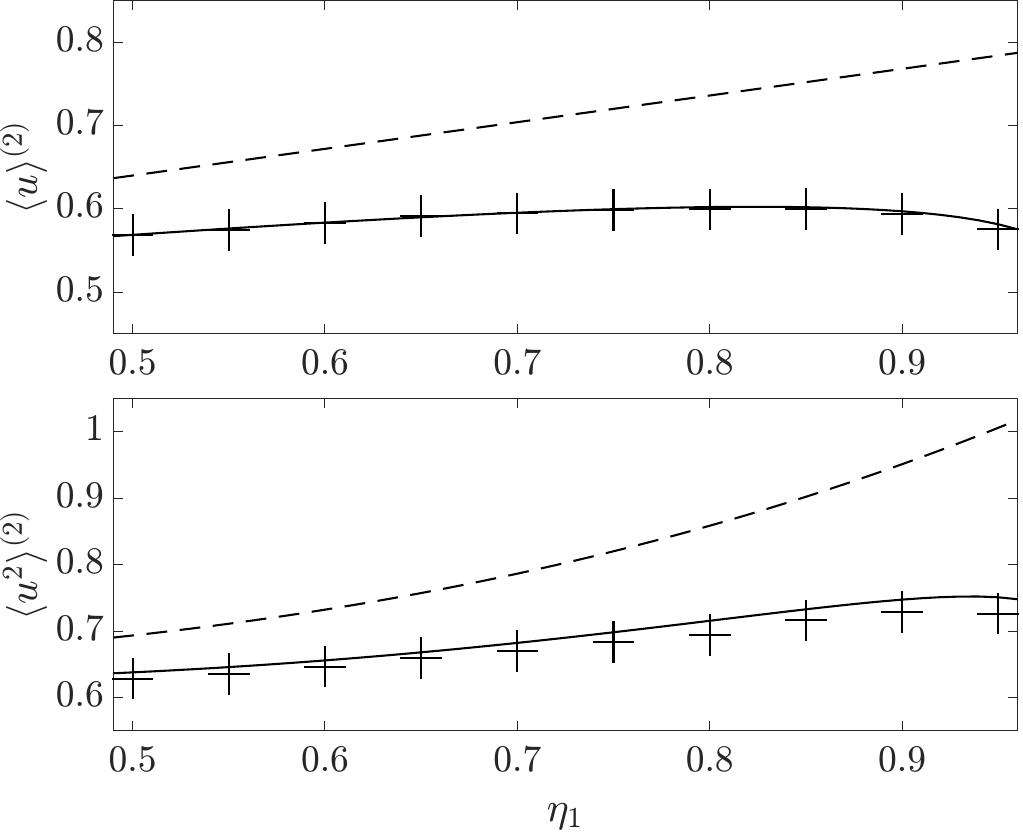}}
\put(0,0.4){\small (a)}
\put(8.2,0.3){\small (c)}
\put(8.2,3.3){\small (b)}
\end{picture}
  \caption{Variation of  (a)  the speeds $c_k$ of contact discontinuities, (b) the  moment $\aver{u}^{(2)}$, and (c) the  moment $\aver{u^2}^{(2)}$ with respect to the spectral parameter $\eta_1$ for the set of parameters [KdV4] of Table \ref{tab:param}. The legend is defined in Fig. \ref{fig:kdv1}.}
  \label{fig:kdv2}
\end{figure}

Fig. \ref{fig:examples_kdv2} displays the moment $\aver{u(x,t)}$ computed numerically with $R=50$ realisations of the Riemann problem for the set of parameters [KdV2] of Table \ref{tab:param}. The discontinuities at $x=c_i t$ are replaced by the lines of finite slopes due to the averaging procedure used to compute the moments, as explained in Appendix \ref{sec:aver}. 
The variation of $\aver{u(x,t)}$ computed numerically is in good agreement with the analytical solution \eqref{eq:sol_moments}, \eqref{eq:c1c2}, \eqref{eq:int2} derived in Sec. \ref{sec:riemann}; for an exact comparison with the numerics, the solution with discontinuities \eqref{eq:sol_moments} are replaced by \eqref{eq:aver_num2} to take into account the averaging procedure. The component densities $w_1^{(2)}$ and $w_2^{(2)}$ in the interaction region are smaller than the initial densities $w_1^{(+)}$ and $w_2^{(-)}$ due to the effective repelling between the solitons, resulting in the moment $\aver{u}^{(2)}$ in the interaction region being smaller than $\aver{u}^{(2)}_{\rm free} = 4\eta_1 w_1^{(+)} + 4\eta_2 w_2^{(-)}$, see \eqref{eq:int2}. Similarly, the moment $\aver{u^2}^{(2)}$ is smaller than $\aver{u^2}^{(2)}_{\rm free} = 16/3\eta_1^3 w_1^{(+)} + 16/3\eta_2^3 w_2^{(-)}$ as shown later in Figs. \ref{fig:kdv1}(c) and \ref{fig:kdv2}(c).

The resolution of the SG Riemann problem in Sec. \ref{sec:riemann} yields the analytical expression of speeds $c_j$ of the constant discontinuities (cf. \eqref{eq:c1c2}) as well as the moments $\aver{u}^{(2)}$ (cf. \eqref{eq:int2}) and $\aver{u^2}^{(2)}$ in the interaction region in terms of the parameters $w_1^{(+)}$, $w_2^{(-)}$, $\eta_1$ and $\eta_2$. Thanks to the new numerical  implementation of SGs described in Sec. \ref{sec:synthesis}, these analytical results can now be verified in dense regime of SGs, with $w_1^{(+)}+w_2^{(-)}$ up to $0.3$, which further the numerical studies of \cite{carbone_macroscopic_2016}.
We thus reproduce the numerical realisations of Figs. \ref{fig:examples_kdv1} and \ref{fig:examples_kdv2} for different values of  the initial densities and spectral parameters, and extract the quantities $c_j$, $\aver{u}^{(2)}$ and $\aver{u^2}^{(2)}$ from the variation of the moments $\aver{u(x,t)}$ and $\aver{u^2(x,t)}$, cf. Appendix \ref{sec:aver}. 

Fig. \ref{fig:kdv1}(a) displays the comparison between the  speeds $c_1$ and $c_2$ of contact discontinuities fitted from the numerical averages with the analytical result \eqref{eq:c1c2} for different values of the initial ``left'' density $w_2^{(-)}$ and the set of parameters [KdV3] of Table \ref{tab:param}.
Since the $\eta_2$-solitons  are only interacting with the $\eta_1$-solitons, their effective velocity $c_2$ is independent from the initial density $w_2^{(-)}$, cf. \eqref{eq:c1c2}, which is confirmed by the simulations. On the contrary, as the density of solitons $\eta_2$ increases, the solitons $\eta_1$ are slowed down i.e. their effective velocity decreases. For the same problem, Figs. \ref{fig:kdv1}(b) and (c) display the comparison between the moments $\aver{u}^{(2)}$ and $\aver{u^2}^{(2)}$ extracted from the numerical solutions with the analytical result \eqref{eq:averu}, \eqref{eq:w12b}. Both moments increase with $w_2^{(-)}$ in good agreement with the analytical expression.

Fig. \ref{fig:kdv2}(a)  displays the variation of the  speeds of the contact discontinuities with the ``right'' spectral parameter $\eta_1$ and the set of parameters [KdV4] of Table \ref{tab:param}.
Since $c_1$ can be seen as the effective velocity of the $\eta_1$-solitons, it trivially increases with the spectral parameter similar to the free soliton velocity $s_{01}=4\eta_1^2$. Besides both discontinuity speeds depend on the phase-shift kernel $G(\eta_1,\eta_2)$ which diverges as $\eta_1$ approaches $\eta_2$. 
This non-trivial dependence on the spectral parameter is more pronounced in Figs. \ref{fig:kdv2}(b) and (c) which display a non-monotonic variation of the moments $\aver{u}^{(2)}$ and $\aver{u^2}^{(2)}$ with $\eta_1$ in the interaction region, again, in good agreement with the formulae \eqref{eq:averu}, \eqref{eq:w12b}. Such variation is not expected in the reference non-interacting case where the moments $\aver{u}_{\rm free}^{(2)}$ and $\aver{u^2}_{\rm free}^{(2)}$   linearly depend on $\eta_1$ or $\eta_1^3$ respectively, cf. for instance \eqref{eq:sol_moments_free}.

\subsubsection{Collision of KdV polychromatic SGs}
\label{sec:bikdv}

We now consider the collision of a bichromatic SG ($M=2$) and a monochromatic SG, implemented with the following initial condition
\begin{equation}
\label{eq:init3}
\left(w_1(x,0),w_2(x,0),w_3(x,0) \right) =
	\begin{cases}
		\left(0,w_2^{(-)},w_3^{(-)} \right), &\text{if } x<0,\\[2mm]
		\left(w_1^{(+)},0,0 \right), &\text{if } x> 0,
	\end{cases}
	\qquad s_{01}<s_{02}<s_{03}.
\end{equation}
This furthers previous numerical works \cite{congy_soliton_2021}, where the interaction between polychromatic gases was investigated numerically for rarefied bidirectional SGs.

\begin{figure}
  \unitlength=1cm
    \begin{picture}(15,9)   
\put(-1,0){\includegraphics[width=17.5cm]{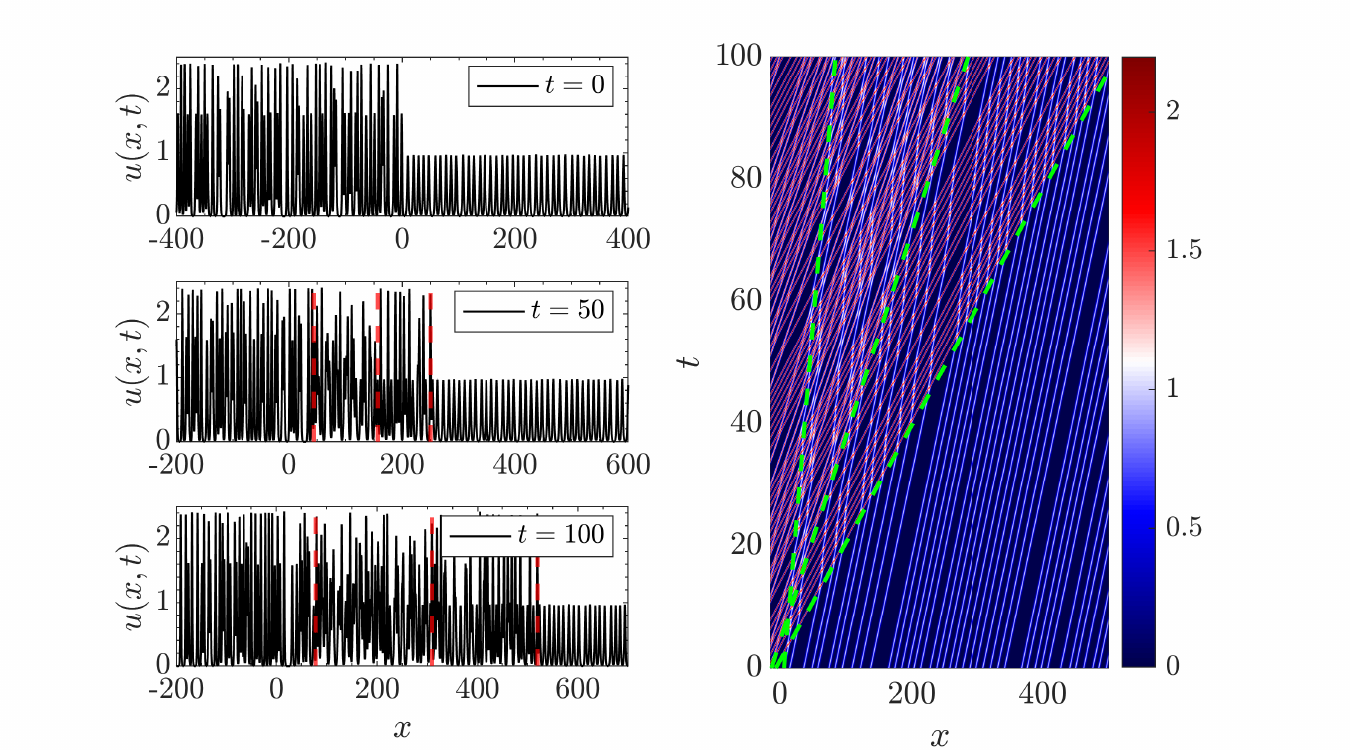}}
\put(0.5,0){\small (a)}
\put(8,0){\small (b)}
\end{picture}
  \caption{(a) Snapshots of a realisation $u(x,t)$ of the SG Riemann problem with $M_-=2$, $M_+=1$ and the set of parameters [KdV5] of Table \ref{tab:param}.
  (b) Corresponding spatio-temporal diagram in the $(x,t)$-plane. 
  The legend is defined in Fig. \ref{fig:examples_kdv1}. }
  \label{fig:examples_kdv3}
\end{figure}

\begin{figure}
  \centering
  \includegraphics[width=0.6\textwidth]{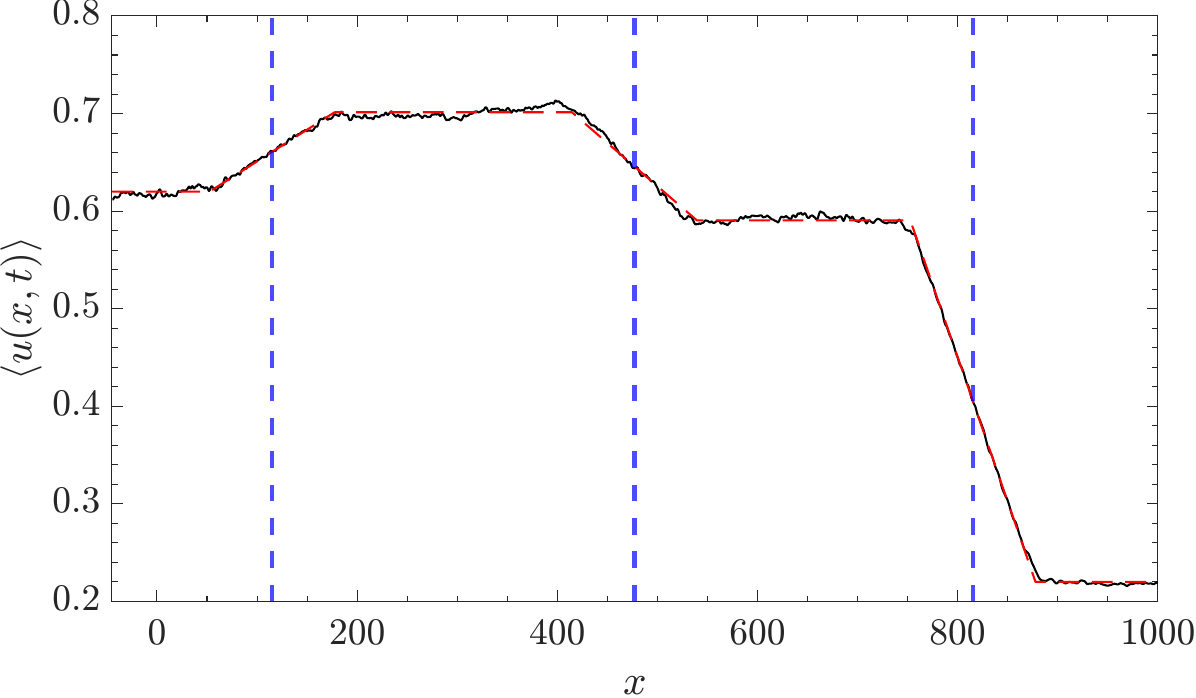}
  \caption{The moment $\aver{u(x,t)}$ in the SG Riemann problem at $t=158$ with the set of parameters [KdV6] of Table \ref{tab:param}. 
  The black solid line corresponds the moment computed numerically with $R=50$ SG realisations and spatial averaging over $L=175$. The red dashed line corresponds to the solution \eqref{eq:sol_moments}, \eqref{eq:sol3comp} augmented by a spatial average, where the coefficients are obtained solving the Rankine-Hugoniot conditions. The dashed blue lines indicate the position of the contact discontinuities $x=c_j t$.}
  \label{fig:examples_kdv4}
\end{figure}

Snapshots of a SG realisation of the Riemann problem with the initial condition \eqref{eq:init3} and the set of parameters [KdV5] of Table \ref{tab:param} are displayed in Fig. \ref{fig:examples_kdv3}(a). 
The SG has now two interactions regions (see the analytical description in Sec. \ref{sec:riemann}):
\begin{itemize}
\item Region (2) ($c_1 t < x < c_2 t$) where the 3 monochromatic components interact. 
In this region the $\eta_1$- and $\eta_2$-solitons have respectively the effective velocities $c_1$ and $c_2$, cf. \eqref{eq:z}.
\item Region (3) ($c_2 t < x < c_3 t$) which only includes 2 monochromatic components $\eta=\eta_1$ and $\eta=\eta_3$. 
In this region the $\eta_3$-solitons  have the effective velocity $c_3$, cf. \eqref{eq:z}.
\end{itemize}
The individual trajectory of each soliton can be followed in the spatio-temporal plot in Fig. \ref{fig:examples_kdv3}(b): with the chosen colour scheme 
the trajectories of the solitons with parameter $\eta_1=0.7$ are depicted by white lines, and the trajectories of the solitons with parameters $\eta_2=0.9$, $\eta_3=1.1$ by red lines. Fig. \ref{fig:examples_kdv4} displays the moment $\aver{u(x,t)}$ computed numerically with $R=50$ realisations of the Riemann problem with the set of parameters [KdV6] of Table \ref{tab:param}, where the 4 plateaus can be clearly identified after averaging. Fig. \ref{fig:examples_kdv4}  also displays the solution \eqref{eq:sol_moments}, \eqref{eq:sol3comp}, augmented by the spatial average \eqref{eq:aver_num2}, where $c_j$ and $w_j^{(k)}$ are obtained by numerically solving the Rankine-Hugoniot conditions \eqref{eq:RH}.

Similar to Sec. \ref{sec:monokdv}, we reproduce the numerical realisations of Fig. \ref{fig:examples_kdv4} for different initial densities and spectral parameters, and extract  the quantities $c_j$, $\aver{u}^{(k)}$ and $\aver{u^2}^{(k)}$ from the variation of the moments $\aver{u(x,t)}$ and $\aver{u^2(x,t)}$.
 We then compare these values with the results  obtained via the numerical resolution of the Rankine-Hugoniot conditions \eqref{eq:RH}, cf. Sec. \ref{sec:riemann}. 

\begin{figure}
\unitlength=1cm
  \begin{picture}(15,4.5)   
\put(0,0){\includegraphics[width=5.2cm]{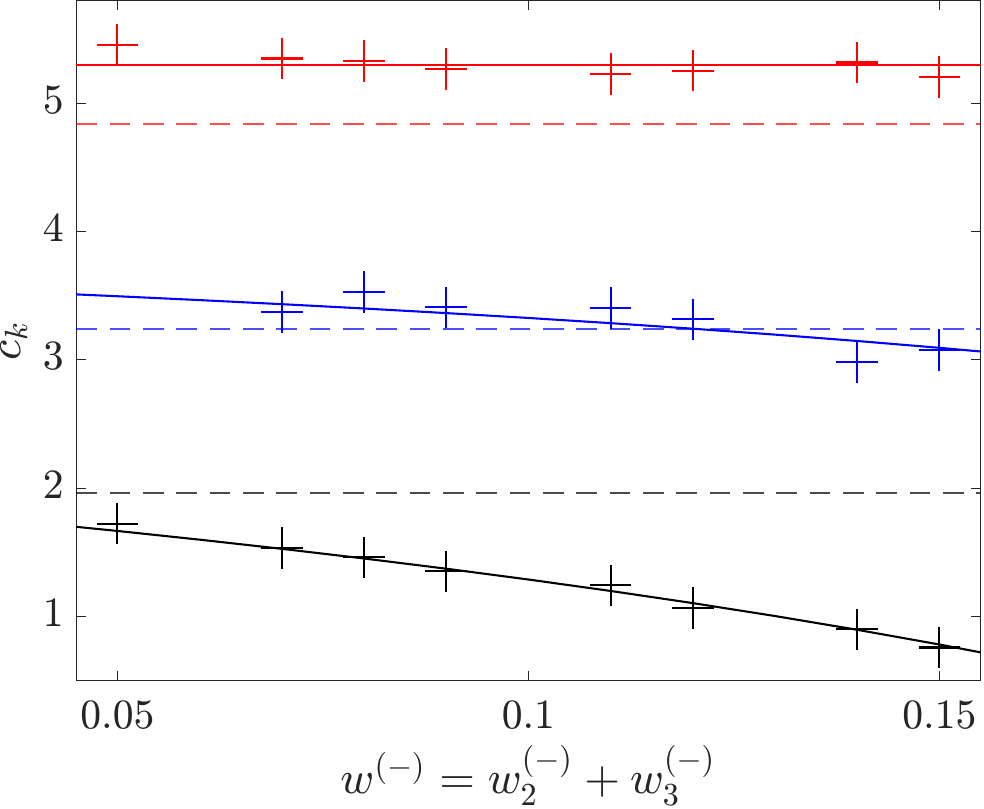}}
\put(5.3,0){\includegraphics[width=5.2cm]{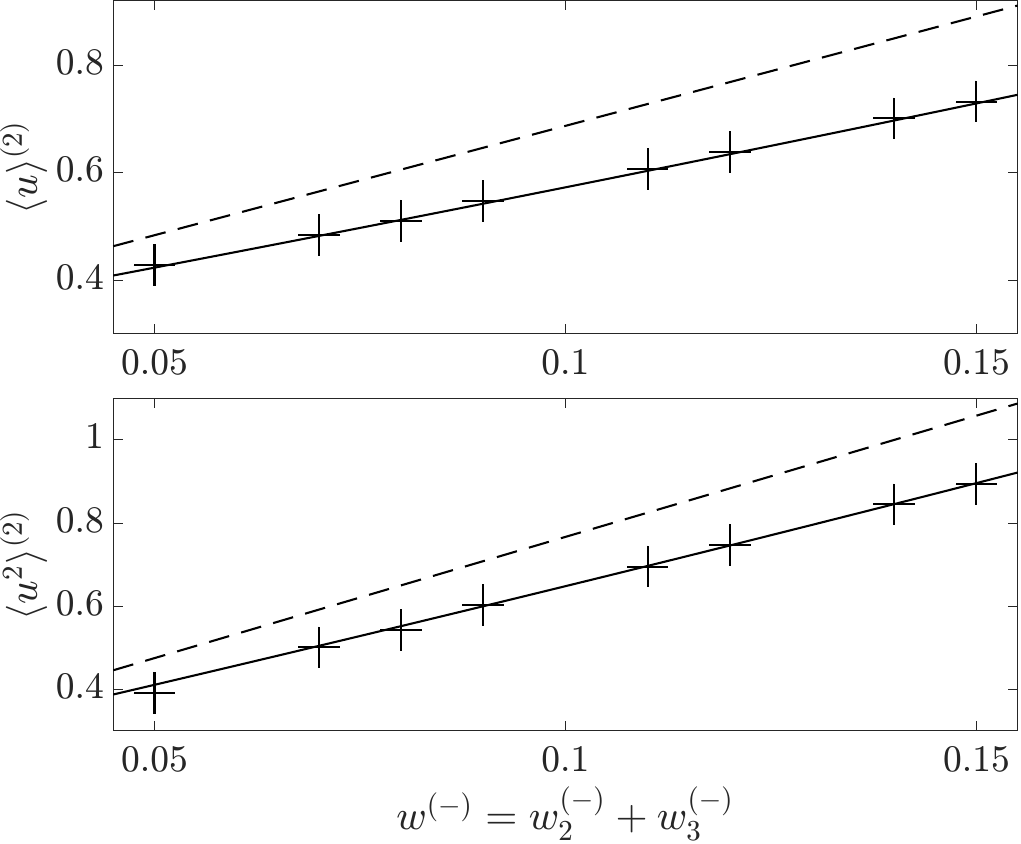}}
\put(10.6,0){\includegraphics[width=5.2cm]{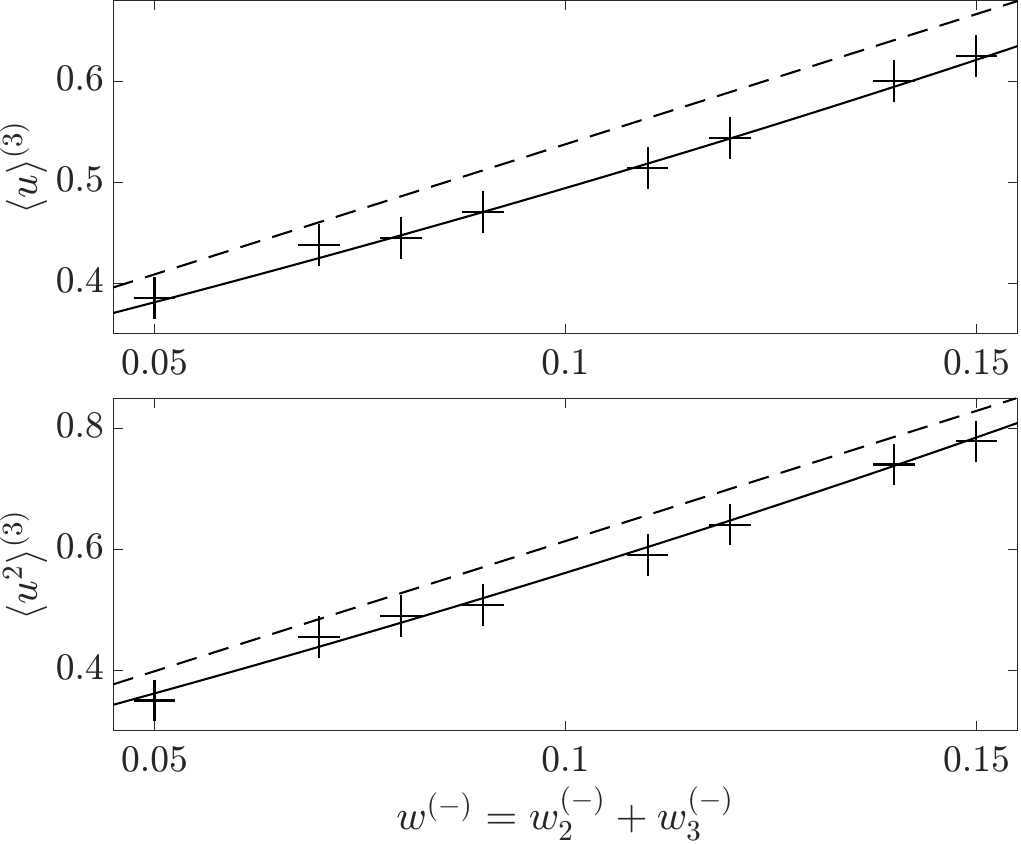}}
\put(0,0){\scriptsize (a)}
\put(5.3,0){\scriptsize (d)}
\put(5.3,2.3){\scriptsize (b)}
\put(10.6,0){\scriptsize (e)}
\put(10.6,2.3){\scriptsize (c)}
\end{picture}
  \caption{Variation of (a) the speeds $c_k$ of contact discontinuities, (b,c) the moments $\aver{u}^{(j)}$, and (d,e) the moments $\aver{u^2}^{(j)}$ with respect to the left total density $w^{(-)}=w_2^{(-)}+w_3^{(-)}$ for the set of parameters [KdV7] of Table \ref{tab:param}. The solid lines are obtained solving the Rankine-Hugoniot conditions and the markers are extracted from the SG numerical simulations. The dashed lines correspond to the same parameters obtained with the reference solution \eqref{eq:sol_free} for non-interacting SGs. 
  }
  \label{fig:kdv3}
\end{figure}

Fig. \ref{fig:kdv3} displays the comparison between the  speeds of discontinuities and the moments obtained numerically and from the Rankine-Hugoniot conditions for different values of the initial left total density $w^{(-)}=w_2^{(-)}+w_3^{(-)}$ and the set of parameters [KdV7] of Table \ref{tab:param}. The speed $c_1$
significantly decreases as $w^{(-)}=w_2^{(-)}+w_3^{(-)}$ increases. Indeed, the positions of the $\eta_1$-solitons  are now phase-shifted backward due to the interactions with the $\eta_2$- and $\eta_3$-solitons. On the contrary, the speed $c_3$
remains unchanged as the density of solitons $\eta_1$ does not change, cf. the analytical expression \eqref{eq:c3}.
Similar to the collision of monochromatic SGs we have $\aver{u}^{(k)}<\aver{u}_{\rm free}^{(k)}$ and $\aver{u^2}^{(k)}<\aver{u^2}_{\rm free}^{(k)}$, and the interaction between the 3 components tends to rarefy the gas. As expected, these moments increase if the initial densities increase, in full agreement with the theory developed in Sec. \ref{sec:riemann}. 

\begin{figure}
\unitlength=1cm
  \begin{picture}(15,4.5)   
\put(0,0){\includegraphics[width=5.2cm]{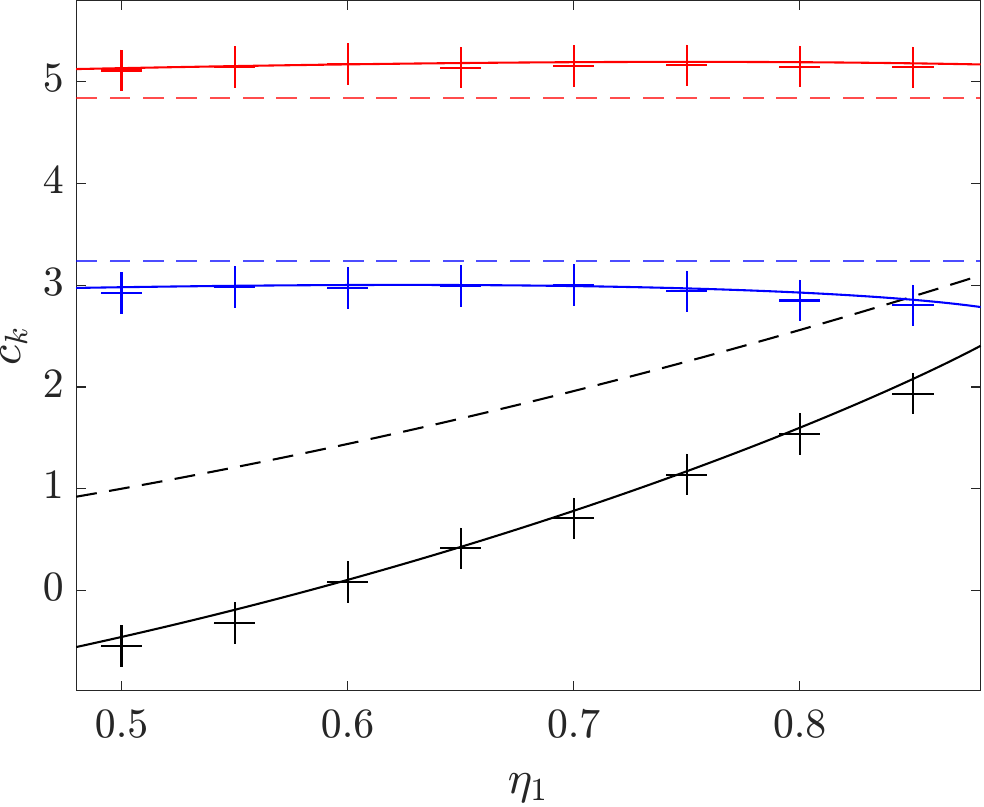}}
\put(5.3,0){\includegraphics[width=5.2cm]{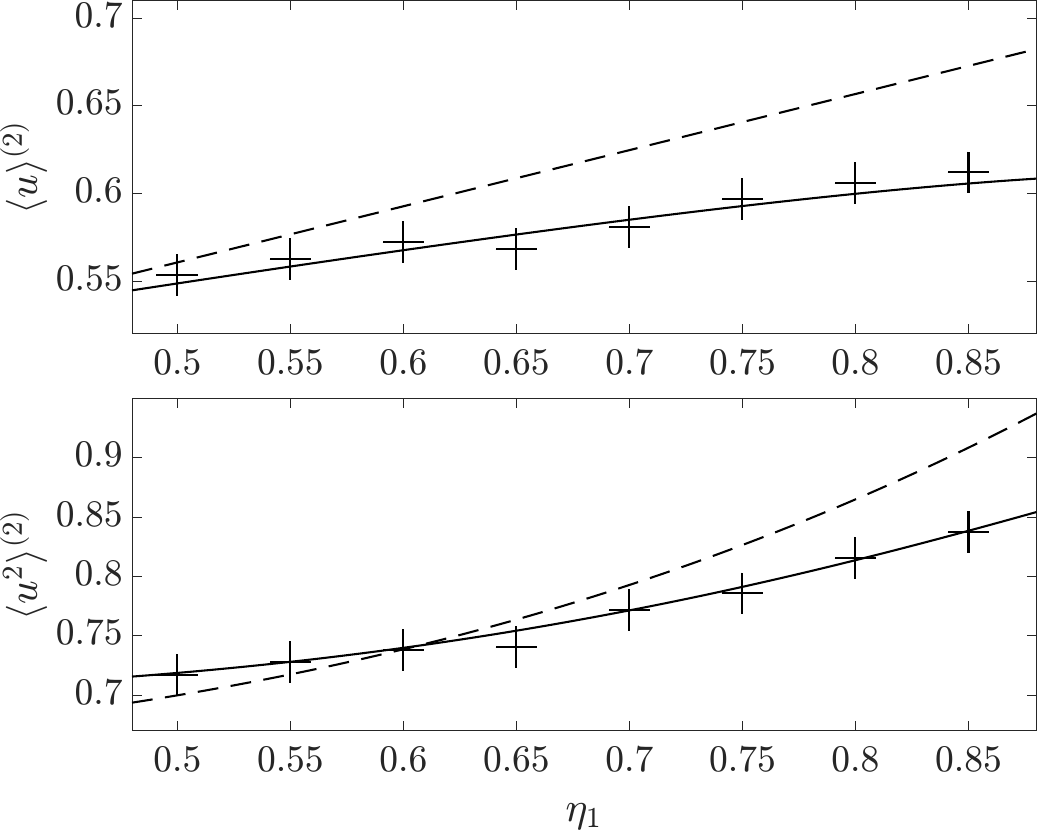}}
\put(10.6,0){\includegraphics[width=5.2cm]{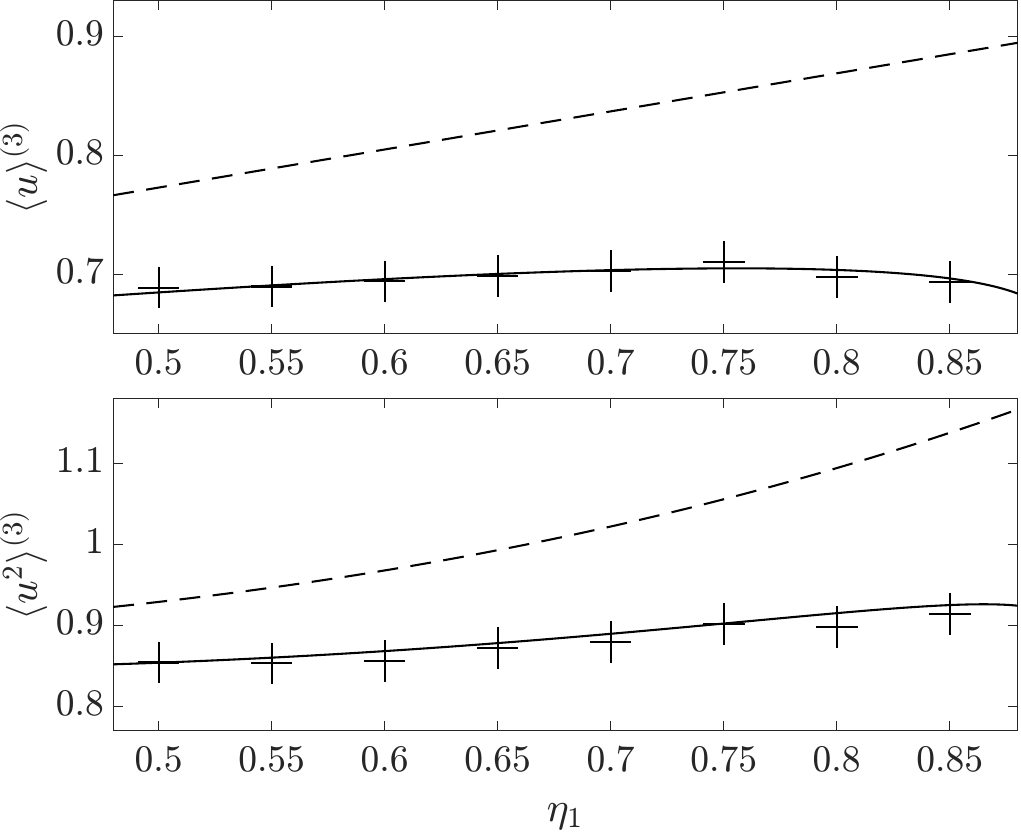}}
\put(0,0){\scriptsize (a)}
\put(5.3,0){\scriptsize (d)}
\put(5.3,2.3){\scriptsize (b)}
\put(10.6,0){\scriptsize (e)}
\put(10.6,2.3){\scriptsize (c)}
\end{picture}
  \caption{Variation of (a)  the speeds $c_k$ of contact discontinuities, (b,c) the moments $\aver{u}^{(j)}$, and (d,e) the moments $\aver{u^2}^{(j)}$ with respect to the spectral parameter $\eta_1$  for the set of parameters [KdV8] of Table \ref{tab:param}. The legend is defined in Fig. \ref{fig:kdv3}.
  }
  \label{fig:kdv4}
\end{figure}
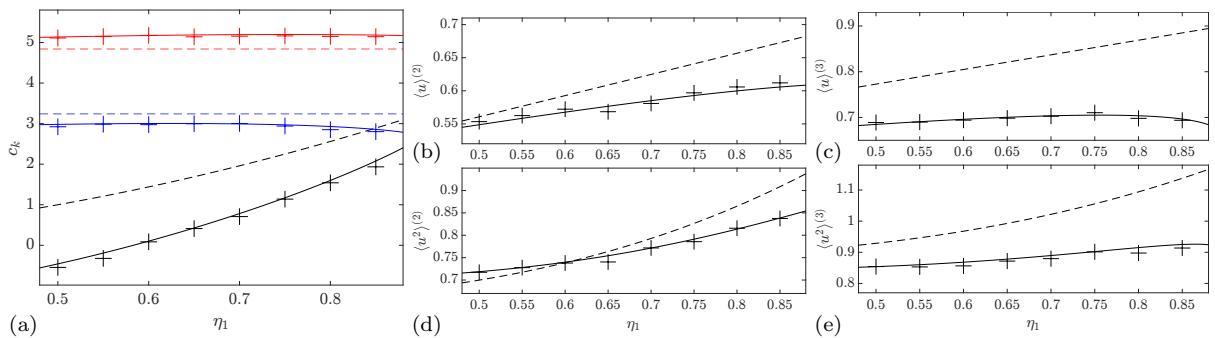

Fig. \ref{fig:kdv4} displays the comparison between the shock speeds and moments obtained numerically and from the Rankine-Hugoniot conditions for different values of the right spectral parameter $\eta_1$, and the set of parameters [KdV8] of Table \ref{tab:param}. 
Although the 3 monochromatic components 
co-exist in the interaction region (2), one can observe that the variation of $c_1$ and $c_2$ qualitatively resembles the variation of speeds computed in the monochromatic collision case, see Fig. \ref{fig:kdv2}(a). $c_3$ is almost unchanged as $\eta_1$ varies, due to the weak, but non-negligible, phase-shift between solitons $\eta_1$ and $\eta_3$.

Once again $G(\eta_1,\eta_2)$ and $G(\eta_1,\eta_3)$ have a non-trivial dependence on $\eta_1$, which is manifest in Figs. \ref{fig:kdv4}(b) displaying the variation of the moments $\aver{u}^{(2)}$ and $\aver{u^2}^{(2)}$ with $\eta_1$ in the interaction region (2). In both regions, the moments extracted numerically compare very well with the theoretical predictions.

\subsection{fNLS}

The parameters of the different initial conditions for the fNLS SG Riemann problem are summarised in the table \ref{tab:param2}.

\begin{table}[h]
\renewcommand{\arraystretch}{1.5}
\centering
\footnotesize
\begin{tabular}{p{1.1cm}p{5.3cm}p{4.8cm}p{2.8cm}}
\hline \hline
&spectral parameters & left densities & right densities \\
\hline
[fNLS1] & $(z_1,z_2)=(0.01+0.4i,0.5i)$ & $w_2^{(-)}=0.08$ & $w_1^{(+)}=0.07$ \\

[fNLS2] & $(z_1,z_2)=(0.01+0.5i,-0.01+0.5i)$ & $w_2^{(-)}=0.08$ & $w_1^{(+)}=0.08$ \\

[fNLS3] & $(z_1,z_2)=(0.5+0.5i,0.5i)$ & $w_2^{(-)}=0.08$ & $w_1^{(+)}\in[0.05,0.08]$ \\

[fNLS4] & $(z_1,z_2)=(\xi_1+0.5i,0.5i)$ \newline with $\xi_1\in[0.3,0.7]$ & $w_2^{(-)}=0.08$ & $w_1^{(+)}=0.08$ \\

[fNLS5] & $(z_1,z_2)=(0.5+\eta_1i,0.5i)$ \newline with $\eta_1\in[0.3,0.7]$ & $w_2^{(-)}=0.08$ & $w_1^{(+)}=0.05$ \\

\hline

[fNLS6] & $(z_1,z_2,z_3)=(0.01+0.9i,0.7i,0.9i)$ & $(w_2^{(-)},w_3^{(-)})=(0.39,0.61)w^{(-)}$ \newline with $w^{(-)} \in[0.15,0.19]$ & $w_1^{(+)}=0.14$ \\

[fNLS7] & $(z_1,z_2,z_3)=(\xi_1+0.9i,0.7i,0.9i)$ \newline with $\xi_1\in[0.01,0.03]$& $(w_2^{(-)},w_3^{(-)})=(0.07,0.11)$ & $w_1^{(+)}=0.14$ \\

[fNLS8] & $(z_1,z_2,z_3)=(0.01+\eta_1i,0.9i,0.7i)$ \newline with $\eta_1\in[0.68,0.72]$& $(w_2^{(-)},w_3^{(-)})=(0.11,0.07)$ & $w_1^{(+)}=0.11$ \\

\hline \hline
\end{tabular}
\caption{Parameters of the fNLS SG Riemann problem with the step initial condition \eqref{eq:init}, \eqref{eq:init2} implemented numerically.}
\label{tab:param2}
\end{table}

\subsubsection{Collision of  monochromatic fNLS SGs}
\label{sec:mononls}

\begin{figure}
  \unitlength=1cm
  \begin{picture}(15,9)   
\put(-1,0){\includegraphics[width=17.5cm]{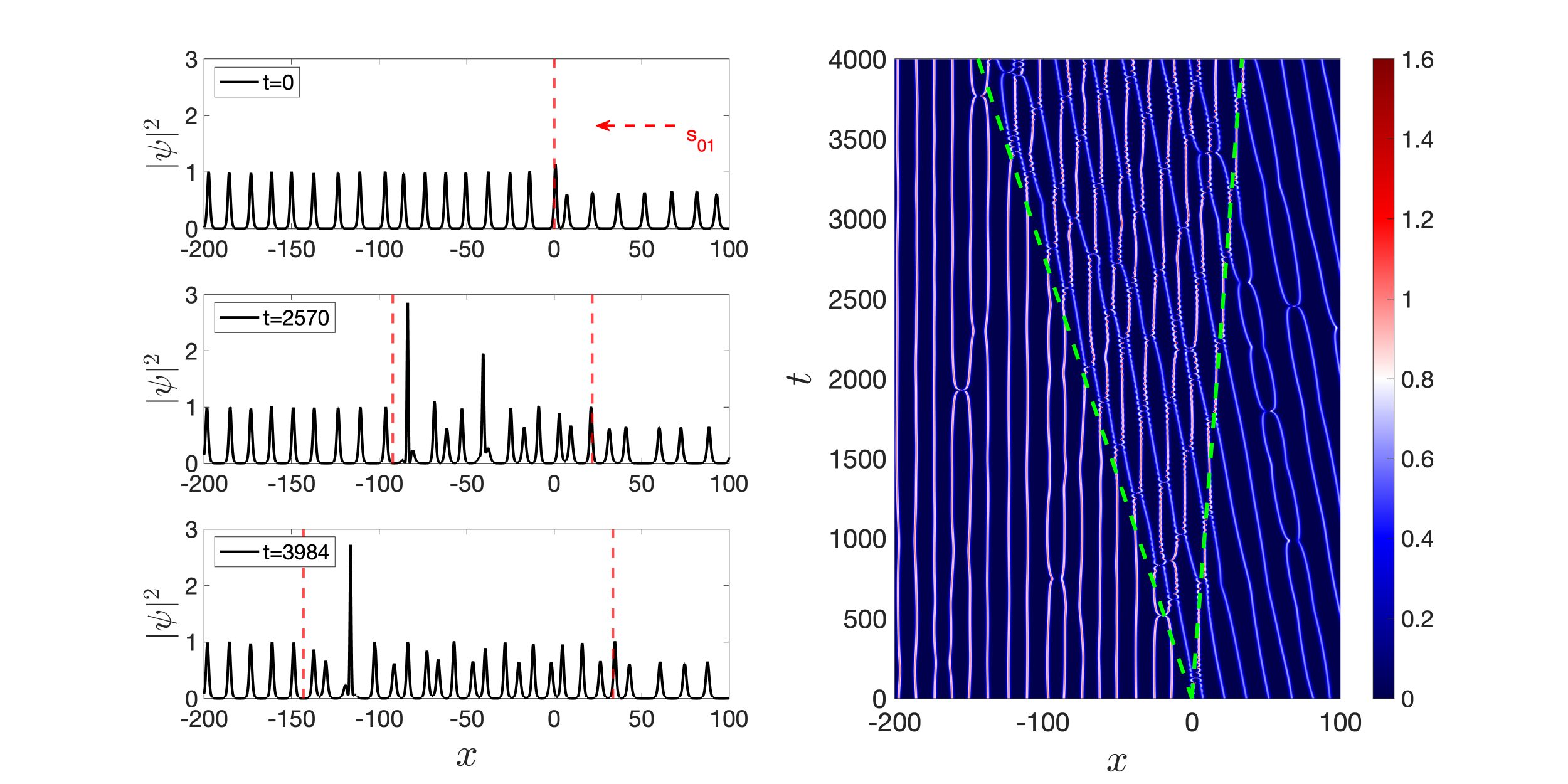}}
\put(0.5,0){\small (a)}
\put(8,0){\small (b)}
\end{picture}
  \caption{(a) Snapshots of a realisation $|\psi(x,t)|^2$ of the SG Riemann problem with $M_-=M_+=1$ and the set of parameters [fNLS1] of Table \ref{tab:param2}. The contact discontinuities separating the 3 plateaus are indicated by red dashed lines. The dashed red arrow indicates the direction of propagation of the solitons.
  (b) Corresponding spatio-temporal diagram in the $(x,t)$-plane. The colour corresponds to the intensity of the wavefield $u(x,t)$. The trajectories $x=c_j t$ of the contact discontinuities are drawn in green dashed lines.}
  \label{fig:examples_NLS}
\end{figure}

\begin{figure}
  \unitlength=1cm
  \begin{picture}(15,8)   
\put(-1,0){\includegraphics[width=17.5cm]{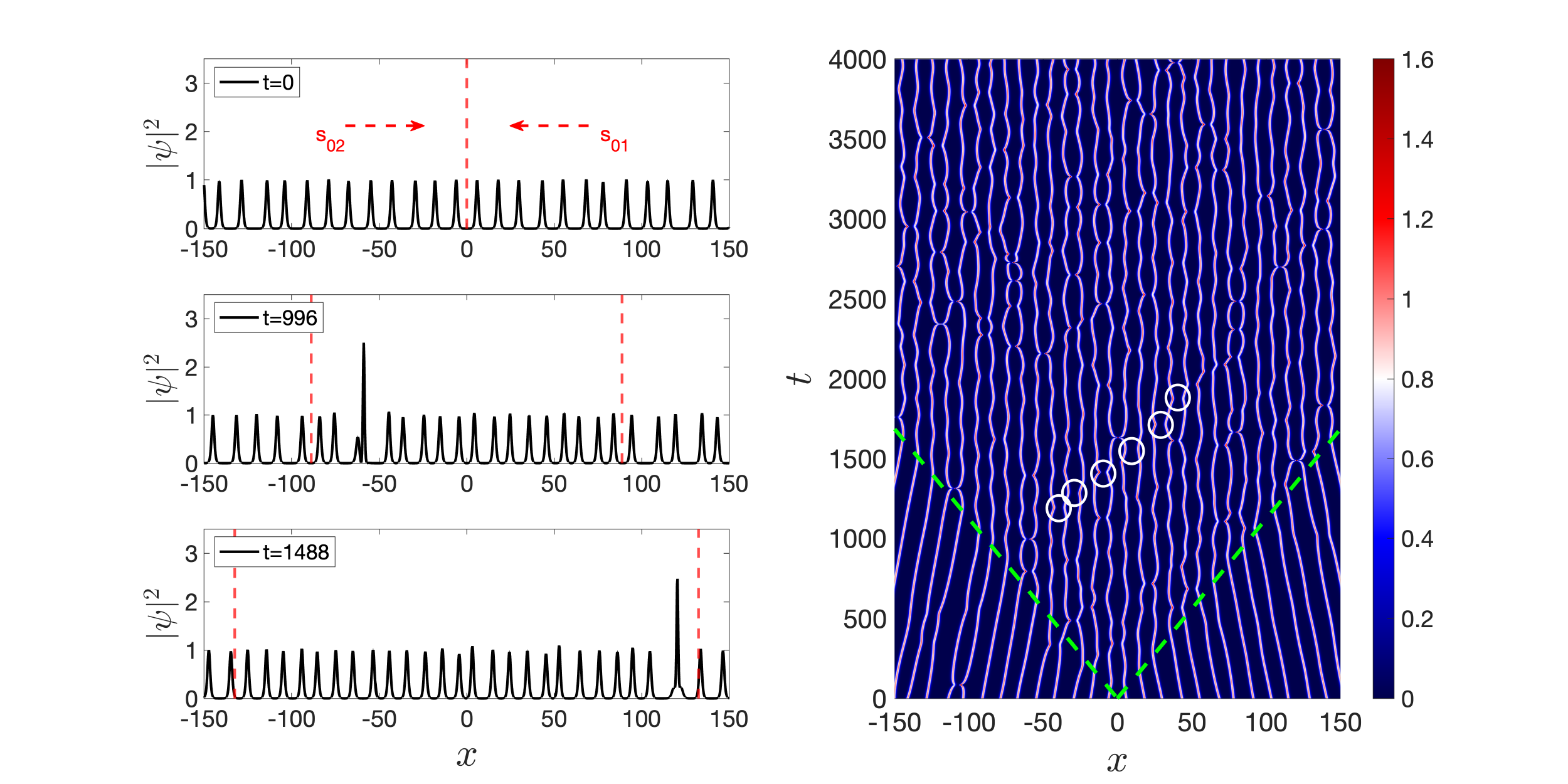}}
\put(0.5,0){\small (a)}
\put(8.2,0){\small (b)}
\end{picture}
 \caption{(a) Snapshots of a realisation $u(x,t)$ of the SG Riemann problem with $M_-=M_+=1$ and the set of parameters [fNLS2] of Table \ref{tab:param2}.
  (b) Corresponding spatio-temporal diagram in the $(x,t)$-plane. The white circles indicate the locations of the successive collisions of one right-propagating soliton with different left-propagating solitons.
  The rest of the legend is defined in Fig. \ref{fig:examples_NLS}. }
  \label{fig:examples_NLS2}
\end{figure}

The spectral parameters $z_1 = \xi_1 + i \eta_1$ and $z_2 = \xi_2 + i \eta_2$ are now complex, allowing for a broader variety of interactions between the solitons compared to KdV SGs: ``amplitudes'' $2\eta_j^2$ can be tuned independently of the soliton free velocities $s_{0j}=-4\xi_j$. Typical evolutions of SG realisations of the Riemann problem with the initial condition \eqref{eq:initmono} for the fNLS equation are displayed in Figs. \ref{fig:examples_NLS} and \ref{fig:examples_NLS2}.

In Fig. \ref{fig:examples_NLS} (the set of parameters [fNLS1] in Table \ref{tab:param2}), a bound state SG ($s_{02}=0$) interacts with a left-propagating SG ($s_{01}<0$). In this case, the slow solitons are the left-propagating ones whereas the fast solitons are the bound states. Similar to the KdV case, the fast solitons, with trajectories depicted by red lines in Fig. \ref{fig:examples_NLS}(b), accelerate, $c_2>s_{02}$ in the interaction region, whereas the slow solitons, with trajectories depicted by light blue lines, decelerate $c_1<s_{01}$. Notice that in that case, even if the $z_2$-component is a bound state SG, it acquires a positive effective velocity due to the interaction with the $z_1$-component. The direct comparison of Fig. \ref{fig:examples_kdv1}(a) (KdV) and Fig. \ref{fig:examples_NLS}(a) (fNLS) shows that the variations of the fields $u(x,t)$ and $|\psi(x,t)|^2$ can be drastically different during the interaction: in the latter case, the local amplitude of the field  can increase quite significantly  during the interaction. Indeed, it is known that interaction of fNLS solitons can lead to a local formation of large-amplitude breathers that are often associated with rogue waves \cite{chabchoub_peregrine_2021, agafontsev_multisoliton_2024} (see also \cite{slunyaev_role_2016} for the related discussion in the focusing mKdV equation context). The fNLS soliton gas interactions then can provide  a possible mechanism of the rogue wave formation.  One can also notice in the spatio-temporal plot of Fig. \ref{fig:examples_NLS}(b) that in the fNLS gas interaction the solitons positions oscillate back and forth during the interaction.

In Fig. \ref{fig:examples_NLS2} (set of parameters [fNLS2] in Table \ref{tab:param2}), two monochromatic SGs with the same amplitude parameter $\eta_1=\eta_2$ but opposite free velocities ($s_{02}=-s_{01}$) are interacting, the case considered theoretically in \cite{el_kinetic_2005} and  realised experimentally in deep water waves in \cite{fache_interaction_2023} (albeit for a different set of parameters).  Similar to the previous case of the fNLS SG interaction, Fig. \ref{fig:examples_NLS2}(a) shows that the  field $|\psi|^2$ can exhibit  local rogue wave type large-amplitude fluctuations  during the interaction. Note that the unusual interaction pattern of the SG rendered in a spatio-temporal plot in Fig. \ref{fig:examples_NLS2}(b) 
could be wrongly  interpreted as non-propagating, bound state gas. This is, of course, not the case. Indeed, following the contact discontinuities (highlighted with the green dash line), one can see the trajectories of the moving solitons initially located at $x=0$: $x=c_j t \neq 0$. Generally, the tracer soliton trajectory inside the interaction region can be identified by following the ``X'' interaction patterns formed during the interaction with other solitons, as highlighted by the sequence of white circles in Fig. \ref{fig:examples_NLS2}(b). The ``stationary'' pattern seen in the spatio-temporal plot is the result of a very strong interaction between counter-propagating fNLS solitons with equal amplitudes and small velocities, the configuration not possible in the KdV case.


\begin{figure}
  \unitlength=1cm
  \begin{picture}(15,16.3)
  \put(0,12){\includegraphics[width=7.5cm]{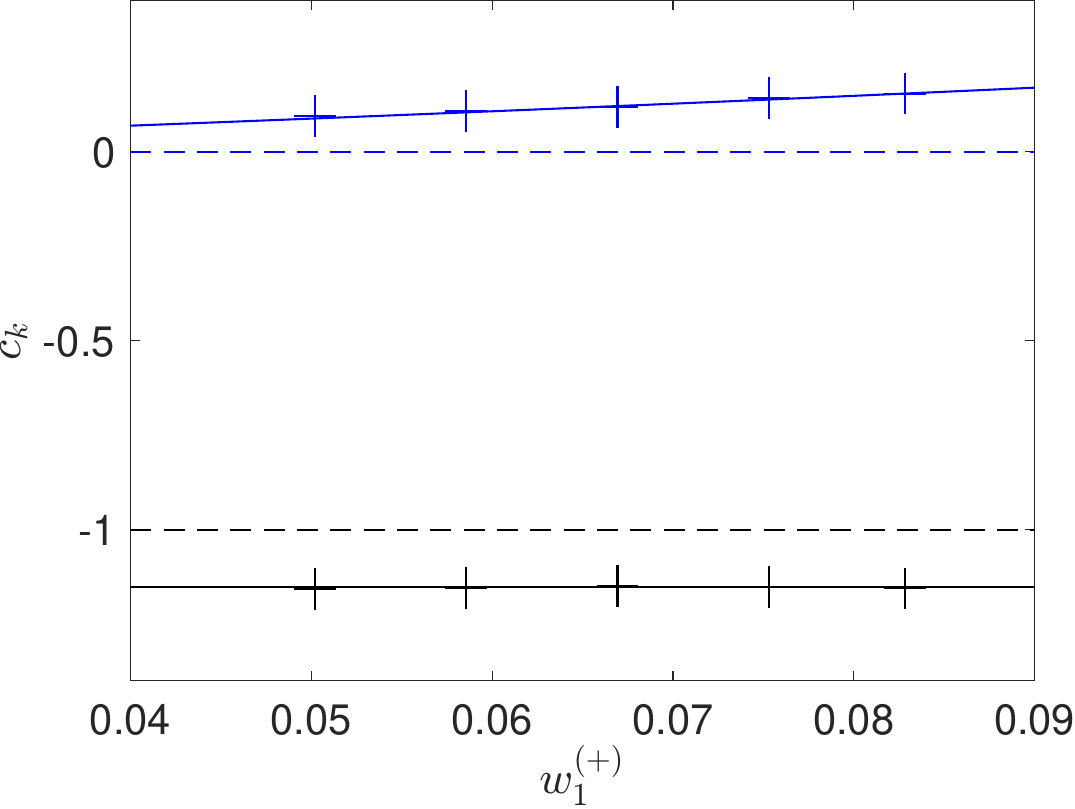}}
\put(8.,12){\includegraphics[width=7.5cm]{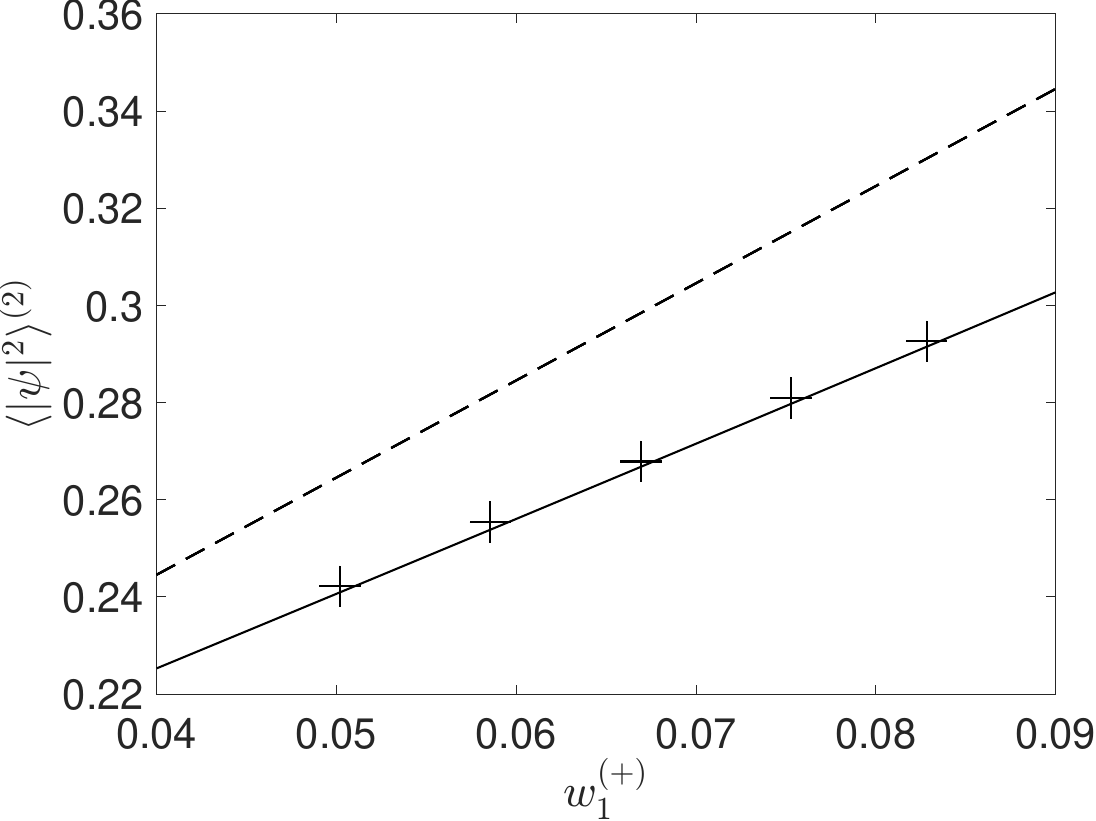}}
\put(0,12){\small (a)}
\put(8.2,12){\small (b)}
  \put(0,6){\includegraphics[width=7.5cm]{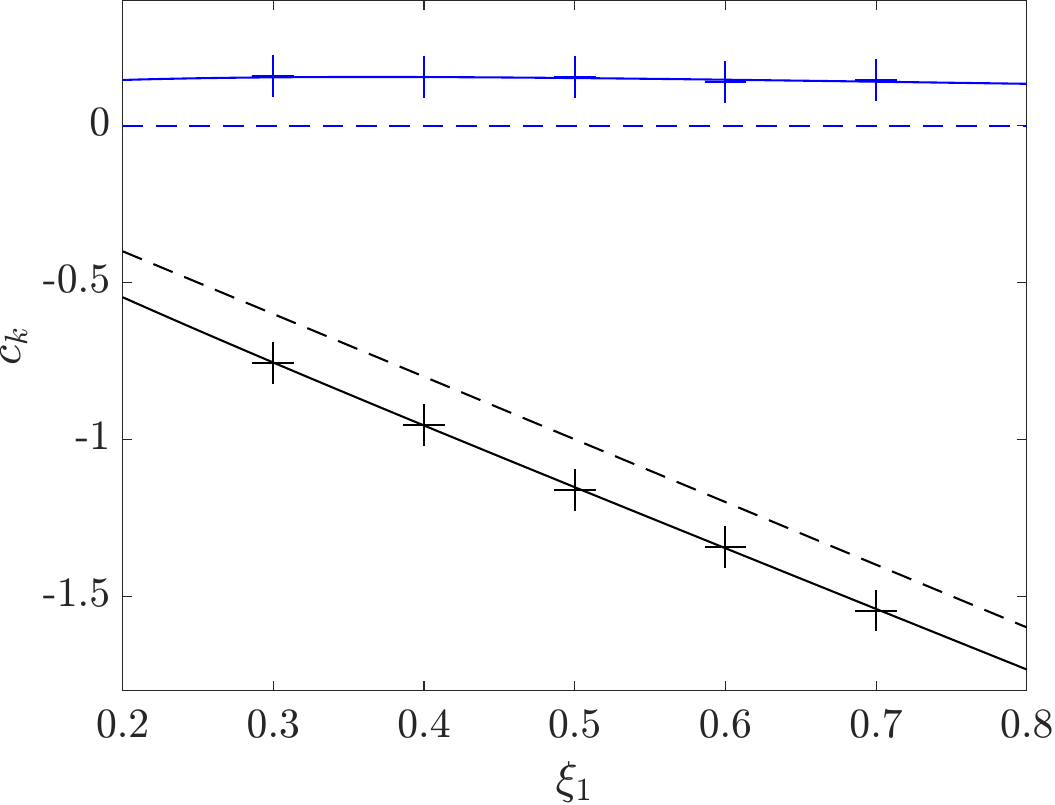}}
\put(8.,6){\includegraphics[width=7.5cm]{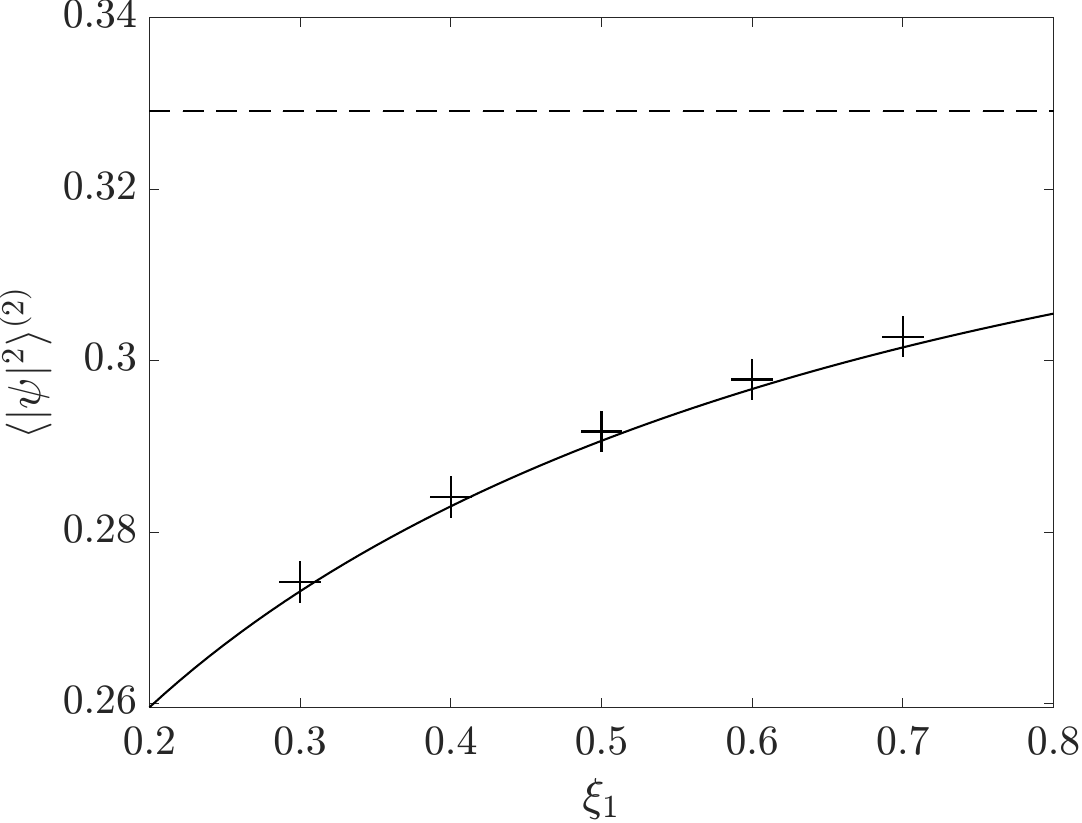}}
\put(0,6){\small (c)}
\put(8.2,6){\small (d)}
\put(0,0){\includegraphics[width=7.5cm]{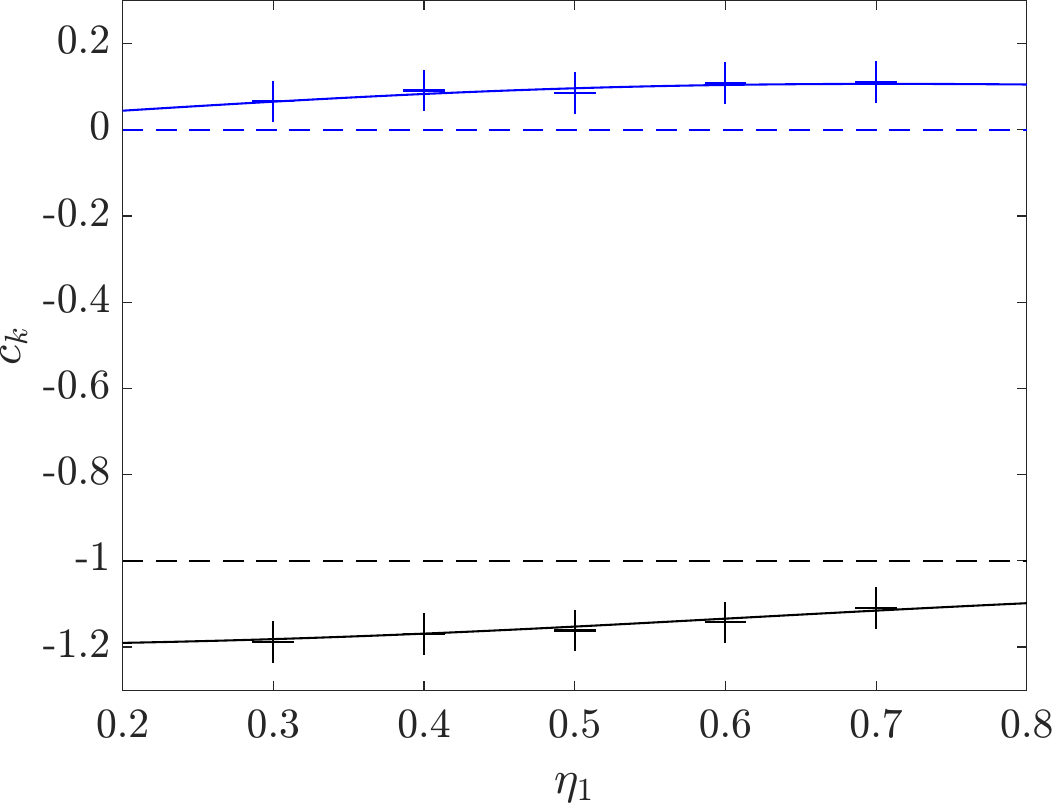}}
\put(8.,0){\includegraphics[width=7.5cm]{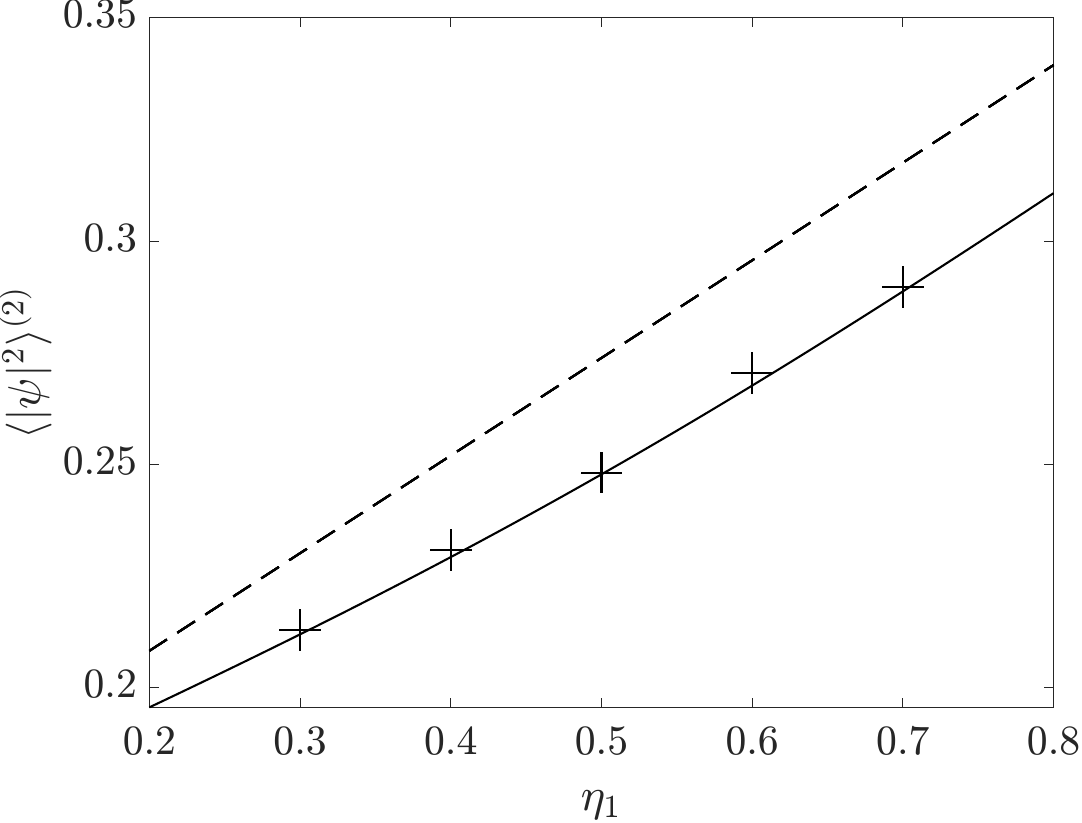}}
\put(0,0){\small (e)}
\put(8.2,0){\small (f)}
\end{picture}
  \caption{Variation of  the speeds $c_k$ of the contact discontinuities and the  moment $\aver{|\psi|^2}^{(2)}$ with respect to (a,b) the initial density $w_1^{(+)}$ for the set of parameters [fNLS3], (c,d) the parameter $\xi_1$ for the set of parameters [fNLS4], and (e,f) the parameter $\eta_1$ for the set of parameters [fNLS5]. The solid lines correspond to the analytical expressions \eqref{eq:c1c2} and \eqref{mom_poly_NLS}, \eqref{eq:w12b} with $s_{0j}=s_0(z_j)$ and $G_{ij}=G_{\rm fNLS}(z_i,z_j)$ given by \eqref{eq:Gnls}. The dashed lines correspond to the same parameters obtained with the reference non-interacting SG solution \eqref{eq:sol_free}, cf. Sec. \ref{sec:riemann}.}
  \label{fig:NLS3}
\end{figure}
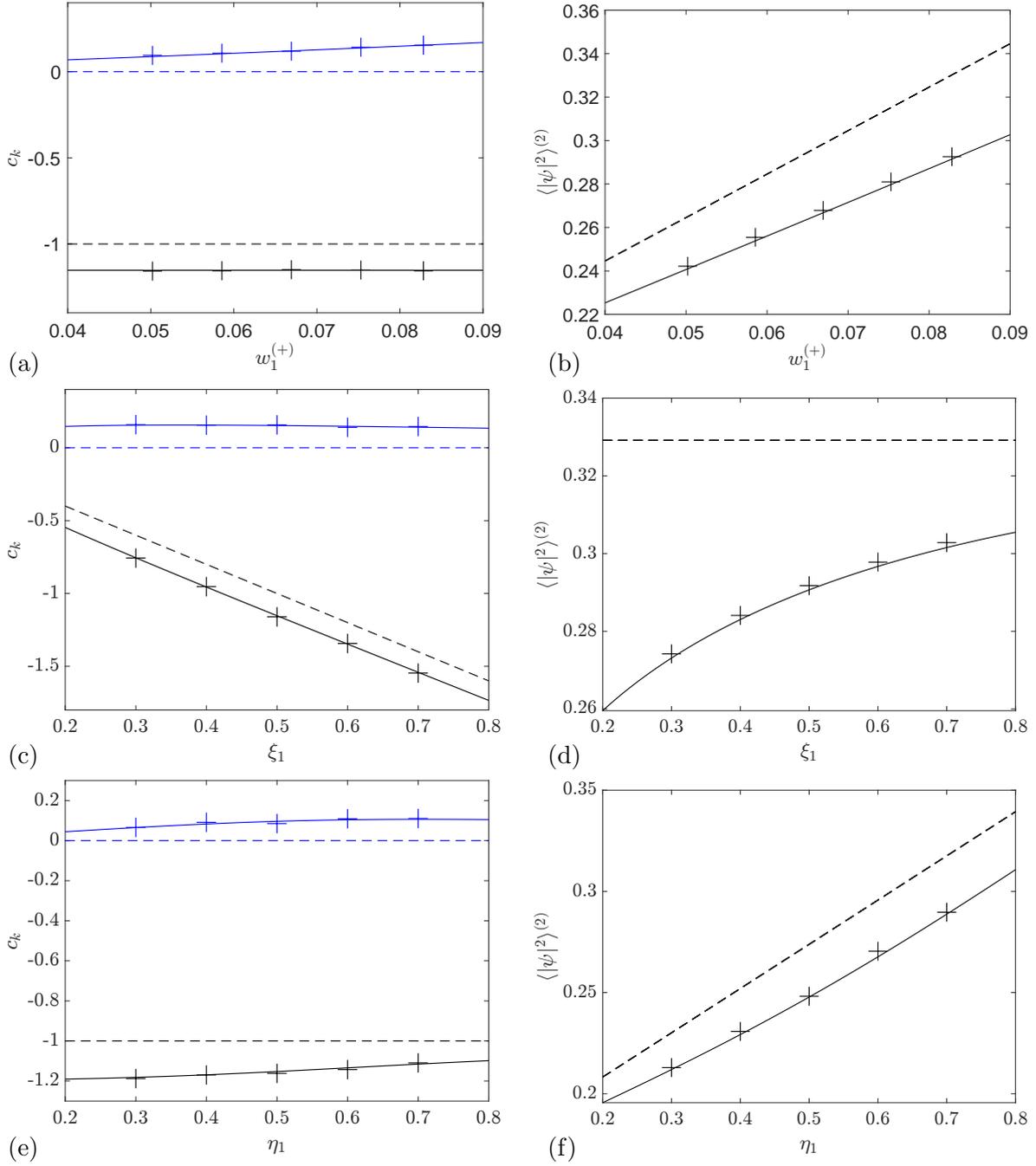

We follow here the procedure described for the KdV problem in Sec. \ref{sec:monokdv}: we first compute the moment $\aver{|\psi(x,t)|^2}$ with $R$ realisations of the SG Riemann problem, and we then extract from this moment the shock speeds $c_j$, as well as the value of moment in the interaction region $\aver{|\psi|^2}^{(2)}$. In order to verify the theoretical predictions, we implement the SG Riemann problem numerically by varying the component's density $w_1^{(+)}$ (set of parameters [fNLS3]), spectral parameters $\xi_1$ (set of parameters [fNLS4]) and $\eta_1$ (set of parameters [fNLS5]). In the three different cases, the SG initially at $x<0$ is a bound state SG ($\xi_2=0$) and the SG initially at $x>0$ is a left-propagating SG ($s_{01} = -4\xi_1 <0$). The analytical expressions of the contact discontinuities speeds are given by \eqref{eq:c1c2} and the statistical moments in the interaction region are given by \eqref{mom_poly_NLS}, \eqref{eq:w12b}, where $G_{12}=G_{\rm fNLS}(z_1, z_2)$ and $G_{21}=G_{\rm fNLS}(z_2, z_1)$ with $G_{\rm fNLS}$ defined in \eqref{eq:Gnls}. 
The comparison between kinetic theory results and numerical results is displayed in Fig. \ref{fig:NLS3}, and shows a good agreement. 

Similar to the KdV case, one can observe in Figs. \ref{fig:NLS3}(b,d,f) that $\aver{|\psi|^2}^{(2)} < \aver{|\psi|^2}_{\rm free}^{(2)}$ where $\aver{|\psi|^2}_{\rm free}^{(2)}$ is obtained by substituting \eqref{eq:sol_free} in \eqref{mom_poly_NLS} (cf. for instance the KdV computation leading to \eqref{eq:sol_moments_free}). Thus 
the interaction between the $z_1$- and $z_2$-solitons also rarefies the fNLS SG in the interaction region $c_1 t<x<c_2 t$. The variations of speeds $c_j$ and moments $\aver{|\psi|^2}^{(2)}$ with density (Figs. \ref{fig:NLS3}(a,b)) and amplitude parameter (Figs. \ref{fig:NLS3}(e,f)) are qualitatively similar to the one determined for the KdV SG Riemann problem (Figs. \ref{fig:kdv1} and \ref{fig:kdv2} respectively). One can notice however in the latter case that the $\aver{|\psi|^2}^{(2)}$ appears to vary monotonically with $\eta_1$ in Fig. \ref{fig:NLS3}(e) contrarily to $\aver{u}^{(2)}$ in Fig. \ref{fig:kdv2}(b).
fNLS SGs offer an additional degree of freedom for the Riemann problem, and the free soliton velocity $s_{01}=-4\xi_1$ can be tuned independently from the parameter $\eta_1$. Interestingly, although the moment $\aver{|\psi|^2}$ does not explicitly depends on $\xi_1$, cf. \eqref{mom_poly_NLS}, Fig. \ref{fig:NLS3}(d) shows that $\aver{|\psi|^2}^{(2)}$ varies with $\xi_1$ due to the interaction between the two components (indeed $w_i^{(2)}$ depends on $G_{ij}$, cf. \eqref{eq:w12b}).

\subsubsection{Collision of  polychromatic fNLS SGs}
\label{sec:binls}

We now consider the collision of a bi-chromatic fNLS SG ($M=2$) and a monochromatic SG implemented with the initial condition \eqref{eq:init3}. As described in Secs. \ref{sec:riemann} and \ref{sec:bikdv}, the Riemann problem solution has now two interactions regions, denoted thereafter (2) and (3). The values of the speeds $c_j$ and the moments $\aver{|\psi|^2}$ predicted by the kinetic theory are given by eqs.~\eqref{eq:z} and \eqref{mom_poly_NLS}, where the densities $w_j^{(k)}$ are obtained by solving the Rankine-Hugoniot conditions \eqref{eq:RH} numerically with the fNLS phase-shift kernel \eqref{eq:Gnls}. 

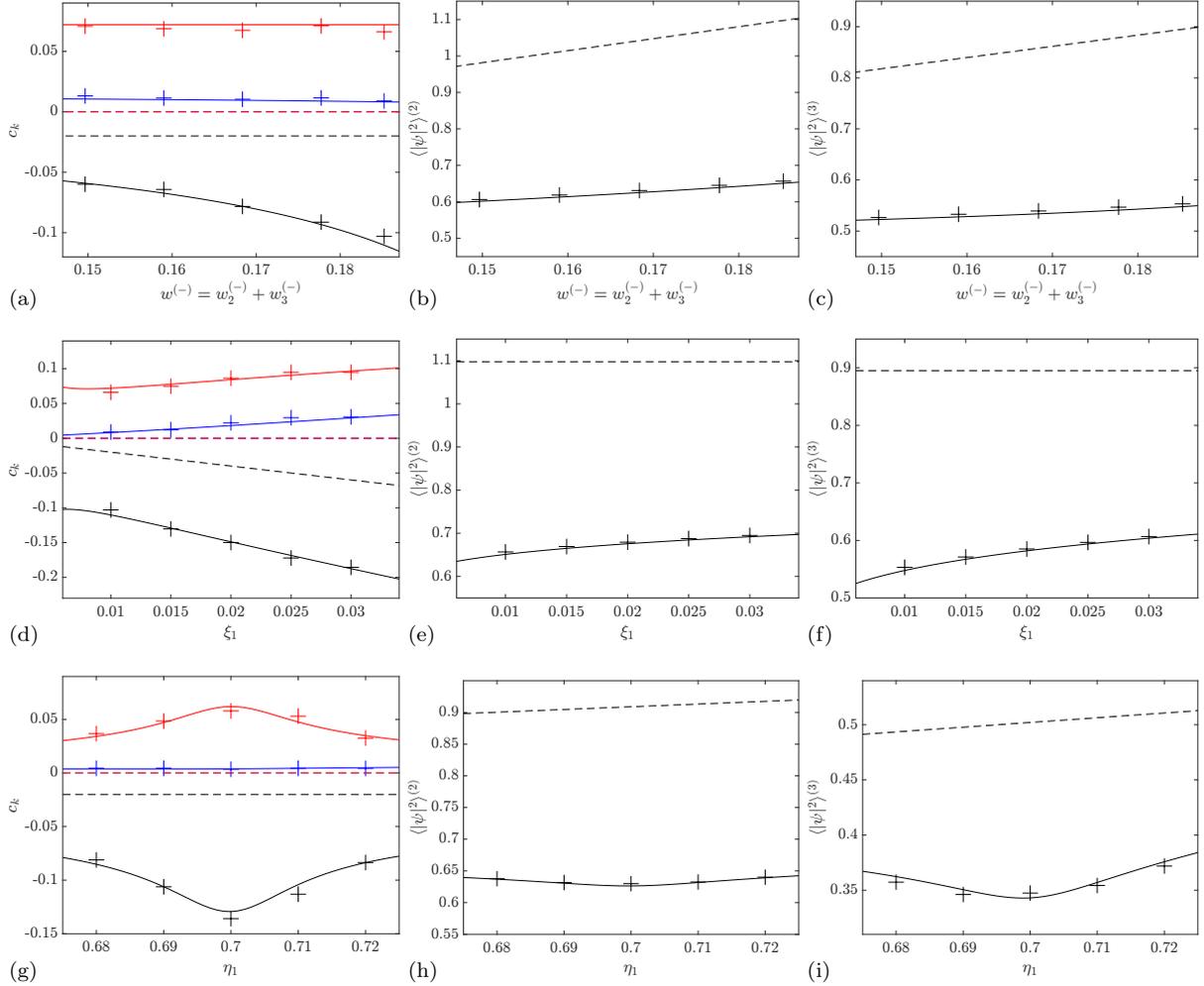
\begin{figure}
  \unitlength=1cm
  \begin{picture}(15,13.5)
\put(0,9){\includegraphics[width=5.2cm]{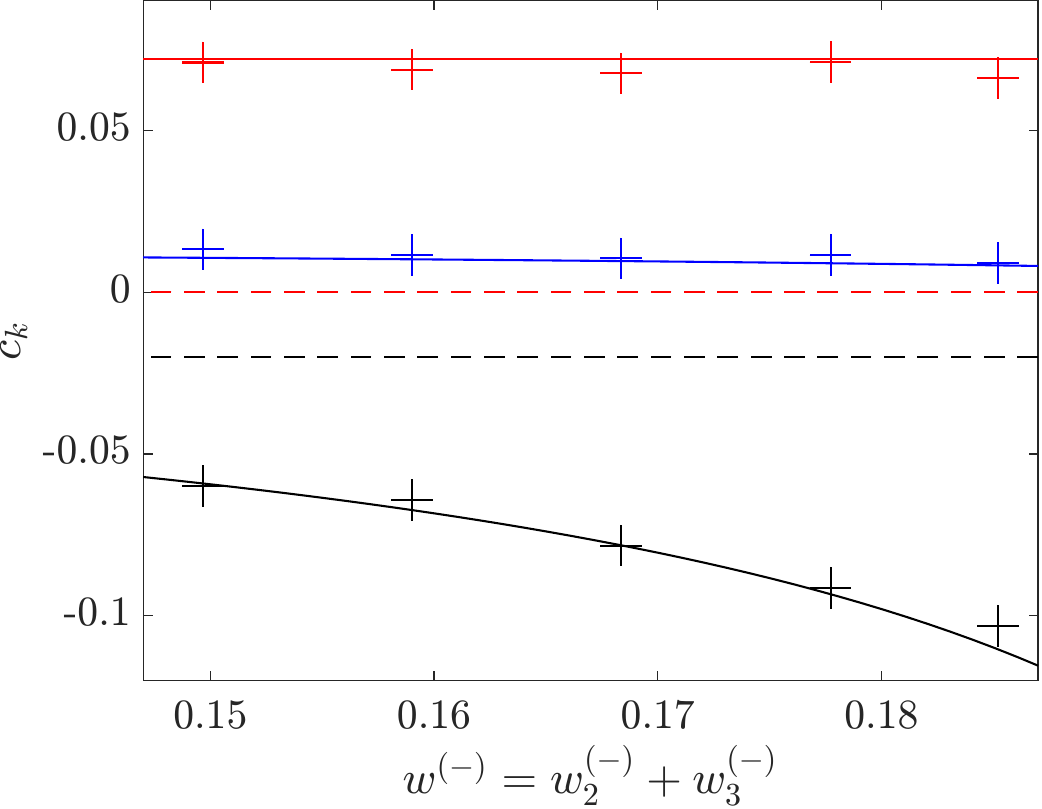}}
\put(5.3,9){\includegraphics[width=5.2cm]{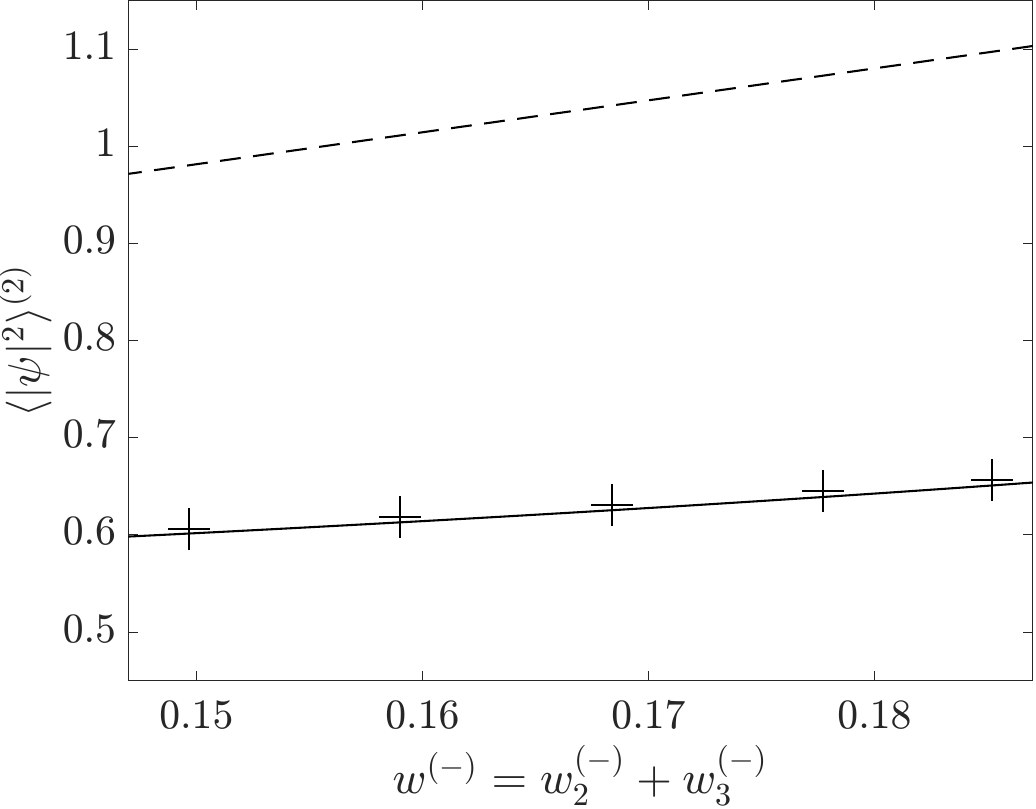}}
\put(10.6,9){\includegraphics[width=5.2cm]{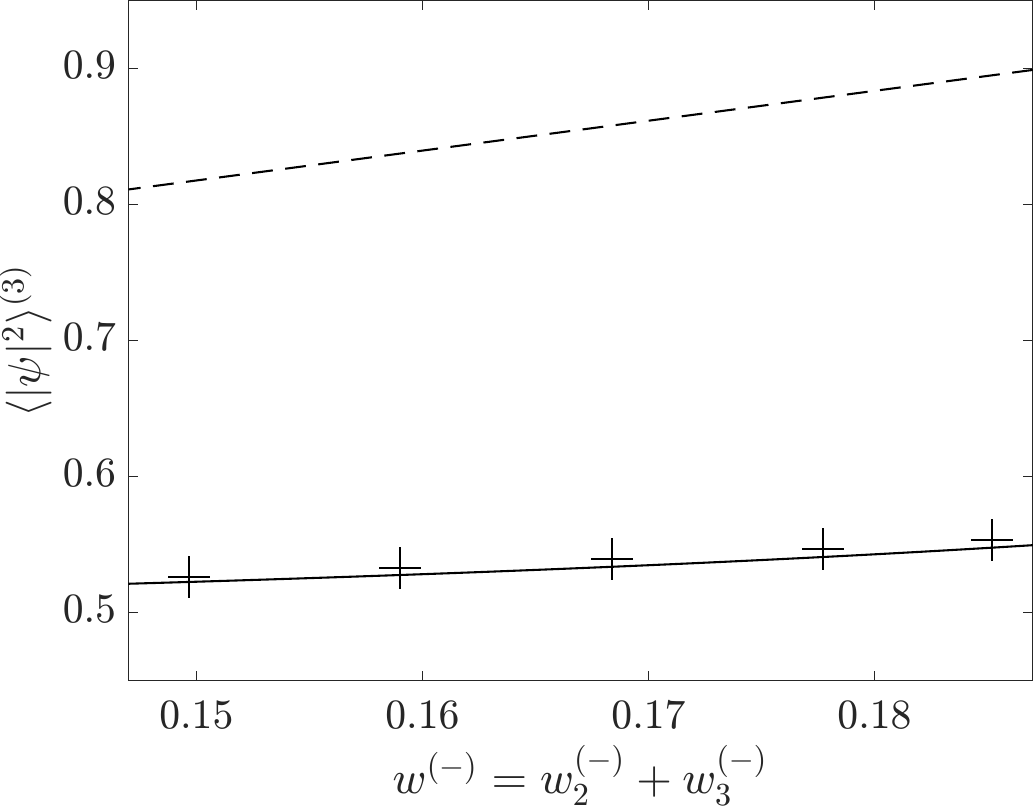}}
\put(0,9){\scriptsize (a)}
\put(5.3,9){\scriptsize (b)}
\put(10.6,9){\scriptsize (c)}
\put(0,4.5){\includegraphics[width=5.2cm]{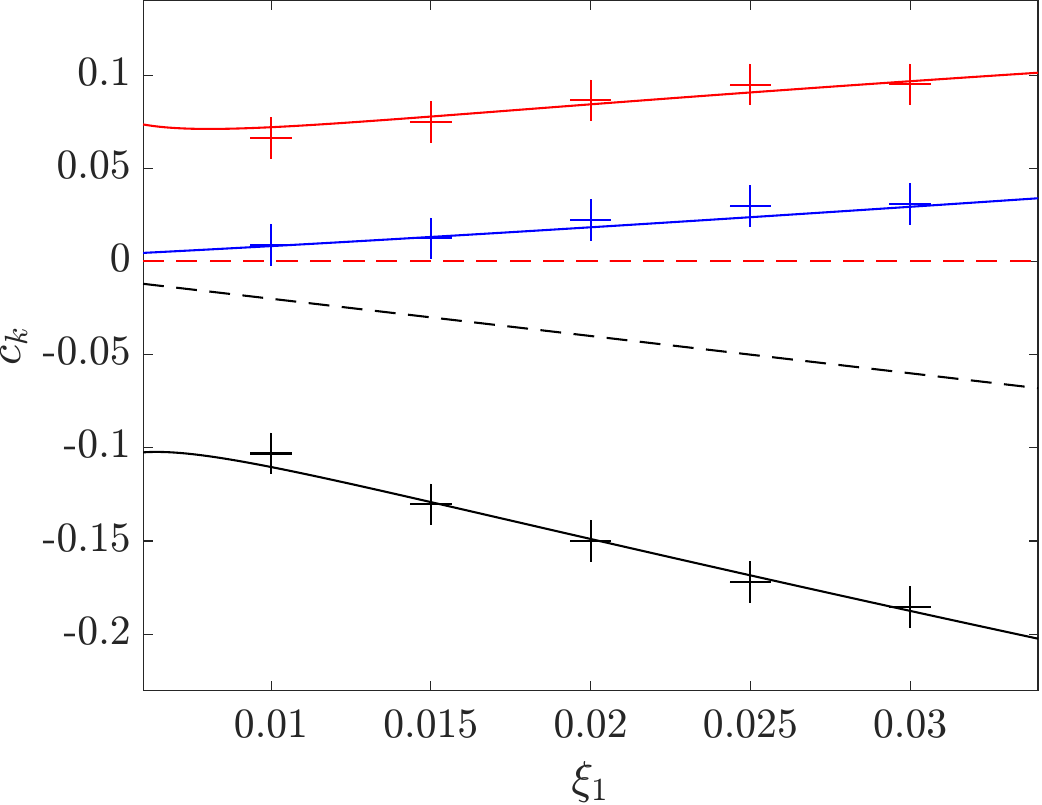}}
\put(5.3,4.5){\includegraphics[width=5.2cm]{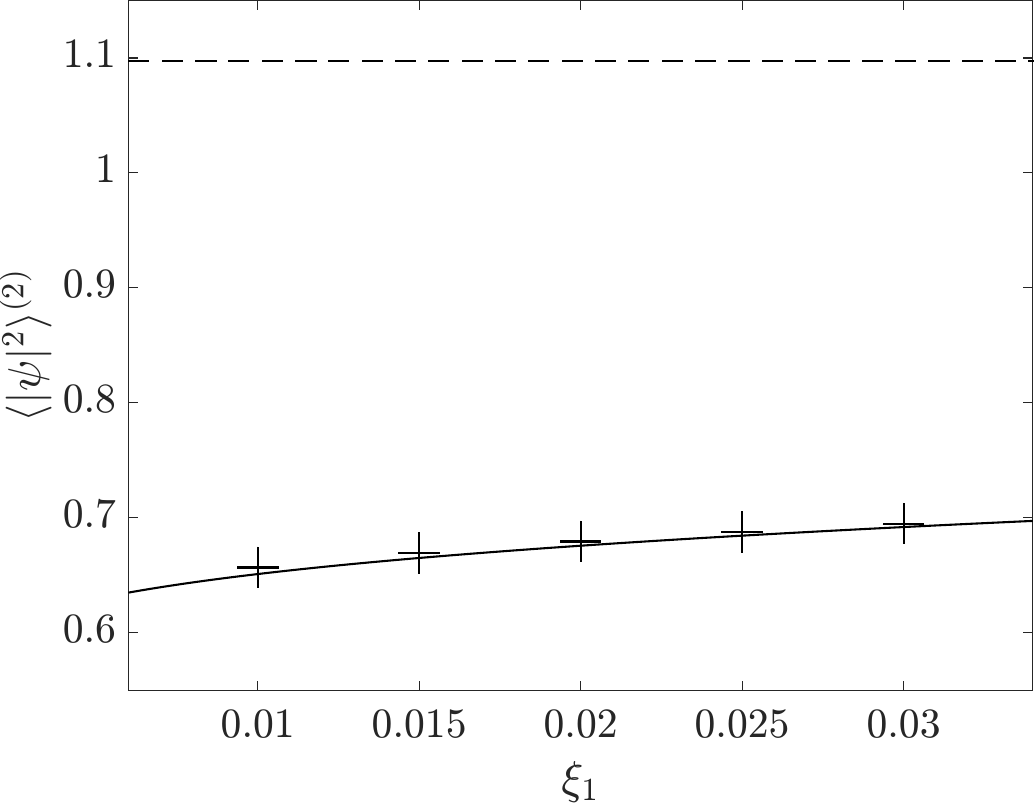}}
\put(10.6,4.5){\includegraphics[width=5.2cm]{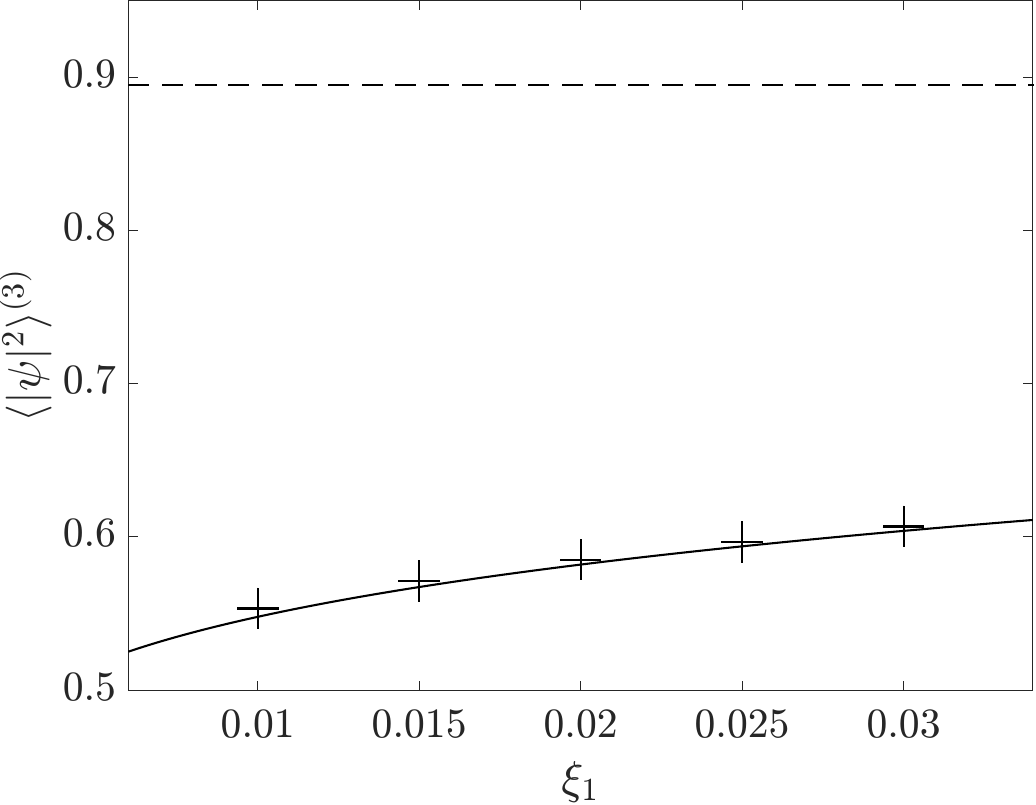}}
\put(0,4.5){\scriptsize (d)}
\put(5.3,4.5){\scriptsize (e)}
\put(10.6,4.5){\scriptsize (f)}
\put(0,0){\includegraphics[width=5.2cm]{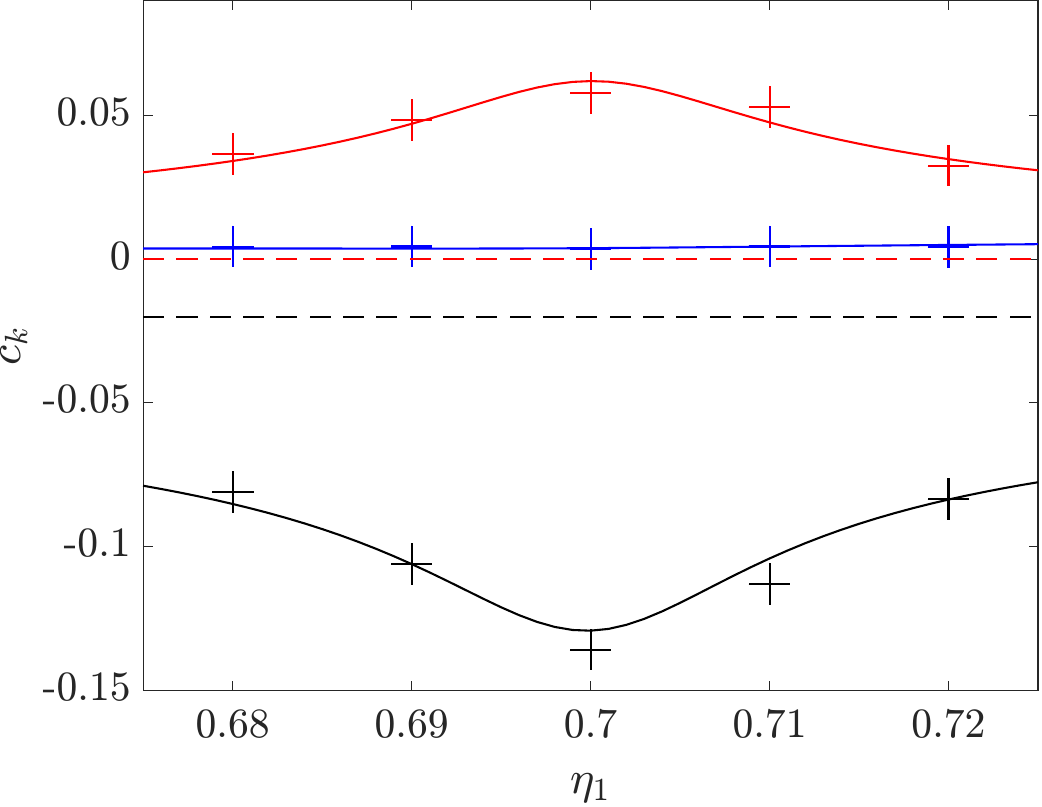}}
\put(5.3,0){\includegraphics[width=5.2cm]{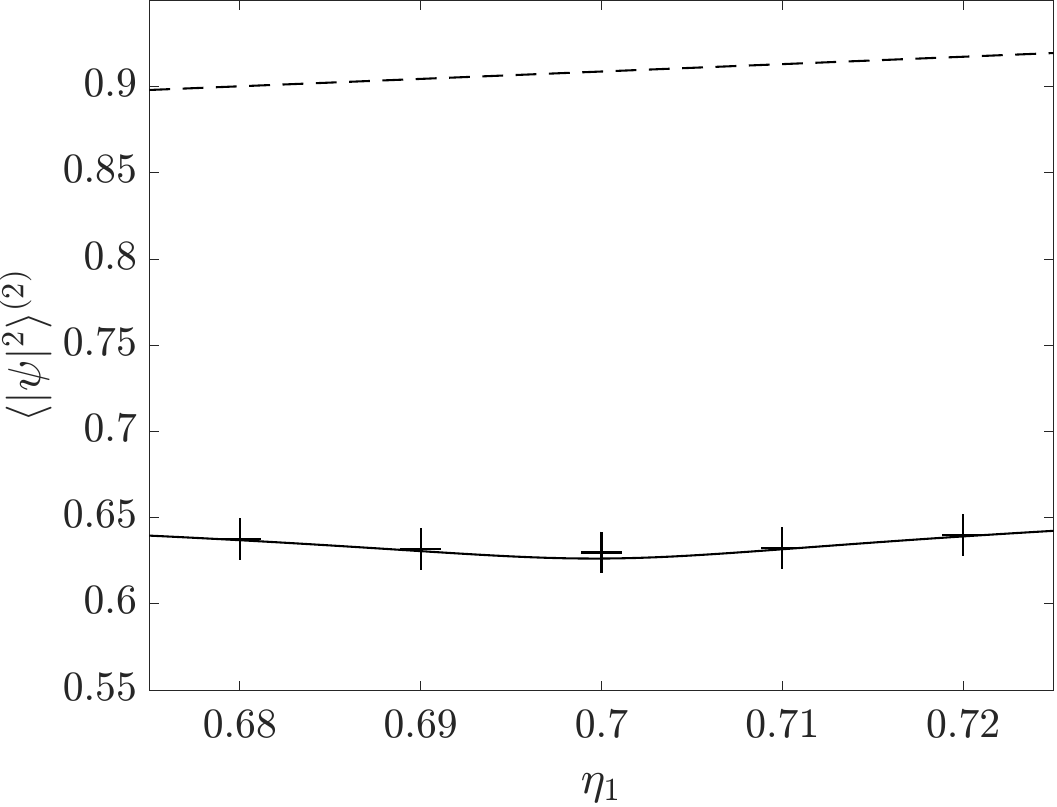}}
\put(10.6,0){\includegraphics[width=5.2cm]{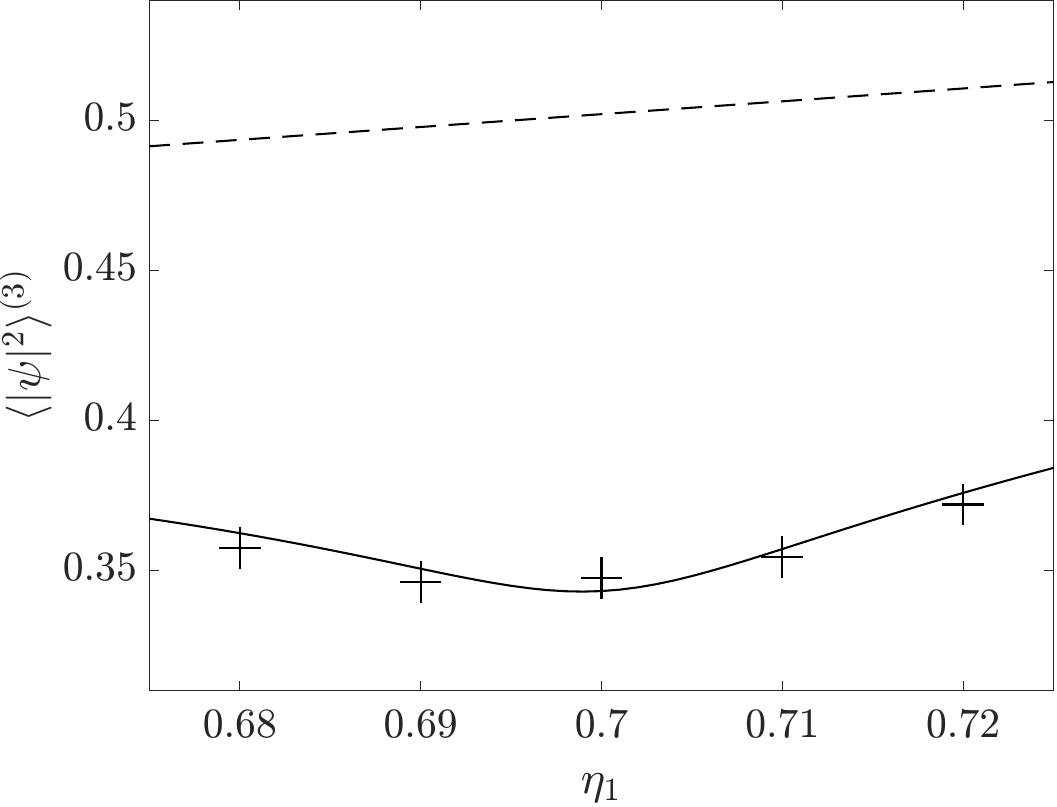}}
\put(0,0){\scriptsize (g)}
\put(5.3,0){\scriptsize (h)}
\put(10.6,0){\scriptsize (i)}
\end{picture}
  \caption{Variation of the contact discontinuities speeds $c_k$ and the  moments $\aver{|\psi|^2}^{(j)}$ with respect to (a,b,c) the initial density $w^{(-)}=w_2^{(-)}+w_3^{(-)}$ for the set of parameters [fNLS6], (d,e,f) the parameter $\xi_1$ for the set of parameters [fNLS7], and (g,h,i) the parameter $\eta_1$ for the set of parameters [fNLS8]. 
 The solid lines are obtained by solving the Rankine-Hugoniot conditions \eqref{eq:RH}, and the markers are extracted from the SG numerical simulations. The dashed lines correspond to the same parameters obtained with the reference solution \eqref{eq:sol_free}; note that the free velocities $s_{02}$ and $s_{03}$ are equal.
 }
 \label{fig:NLS4}
\end{figure}

We implement the SG Riemann problem numerically by varying here the density $w^{(-)}=w_2^{(-)}+w_3^{(-)}$ (set of parameters [fNLS6]), the spectral parameters $\xi_1$ (set of parameters [fNLS7]) and $\eta_1$ (set of parameters [fNLS8]). 
In the three different cases, the SG initially at $x<0$ is a bound state SG ($\xi_2=\xi_3=0$) and the SG initially at $x>0$ is a left-propagating SG ($s_{01} = -4\xi_1 <0$). In particular, the inequality $s_{02}<s_{03}$ no longer holds, i.e. the $z_2$- and $z_3$-solitons have the same free velocity. However, these two components acquire different effective velocities because of the interactions between the solitons, and the labelling $z_2$ and $z_3$ is chosen such that the ordering prescribed by the effective velocities in \eqref{eq:ordering} still holds.  
The comparison between the  results of kinetic theory and numerical results is displayed in Fig. \ref{fig:NLS4}, and shows, once again, a very good agreement. 

One can observe in Figs. \ref{fig:NLS4}(a,d,g) that the interaction between the solitons removes the ``velocity degeneracy'' $s_{02}=s_{03}$ between the two bound state components $z=z_2$ and $z=z_3$ as indicated above. In particular, we have $0<c_2<c_3$, showing that the effective velocity of the $z_2$-solitons in the interactions region (2) is positive but smaller than the effective velocity of the $z_3$-solitons in the interactions region (3), cf. Sec. \ref{sec:riemann}. More generally, we have $G_{31} > G_{21}>0$ for the values of the spectral parameters chosen here, and the phase-shift due to the interaction with $z_1$-solitons is larger for the $z_3$-solitons than for the $z_2$-solitons, resulting in the effective velocity $s_3$ higher than $s_2$.

The variation of the Riemann problem solution's parameters  with the density displayed in Figs. \ref{fig:NLS4}(a,b) is qualitatively similar to the one observed for KdV SGs in Fig. \ref{fig:kdv3}. We observe again in Figs. \ref{fig:NLS4}(e,f) that the moments $\aver{|\psi|^2}^{(j)}$ depend non-trivially on $\xi_1$ due to the interaction between the components. Finally we notice in the last numerical experiment (cf. Figs. \ref{fig:NLS4}(g,h,i)) that the interaction effects are maximised when $\eta_1=0.7=\eta_3$: indeed the distance between the two spectral parameters $z_1$ and $z_3$ is minimal at this value, yielding a larger phase-shift for the interaction between the $z_1$- and $z_3$-solitons.

	\section{Conclusions}
	\label{sec:concl}
In this paper we have performed a systematic comparison of some benchmark results of the kinetic theory of dense soliton gases (SGs) for the KdV and the focusing NLS equation with the results of direct numerical implementation of the relevant SG configurations. Specifically, we considered a Riemann problem for the so-called polychromatic SGs consisting of finite number of  ``monochromatic'' components, each characterised  by nearly identical spectral parameters of the component's solitons. Within the kinetic theory, such polychromatic SGs admit a particularly efficient modelling  by representing the SG density of states (DOS) by a linear superposition of delta functions centered at distinct spectral points (eq.\eqref{u_delta1}) so that the kinetic equation reduces to a  system of hydrodynamic conservation laws \eqref{cont_1}, \eqref{s_alg} with well-defined mathematical properties (Riemann invariants, linear degeneracy, hydrodynamic integrability, see \cite{el_kinetic_2011}). The Riemann problem solution for such a system consists of a finite number of constant states separated by propagating contact discontinuities satisfying appropriate Rankine-Hugoniot conditions. In this paper, we have derived the full system of conditions  defining the weak solutions to the Riemann problems for the collision of two polychromatic SGs with an arbitrary number of components. This (algebraic) system is explicitly resolved for the simplest case of the collision of two monochromatic SGs \cite{el_kinetic_2005, carbone_macroscopic_2016}; in the general case considered here it is solved numerically.

The main contribution of the paper is the direct numerical verification of the spectral kinetic  theory of dense SGs by comparing the predictions for the average field parameters  and the speeds of contact discontinuities in the weak solutions of polychromatic Riemann problems for the SG kinetic equation with the respective physical parameters extracted from the exact (numerical) solutions of the  ``microscopic'', dispersive models---the KdV and the fNLS equations. Although there have been a limited number of previous works dedicated to the numerical validation of the SG kinetic theory (see e.g. \cite{carbone_macroscopic_2016, congy_soliton_2021}), they were all performed in a rarefied gas regime, not providing thus a robust quantitative confirmation of the validity of the kinetic theory in a sufficiently broad range of spectral and density parameters. Given the current rapidly growing interest in SG theory  and applications (see \cite{suret1_soliton_2023} for the broad review of the state-of-the-art in the topic), such a validation is of a paramount importance. 

To achieve a numerical implementation of polychromatic  SGs with $n$-soliton solutions we have developed a novel efficient IST-based algorithm for the synthesis of dense, spatially uniform $n$-soliton ensembles with large $n$ and narrow distributions of the discrete spectrum eigenvalues that model the delta-function DOS ansatz \eqref{u_delta1}. The algorithm employs the previous methods \cite{huang1992darboux, gelash2018strongly} based on the Darboux transformation and the use of high
precision arithmetic routine, combined with the recent developments of generalised hydrodynamics \cite{bonnemain_generalized_2022} and the theory of soliton condensates \cite{congy_dispersive_2023}.  An important technical advantage of the method of the numerical SG synthesis used here is that the evolution in time of SG realisations does not rely on numerical approximations of the KdV or fNLS equations (e.g. with finite difference or spectral methods).  The time variable $t$ plays the role of a parameter in the exact $n$-soliton solution, which is computed, without propagation of errors, at any value of $t$.  

The numerical simulations performed in our work have demonstrated an excellent agreement with the predictions of the spectral theory of dense SGs for the KdV and fNLS equations providing thus a  confirmation of the robustness and the applicability of this theory to physically relevant problems. Indeed,  very recent physical experiments on the interaction of deep water monochromatic SGs \cite{fache_interaction_2023} and on the optical soliton refraction by a SG \cite{suret_soliton_2023} showed some promising applications of the spectral kinetic theory, which could be extended to other physical systems, e.g. Bose-Einstein condensates. Another aspect of SG theory highlighted by our work is that, although the spectral description of the evolution of polychromatic SGs is universal at the level of the hydrodynamic reductions of the kinetic equation (eqs. \eqref{cont_1}, \eqref{s_alg}), the physical wave field behavior in such gases, as revealed by direct numerical simulations of SG solutions to the original dispersive PDEs, can be drastically different for different integrable equations (cf. Fig. \ref{fig:examples_kdv1} (KdV SG) and Fig. \ref{fig:examples_NLS} (fNLS SG)).

The theory and the  method for the numerical synthesis of dense SGs developed in this work can be applied to various SG configurations. One of the perspective areas of the application of the results of this paper  is the study of the interaction of  SGs with dispersive hydrodynamic mean flows such as rarefaction waves and dispersive shock waves---a natural (but  non-trivial) extension of the theory developed in \cite{maiden_solitonic_2018, sprenger_hydrodynamic_2018, ablowitz_solitonmean_2023}. The effects of hydrodynamic  tunnelling   and trapping for individual solitons studied in these papers could be investigated in the context of SGs, where the interactions between solitons could lead to significant modifications of the known soliton-mean flow interaction patterns.  The  theory and the numerical algorithms developed in this paper can prove useful for such a study. Another direction suggested by our work is the study of the rogue wave formation due to the interaction of SGs.

\appendix

\section{Algorithm for the $\bs{n}$-soliton solution }
\label{sec:app_algo}

The algorithm generating the exact $n$-soliton \eqref{eq:soliton-sol}, originally developed in \cite{huang1992darboux},
relies on the Darboux transformation.  This scheme is subject to
roundoff errors during summation of exponentially small and large
values for a large number of solitons $n$. We improve it following
\cite{gelash2018strongly}, with the implementation of  high
precision arithmetic routine to overcome the numerical accuracy problems and
generate solutions with a number of solitons $n \gtrsim 10$.

In order to simplify the algorithm, it is suggested to consider
simultaneously the KdV equation \eqref{eq:kdv} and its equivalent form
\begin{equation}
  u_t-6uu_x+u_{xxx}=0
  \label{eq:kvd_2}
\end{equation}
obtained from  \eqref{eq:kdv} by the reflection $u \to - u$.
The Darboux transformation presented here relates the Jost solution
associated with the $(n-1)$-soliton solution of one equation, to the
$n$-soliton solution of the other equation. 

Considering the direct scattering problem for the Lax pair in the matrix  form
\begin{equation}
  \Phi_{x}=\left(\begin{array}{cc}
      \eta    & \mp 1  \\
      \varphi & - \eta
    \end{array}\right) \Phi,
\end{equation}
with $-1$ corresponding to \eqref{eq:kdv} and $+1$ to \eqref{eq:kvd_2}, the Jost solutions $J,\bar{J}\in \mathbb{R}^{2 \times 2}$ are defined recursively by the Darboux transformations $D(\eta)$ and $\bar{D}(\eta)$ such that:
\begin{equation}
  J_{n}(\eta)=D_{n}(\eta) J_{n-1}(\eta),\quad \textrm{with:} \quad D_{n}(\eta)=I+\frac{2\tilde \eta_{n}}{\eta-\tilde \eta_{n}} P_{n},
\end{equation}
\begin{equation}
  \bar{J}_{n}(\eta)=\bar{D}_{n}(\eta) \bar{J}_{n-1}(\eta), \quad \textrm{with:} \quad \bar{D}_{n}(\eta)=I-\frac{2\tilde \eta_{n}}{\eta+\tilde \eta_{n}} \bar{P}_{n}.
\end{equation}
$P_n(x,t)$ and $\bar{P}_n(x,t)$ are independent of $\eta$ and
have the form:
\begin{equation}
  P_{n}=\sigma_2\bar{P}^{ \textrm{T}}_{n}\sigma_2=\frac{J_{n-1}\left(-\tilde \eta_{n}\right)\left(\begin{array}{c}
        -b_n \\
        1
      \end{array}\right)\left(\begin{array}{ll}
        b_n & 1
      \end{array}\right) \bar{J}_{n-1}^{-1}\left(\tilde \eta_{n}\right)}{\left(\begin{array}{ll}
        b_n & 1
      \end{array}\right) \bar{J}_{n-1}^{-1}\left(\tilde \eta_{n}\right) J_{n-1}\left(-\tilde \eta_{n}\right)\left(\begin{array}{c}
        -b_n \\
        1
      \end{array}\right)},
\end{equation}
with the real constants $b_n$ depending on the spatial phases $x_n^0$ defined in Sec. \ref{sec:nsol} 
\begin{equation}\label{norm_const}
  b_n=\left(-1\right)^{n}\exp\left(2\tilde \eta_n x^{0}_{n}\right).
\end{equation}
The Jost solutions for the initial seed solution $\varphi_0=0$ are
given by
\begin{equation}
  J_{0}(\eta)=\bar{J}_{0}(\eta)=\left(\begin{array}{cc}
      \exp \left[\eta x-4 \eta^{3} t\right] & -\exp \left[-\eta x+4 \eta^{3} t\right]         \\
      0                                     & - 2 \eta \exp \left[-\eta x+4 \eta^{3} t\right]
    \end{array}\right),
\end{equation}
and one can show that at each recursion step
\begin{equation}
  u_{n}=u_{n-1}+4\tilde \eta_{n}\left(P_{n}\right)_{21},
\end{equation}
where $u_n$ is the $n$-soliton solution of \eqref{eq:kdv} for $n$
even and solution of \eqref{eq:kvd_2} for $n$ odd. Recently,  a more efficient and accurate  algorithm has been proposed in \cite{prins_accurate_2021} to generate the $n$-soliton KdV solution employing a $2$-fold Crum transform.

\section{Soliton condensates}
\label{sec:condensate}
The notion of soliton condensate was first introduced in \cite{el_spectral_2020} for the fNLS equation and then thoroughly studied in \cite{congy_dispersive_2023} for the KdV equation. Spectrally, a soliton condensate is realised by vanishing the spectral scaling function $\sigma(\eta)$ \eqref{eq:sigma} so that in the KdV case the condition $\sigma =0$ yields the soliton condensate NDR:
\begin{equation} \label{NDR_cond}
  \int_\Gamma \ln \left|\frac{\mu+\eta}{\mu-\eta}\right|
  f(\mu)d\mu = \eta,
\end{equation}
which is the limit of the basic constraint \eqref{dos_constr}. As follows from eq. \eqref{G_kdv_nls} in Sec. \ref{sec:fNLS}  the  NDR for the special class of the bound state fNLS condensates has the same form \eqref{NDR_cond} with  $\eta$  corresponding to the imaginary part of the fNLS complex spectral parameter $z = \xi_1 + i \eta$, where $\eta \in \Gamma$ and a fixed $\xi_1$  determines the transport velocity $s_0=-4\xi_1$ of the condensate as a whole.

Generally, the spectral support $\Gamma$ in \eqref{NDR_cond} is given by  a union of $M+1$ finite disjoint intervals, $\Gamma =[0,\la_1]\cup [\la_2,\la_3]\cup \dots \cup [\la_{2M}, \la_{2M+1}]$. The solutions $f^{(M)}(\eta; {\bs \la})$ of the condensate NDR \eqref{NDR_cond}   for an arbitrary $M \in \mathbb{N}$ were constructed in \cite{congy_dispersive_2023}. E.g. for the simplest case of genus $0$  condensate one has $\Gamma=[0,\la_1]$ and the NDR \eqref{NDR_cond} is solved by
\begin{equation}\label{uv0}
  f^{(0)}(\eta; \la_1)= \frac{\eta}{\pi\sqrt{\la_1^2-\eta^2}}
\end{equation}
as readily verified by direct substitution. For $M=1$ the condensate DOS is expressed  by the formula
\begin{equation}\label{f1}
f^{(1)}(\eta;\lambda_1,\lambda_2,\lambda_3) = \frac{i\eta(\eta^2-w^2)}{\pi \sqrt{(\eta^2-\lambda_1^2)(\eta^2-\lambda_2^2)(\eta^2-\lambda_3^2)}},
\end{equation}
where
\begin{equation}
 w^2=\lambda_3^2-(\lambda_3^2-\lambda_1^2) \frac{{\rm E} \left(m \right)}{{\rm
        K} \left(m \right)}, \quad
  m = \frac{\lambda_2^2-\lambda_1^2}{\lambda_3^2-\lambda_1^2},
  \end{equation}
and   ${\rm
        K}(m)$, ${\rm
        E}(m) $ are the complete elliptic integrals of the first and the second kind respectively.
        
In the general case the DOS for the genus $M$ soliton condensate is found from the formula (see \cite{congy_dispersive_2023, kuijlaars_minimal_2021} for details)
\begin{equation}\label{fn}
f^{(M)}(\eta; \la_1, \dots \la_{2N+1}) = \frac{iP(\eta)}{2\pi R(\eta)}, 
\end{equation}
where
\begin{equation}
 R(\eta)=\sqrt{ (\eta^2-\lambda_1^2)(\eta^2-\lambda_2^2)\dots (\eta^2-\lambda_{2M+1}^2)}
 \end{equation}
and $P(\eta)$ is an odd monic polynomial of degree $2M+1$  that is chosen so
that 
\begin{equation} \label{dpdq_norm}
  \int \limits_{\la_{2j-1}}^{\la_{2j}} \frac{P(\eta)}{ R(\eta)} d \eta = 0, \quad j=1, \dots, M.
\end{equation}

Typical plots of the  DOS for  the genus $1$ and genus $2$ soliton condensates are shown in Fig. \ref{fig:examples_dos}.
\begin{figure}
  \unitlength=1cm
  \begin{picture}(15,5.5)   
\put(0,0){\includegraphics[width=7.2cm]{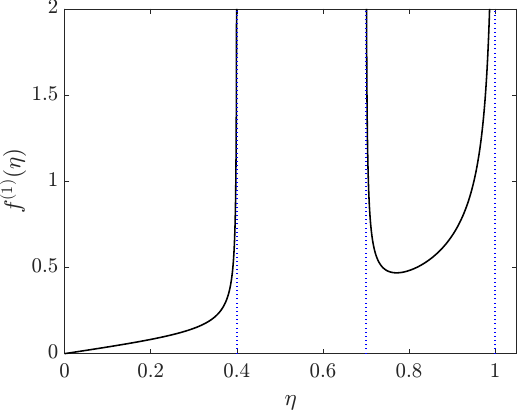}}
\put(8.5,0){\includegraphics[width=7.2cm]{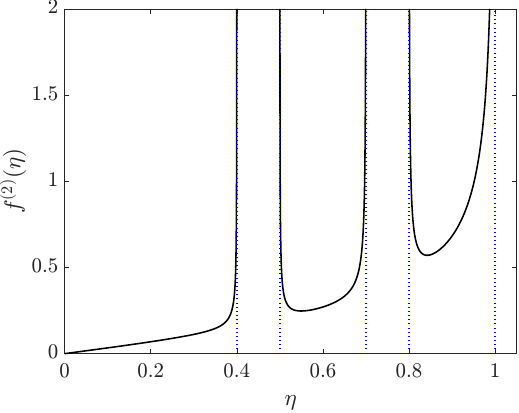}}
\put(0,0){\small (a)}
\put(8.5,0){\small (b)}
\end{picture}
  \caption{Example of DOS for (a) genus 1 condensate with $(\lambda_1,\lambda_2,\lambda_3)=(0.4,0.7,1)$, and (b) genus 2 condensate with $(\lambda_1,\lambda_2,\lambda_3,\lambda_4,\lambda_5)=(0.4,0.5,0.7,0.8,1)$. The blue dashed lines indicate the position $\eta=\lambda_i$.}
  \label{fig:examples_dos}
\end{figure}

One can use the DOS   $f^{(M)}(\eta; {\bs \lambda})$ from \eqref{fn} to compute the ensemble averages $\langle u \rangle$ and 
$\langle u^2 \rangle$ in the genus $M$ KdV soliton condensate using formulae \eqref{eq:averu} (for the fNLS case the  formulae for the ensemble averages of the two first conserved densities are given by \eqref{mom_poly_NLS_bound}). The result is that for the genus zero KdV soliton condensate the variance of the KdV random wave field $\langle u^2 \rangle - \langle u \rangle^2 =0$, which implies that the KdV solution  for the genus zero soliton condensate is almost surely a constant, $u(x,t) = \langle u \rangle = const$ \cite{congy_dispersive_2023} (note that this is not the case for the fNLS soliton condensate, see \cite{gelash2019bound}). The general conjecture formulated in \cite{congy_dispersive_2023} and supported by careful numerical simulations is that  any realisation of the KdV soliton condensate of genus $M$ with the DOS $f^{(M)}(\eta; {\bs \lambda})$ given by \eqref{fn} almost surely coincides  with  $n$-gap  KdV solution associated with the hyperelliptic Riemann surface of $\sqrt{(\eta^2-\lambda_1^2)(\eta^2-\lambda_2^2)\dots(\eta^2-\lambda_{2M+1}^2)}$.

\section{Comparison between condensate DOS and uniform DOS}
\label{sec:comparison}

We compare here the implementation of the DOS $f(\eta)=w_1 \delta(\eta-\eta_1)$ for KdV SGs via the two DOS: 
\begin{align}
	\label{eq:fcond}
	&f_{\rm cond.}(\eta) = C f^{(1)}(\eta;\lambda_1=0,\lambda_2=\eta_1-\varepsilon,\lambda_3=\eta_1+\varepsilon),\quad 0<C<1,\\
	\label{eq:funif}
	&f_{\rm unif.}(\eta) = w_1\begin{cases}
		1/(2\varepsilon), &|\eta-\eta_1|<\varepsilon,\\
		0, &|\eta-\eta_1|>\varepsilon.
	\end{cases}
\end{align}
For monochromatic SGs, the spatial density is equal to the component density: $\kappa=\int_\Gamma f(\eta) \rmd \eta = w_1$.
Both DOS have the same  compact support $\Gamma = [\eta_1-\varepsilon,\eta_1+\varepsilon]$. 

\subsection{Numerical implementations}
\label{sec:implementation}

The numerical implementation of the monochromatic SG using the condensate DOS \eqref{eq:fcond} is described in detail in Sec. \ref{sec:RP_num}. We have:
\begin{equation}
\kappa = C \kappa^{(1)},\quad \phi_{\rm cond.}(\eta)= \frac{f^{(1)}(\eta)}{\kappa^{(1)}},
\end{equation}
where
\begin{equation}
	\label{fcondmax}
	\kappa^{(1)} = \frac{\eta_1 +\varepsilon }{\pi} \left[{\rm K}\left(\frac{4 \varepsilon  \eta_1 }{(\varepsilon +\eta_1 )^2}\right) \left(\frac{{\rm E}\left(\frac{(\varepsilon -\eta_1 )^2}{(\varepsilon +\eta_1 )^2}\right)}{{\rm K}\left(\frac{(\varepsilon -\eta_1 )^2}{(\varepsilon +\eta_1 )^2}\right)}-1\right)+{\rm E}\left(\frac{4 \varepsilon  \eta_1 }{(\varepsilon +\eta_1 )^2}\right)\right].
\end{equation}
In particular, we obtain the relation $f_{\rm cond.}(\eta) = \kappa \phi_{\rm cond.}(\eta)$.

\begin{figure}
	\centering
	\includegraphics[width=8cm]{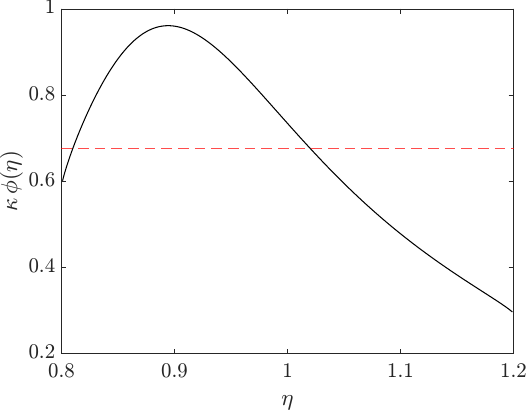}
	\caption{Comparison between $\kappa \,\phi_{\rm unif.}(\eta)$ \eqref{eq:ksunif}, \eqref{eq:phiunif} in solid black line and the DOS $f_{\rm unif.}(\eta)$ \eqref{eq:funif} in dashed red line for $\eta_1=1$ and $\varepsilon=0.2$.}
	\label{fig:comparison2}
\end{figure}

This equality no longer holds for the implementation via the uniform distribution \eqref{eq:funif}: substituting \eqref{eq:funif} in \eqref{eq:kappas}, \eqref{eq:sigma}, we obtain  density of the spatial phases
\begin{equation}
\label{eq:ksunif}
\kappa_s = \frac{\kappa}{2\varepsilon} \int_\Gamma \frac{\rmd \eta}{1-\kappa g(\eta)} ,
\end{equation}
which can be expressed analytically using the identity:
\begin{equation}
	\label{intg}
	g(\eta) = \frac{1}{2\varepsilon}\int_{a}^{b} G(\eta,\mu) \rmd\mu = \frac{1}{2\varepsilon} \left[2 \ln\left( \frac{b-\eta}
	{\eta-a} \right)+ \frac{\eta+b}{\eta} \ln\left( \frac{\eta+b}
	{b-\eta} \right) -  \frac{\eta+a}{\eta} \ln\left( \frac{\eta+a}
	{\eta-a} \right) \right],
\end{equation}
where $a=\eta_1 - \varepsilon$ and $b=\eta_1 + \varepsilon$.
The distribution for the spectral parameters $\tilde \eta_i$ of the $n$-soliton solution is given by \eqref{eq:rho}:
\begin{equation}
\label{eq:phiunif}
\phi_{\rm unif.}(\eta) = \left(\int_\Gamma \frac{ \rmd \mu}{1-\kappa g(\mu)} \right)^{-1} \frac{1}{1-\kappa g(\eta)} .
\end{equation}
In this case, $f_{\rm unif.}(\eta) = w_1/2\varepsilon \neq \kappa \phi_{\rm unif.}(\eta)$ as shown in Fig. \ref{fig:comparison2} where $\eta_1=1$ and $\varepsilon=0.2$ is chosen not too small to highlight the discrepancy between the two distributions.

\begin{figure}
\unitlength=1cm
  \begin{picture}(15,6.3)   
\put(0.3,0){\includegraphics[width=7.5cm]{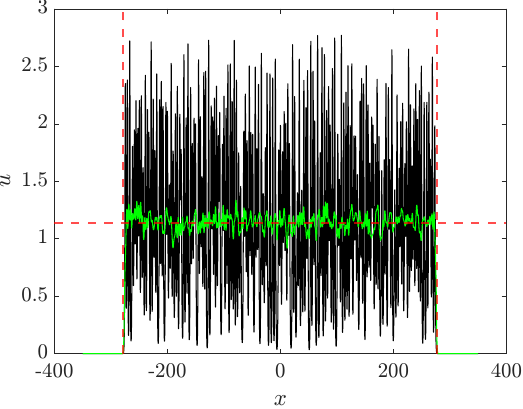}}
\put(8.3,0){\includegraphics[width=7.5cm]{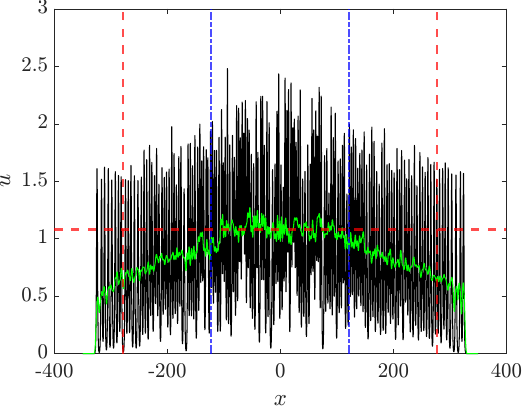}}
\put(0,0){\small (a)}
\put(8.3,0){\small (b)}
\end{picture}
	\caption{Numerical implementation of SGs with the DOS defined by \eqref{eq:fcond} (a) and \eqref{eq:funif} (b) with $\varepsilon=0.2$ and $\kappa=0.27$. The solid black lines correspond to one numerical realisation, i.e. $n$-soliton solution with $n=150$. The green lines correspond to the moment $\aver{u}$ obtained with an average of $10$ realisations and a spatial average of width $L=5$, cf. Appendix \ref{sec:aver}. The dashed horizontal red line red lines correspond to the moment $\aver{u}$ computed analytically with \eqref{eq:averu}. The dashed vertical red line correspond to the positions $|x|=\ell/2=\kappa^{-1}n/2 $. The dash-dotted vertical blue lines correspond to $|x|=\overline{\ell}/2$, cf. \eqref{eq:l2}, and enclose the region where the distribution \eqref{eq:funif} is correctly implemented numerically.}
	\label{fig:comparison3}
\end{figure}

Fig. \ref{fig:comparison3} displays the comparison between the numerical implementation of the SG with the dilute condensate DOS \eqref{eq:fcond} and the uniform DOS \eqref{eq:funif}. In both cases the gases are dense with a density $\kappa=w_1=0.27$ close to the critical density defined in the section \ref{sec:critical}, in order to highlight the differences between the two implementations. To validate the numerical implementations, we compare the moments $\aver{u(x,t)}$ obtained numerically with their analytical values \eqref{eq:averu}, \eqref{eq:fcond} and \eqref{eq:averu}, \eqref{eq:funif}; in particular the moments should be spatially uniform since the DOS are spatially uniform. In the implementation of the diluted condensate DOS, the moment compares well with the analytical value on the entire interval $[-\ell/2,\ell/2]$ with $\ell = \kappa^{-1} n$; the $n$-soliton solution exponentially decays to~$0$ outside this interval. 
This is no longer the case in the implementation of the uniform DOS \eqref{eq:funif}, where $\aver{u}$ compares to the analytical value on a smaller interval $[-\overline{\ell}/2,\overline{\ell}/2]$ with $\overline{\ell}<\ell$, and decreases outside this interval. 

This discrepency can be explained by the difference between $f_{\rm unif.}(\eta)$ and $\kappa \phi_{\rm unif.}(\eta)$, as depicted in Fig. \ref{fig:comparison2}. By construction of the $n$-soliton solution (cf. Sec. \ref{sec:nsol}), the number of solitons with a spectral parameter $\tilde \eta_i \in [\eta_0,\eta_0+\rmd \eta]$  implemented numerically is given by: $n\, \phi(\eta_0) \rmd \eta$. 
This number coincides with the SG quantity $\ell f(\eta_0) \rmd \eta$, corresponding to the number of solitons in a spatial interval of size $\ell = \kappa^{-1} n$ with a parameter  $\in [\eta_0,\eta_0+\rmd \eta]$, only if $\phi(\eta_0) = \kappa^{-1}f(\eta_0)$. In the example presented here, $\phi_{\rm unif.}(\eta_0)$ is significantly lower than $\kappa^{-1}f(\eta_0)$ for $\eta_0$ close to $\eta_{\min}=1.2$ (cf. Fig. \ref{fig:comparison2}), and we can assume that there is not enough solitons with $\tilde \eta_i \sim \eta_{\min}$ to approximate the SG on the whole interval $[-\ell/2,\ell/2]$. 
We can estimate the size $\overline{\ell}$ by supposing that there are no solitons with a spectral parameter $\tilde \eta_i \in[\eta_{\min}-\rmd \eta,\eta_{\min}]$, i.e. the lowest probable spectral parameter, outside the spatial interval  $[-\overline{\ell}/2,\overline{\ell}/2]$. Inside the interval $[-\overline{\ell}/2,\overline{\ell}/2]$, the uniform SG is correctly implemented, and we have the equality $n\, \phi(\eta_{\rm min}) \rmd \eta = \overline{\ell} f(\eta_{\rm min}) \rmd \eta$, which yields the expression
\begin{equation}
\label{eq:l2}
\overline{\ell} = \frac{\phi(\eta_{\rm min})}{f(\eta_{\rm min}) } n,
\end{equation}
where $\eta_{\rm min}$ is the location of the minimum of $\phi(\eta)$. This estimate is in good agreement with the numerical observation as depicted in Fig. \ref{fig:comparison3}. If $f(\eta)= \kappa \phi(\eta)$, $\overline{\ell}$ trivially reduces to $\ell$, which is the case for the soliton condensate \eqref{eq:fcond}. This motivates here the choice of the condensate DOS for the implementation of the Riemann problem as highlighted in Sec. \ref{sec:RP_num}.

Note that the non-uniformity observed in the implementation of the $n$-soliton solution is only manifested when the gas is very dense, i.e. when the denominator $1-\int_\Gamma G(\eta,\mu) f(\mu) \rmd \mu$ in \eqref{eq:rho} is close to $0$ for certain values of $\eta$. If $f(\eta)$ is sufficiently small on the whole spectral support $\Gamma$, \eqref{eq:rho} can be expanded with respect to $f(\eta)$ such that $\phi(\eta)$ is proportional to $f(\eta)+ {\cal O}(f(\eta)^2)$, and the implemented SG is almost uniform on $[-\ell/2,\ell/2]$.

\subsection{Critical density}
\label{sec:critical}

\begin{figure}
	\centering
	\includegraphics[width=8cm]{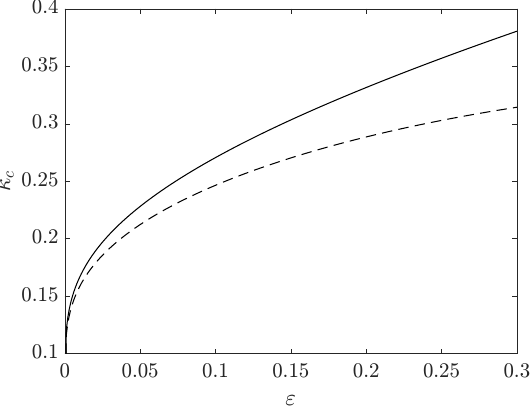}
	\caption{Variation of the maximum densities $\kappa=\kappa^{(1)}$ (solid line) and $\kappa=1/g(\eta_{\rm max})$ with the width of the spectral support $\varepsilon$; $\eta_1=1$ in this example.}
	\label{fig:comparison1}
\end{figure}

Since the DOS $f(\eta)$ is bounded by the inequality \eqref{dos_constr}, SG cannot be arbitrarily dense and the spatial density $\kappa$ is bounded by a critical, maximal spatial density $\kappa_c$. 
For the condensate distribution \eqref{eq:fcond}, the critical density is by definition $\kappa_c=\kappa^{(1)}$, which is obtained when $C=1$ (i.e. soliton condensate).
For the uniform distribution \eqref{eq:funif}, the critical density is given by
\begin{equation}
	\label{maxdens}
	\kappa_c =\frac{1}{g(\eta_{\max})},
\end{equation}
where $\eta_{\max}$ is the location of the maximum of the function $g(\eta)$ defined by \eqref{intg}; $\eta_{\max}$ solves the nonlinear equation:
\begin{equation}
	\label{etamax}
	\left(\frac{\eta_{\rm max}+a}{\eta_{\rm max}-a}\right)^a=\left(\frac{\eta_{\rm max}+b}{b-\eta_{\rm max}}\right)^b.
\end{equation}
Fig. \ref{fig:comparison1} displays the variation of the critical densities for the two DOS \eqref{eq:fcond} and \eqref{eq:funif} with $\varepsilon$ at $\eta_1=1$ fixed. One can notice that we always have $1/g(\eta_{\rm \max})<\kappa^{(1)}$, i.e. the highest possible spatial density $\kappa$ is reached with the dilute condensate implementation.

\section{Averaging procedure}
\label{sec:aver}

We illustrate the averaging procedure for the moment $\aver{u}$ of the KdV SG; the same  technique is used for the moments of the fNLS SG, replacing the KdV realisations average of \eqref{eq:averu} by the fNLS realisations average \eqref{mom_poly_NLS}.
On a sufficiently large spatial scale $L$ (much larger than the typical soliton width), the nonlinear wave field in a SG represents an ergodic random function. The ergodicity property implies that ensemble
averages can be replaced by spatial averages. In practice, the statistical moments such as
$\aver{u}$ are obtained via the ``double averaging''
\begin{equation}
	\label{eq:aver_num}
	\aver{u(x,t)} = \frac{1}{L} \int_{x-L/2}^{x+L/2}\left( \frac{1}{R}\sum_{i=1}^R u_n(x,t;\tilde{\bs{\eta}},\bs{x}^0_i) \right) \rmd x,
\end{equation}
with $\tilde{\bs{\eta}}= (\eta_1,\dots,\tilde \eta_n)$ the set of spectral parameters defined by \eqref{eq:phi},  $\bs{x}^0_i = (x^0_{i,1},\dots,x^0_{i,n})$ the set of random spatial phases uniformly distributed in the interval $I_s$ defined in \eqref{eq:kappas} and $R$ the number of realisations. As discussed in Sec. \ref{sec:nsol}, we made the choice to have the same set of spectral parameters $\tilde{\bs{\eta}}$ for each realisation, whereas the spatial phases in $\bs{x}^0_i$ are randomly distributed for each $i \in \{1,\dots,R\}$. In addition, the components of angular phase vector $\bs{\theta}^0_i = (\theta^0_{i,1},\dots,\theta^0_{i,n})$ (see Sec. \ref{sec:fNLS}) are also randomly distributed on $[0,2\pi)$ for each realisation of the fNLS SG.

Combining \eqref{eq:averu} and \eqref{eq:sol}, the moment $\aver{u}$ should have $M=M_-+M_+$ discontinuities. Because of the additional spatial averaging, the discontinuities at $x=c_k t$ are replaced by lines of finite slope in the region $x \in [c_k\, t-L/2,c_k\, t+L/2]$:
\begin{equation}
	\label{eq:aver_num2}
		\aver{u(x,t)} =
		\begin{cases}
			\aver{u^{(-)}} = \aver{u^{(1)}} , &x<c_1 \,t,\\[2mm]
			\dots\\[2mm]
			\dfrac{\aver{u^{(k)}}+\aver{u^{(k-1)}}}{2}+\dfrac{\aver{u^{(k)}}-\aver{u^{(k-1)}}}{L} (x-c_{k-1}\,t), & c_{k-1}\, t- \dfrac{L}{2} < x < c_{k-1}\, t+\dfrac{L}{2},\\[2mm]
			\aver{u^{(k)}}, & c_{k-1}\, t +\dfrac{L}{2} < x < c_k\, t-\dfrac{L}{2},\\[2mm]
			\dfrac{\aver{u^{(k+1)}}+\aver{u^{(k)}}}{2}+\dfrac{\aver{u^{(k+1)}}-\aver{u^{(k)}}}{L} (x-c_{k}\,t), & c_{k}\, t-\dfrac{L}{2} < x < c_{k}\, t+\dfrac{L}{2},\\[2mm]
			\dots\\[2mm]
			\aver{u^{(+)}} = \aver{u^{(M+1)}}, &c_M \,t < x,
		\end{cases}\quad
\end{equation}
where $\aver{u^{(k)}} = \sum_{j=1}^M 4 \eta_j w_j^{(k)}$. Although $L$ should be sufficiently large to achieve ergodicity, $L< \min_k (c_k-c_{k-1}) t$ so the ``discontinuities'' in \eqref{eq:aver_num2} can still be identified after the averaging procedure; in practice we choose $L={\cal O}(10)$.

In order to compute the moments $\aver{u^{(k)}}$ and velocities $c_k$ numerically, the average \eqref{eq:aver_num} is evaluated in the long time regime where the time $t$ is defined by: 
\begin{equation}
	\label{eq:times}
	(c_M-c_1)t \in \{ 100,200 \}.
\end{equation}
We choose here to fix the width of the interaction region $x \in [c_1\, t,c_M\, t]$ rather the time $t$ to ensure that there are sufficiently many soliton collisions.
We then extract the moments and speeds using the least square fitting with the ansatz \eqref{eq:aver_num2} where $\aver{u^{(k)}}$ and $x_k(t) = c_k\, t$ are the fitting parameters.
Because of the discrepancy in the position of the initial step of the order $\max_j \eta_j^{-1}$ (see Sec. \ref{sec:RP_num2}), the extracted position of the discontinuity $x_k(t)$ has an error of the same order. The speed $c_k$ is thus evaluated by computing the positions $x_k(t)$ at two different large times $1 \ll t_1 < t_2$: 
\begin{equation}
c_k = \frac{x_k(t_2)-x_k(t_1)}{t_2-t_1}.
\end{equation} 
$t_1$ and $t_2$ are given by \eqref{eq:times} to eliminate the initial position discrepancy.

\end{document}